%% file: main.tex
\documentclass{article}[a4paper]
\usepackage{array, tabularx, caption, boldline}
\usepackage[dvipsnames]{xcolor}
\usepackage{graphicx}
\usepackage{makecell}
\usepackage{cellspace}
\usepackage[utf8]{inputenc}
\usepackage{lmodern,textcomp}
\usepackage{hyperref}
\usepackage{multicol}
\hypersetup{
    colorlinks=true,
    linkcolor=blue,
    filecolor=magenta,      
    urlcolor=blue,
    citecolor=blue
}

\usepackage{float}
\usepackage{cite}
\usepackage{ifthen, mciteplus} 
\usepackage{geometry}

\usepackage{amssymb}

\usepackage{url}
\usepackage{placeins}
\newboolean{pdflatex}
\setboolean{pdflatex}{true} 

\newboolean{articletitles}
\setboolean{articletitles}{true} 
\newboolean{uprightparticles}
\setboolean{uprightparticles}{false} 
\newboolean{inbibliography}
\setboolean{inbibliography}{true} 

\usepackage{verbatim}                             
\usepackage{microtype}
\usepackage{lineno}  
\usepackage{xspace} 
\usepackage{caption} 

\newcommand{\BFBcTauNu}{\mathcal{B}(B_c^+ \to \tau^+ \nu_\tau)}
\newcommand{\BFBTauNu}{\mathcal{B}(B^+ \to \tau^+ \nu_\tau)}
\newcommand{\BTauNu}{B^+ \to \tau^+ \nu_\tau}
\newcommand{\BcTauNu}{B_c^+ \to \tau^+ \nu_\tau}
\newcommand{\TauThreePi}{\tau^+ \to \pi^+ \pi^+ \pi^- \bar{\nu}_\tau}

\newcommand{\BkgFluc}{\sigma_\text{bkg}^\text{fluc}}

\newcommand{\revision}[1]{\textcolor{black}{#1}}
\newcommand{\rerevision}[1]{\textcolor{black}{#1}}
 
\title{Prospects for $B_c^+$ and $B^+\to \tau^+ \nu_\tau$ at FCC-ee}
\author{Xunwu Zuo$^{1}$, Marco Fedele$^{2}$, Cl\'ement Helsens$^{1,3}$\footnote{This author was affiliated with these institutes during the preparation of the paper, now affiliated with Institute of Bioengineering, Ecole Polytechnique Fédérale de Lausanne, Lausanne, Switzerland.}, Donal Hill$^{4}$, Syuhei Iguro$^{2,5}$, \\Markus Klute$^{1}$ }
\date{}

\include{symbols}

\begin{document}

\maketitle

\begin{center}{\footnotesize \it
\noindent
$^{1}$ Institut f\"ur Experimentelle Teilchenphysik (ETP), Karlsruhe Institute of Technology, D-76131 Karlsruhe, Germany\\
$^{2}$ Institut f\"ur Theoretische Teilchenphysik (TTP), Karlsruhe Institute of Technology, D-76131 Karlsruhe, Germany\\
$^{3}$ European Organization for Nuclear Research (CERN), Geneva, Switzerland \\
$^{4}$ Institute of Physics, École Polytechnique Fédérale de Lausanne (EPFL), Lausanne, Switzerland\\
$^{5}$ Institut f\"ur Astroteilchenphysik (IAP),
Karlsruhe Institute of Technology, D-76344 Eggenstein-Leopoldshafen, Germany}
\vspace{0.5cm}

{\footnotesize
{{Email:~}}{\bf\color{blue} 
marco.fedele@kit.edu,
clement.helsens@epfl.ch, 
donal.hill@epfl.ch,
igurosyuhei@gmail.com,
markus.klute@cern.ch,
xunwu.zuo@cern.ch}
}
\end{center}

\begin{flushleft}
{\footnotesize
{Preprint numbers:~}
TTP23-015, 
P3H-23-28 
}
\end{flushleft}
\bigskip

\hrule
\begin{abstract}\noindent
The prospects are presented for precise measurements of the branching ratios of the purely leptonic $\BcTauNu$ and $\BTauNu$ decays at the Future Circular Collider (FCC).    
This work is focused on the hadronic $\tau^{+} \to \pi^+ \pi^+ \pi^- \bar{\nu}_\tau$ decay in both $\BcTauNu$ and $\BTauNu$ processes.
Events are selected with two Boosted Decision Tree algorithms to optimise the separation between the two signal processes as well as the generic hadronic $Z$ decay backgrounds.
The range of the expected precision for both signals are evaluated in different scenarios of non-ideal background modelling.
This paper demonstrates, for the first time, that the $\BTauNu$ decay can be well separated from both $\BcTauNu$ and generic $Z\to b\bar{b}$ processes in the FCC-ee collision environment and proposes the corresponding branching ratio measurement as a novel way to determine the CKM matrix element $|V_{ub}|$.  
The theoretical impacts of both $\BTauNu$ and $\BcTauNu$ measurements on New Physics cases are discussed for interpretations in the generic Two-Higgs-doublet model and leptoquark models. 
\end{abstract}
\hrule




\input{introduction}
\input{analysis}

\input{interpretation}

\input{conclusion}

\input{internal}

\bibliographystyle{LHCb}
\bibliography{references}

\end{document}

%% file: symbols.tex
\usepackage{xspace} 
\usepackage{upgreek}
\usepackage{amsmath}
\usepackage{tikz}

\def\delphes {\mbox{\textsc{DELPHES}}\xspace}

\def\MagUp {\mbox{\em Mag\kern -0.05em Up}\xspace}

\ifthenelse{\boolean{uprightparticles}}%
{

 \def\Pnu         {\ensuremath{\upnu}\xspace}

 \def\Ptau        {\ensuremath{\uptau}\xspace}

 \def\PDelta      {\ensuremath{\Delta}\xspace}                 
 \def\PXi         {\ensuremath{\Xi}\xspace}                 
 \def\PLambda     {\ensuremath{\Lambda}\xspace}                 
 \def\PSigma      {\ensuremath{\Sigma}\xspace}                 
 \def\POmega      {\ensuremath{\Omega}\xspace}                 
 \def\PUpsilon    {\ensuremath{\Upsilon}\xspace}

 \def\PB      {\ensuremath{\mathrm{B}}\xspace}                 
                  
 \def\PD      {\ensuremath{\mathrm{D}}\xspace}

 \def\PK      {\ensuremath{\mathrm{K}}\xspace}

 \def\Pc      {\ensuremath{\mathrm{c}}\xspace}

 \def\Pi      {\ensuremath{\mathrm{i}}\xspace}

 \def\Ps      {\ensuremath{\mathrm{s}}\xspace}

 \def\thebaroffset{0.0em}
}
{

 \def\Pnu         {\ensuremath{\nu}\xspace}

 \def\Ptau        {\ensuremath{\tau}\xspace}

 \mathchardef\PDelta="7101
 \mathchardef\PXi="7104
 \mathchardef\PLambda="7103
 \mathchardef\PSigma="7106
 \mathchardef\POmega="710A
 \mathchardef\PUpsilon="7107
                  
 \def\PB      {\ensuremath{B}\xspace}                 
                  
 \def\PD      {\ensuremath{D}\xspace}

 \def\PK      {\ensuremath{K}\xspace}

 \def\Pc      {\ensuremath{c}\xspace}

 \def\Pi      {\ensuremath{i}\xspace}

 \def\Ps      {\ensuremath{s}\xspace}

 \def\thebaroffset{0.18em}
}
\newcommand{\offsetoverline}[2][\thebaroffset]{\kern #1\overline{\kern -#1 #2}}%

\makeatletter
\ifcase \@ptsize \relax
  \newcommand{\miniscule}{\@setfontsize\miniscule{4}{5}}
\or
  \newcommand{\miniscule}{\@setfontsize\miniscule{5}{6}}
\or
  \newcommand{\miniscule}{\@setfontsize\miniscule{5}{6}}
\fi
\makeatother

\DeclareRobustCommand{\optbar}[1]{\shortstack{{\miniscule (\rule[.5ex]{1.25em}{.18mm})}
  \\ [-.7ex] $#1$}}

\def\taup       {{\ensuremath{\Ptau^+}}\xspace}

\def\neu        {{\ensuremath{\Pnu}}\xspace}

\def\neut       {{\ensuremath{\neu_\tau}}\xspace}

\def\squark    {{\ensuremath{\Ps}}\xspace}

\def\cquark    {{\ensuremath{\Pc}}\xspace}

\def\KorKbar {\kern \thebaroffset\optbar{\kern -\thebaroffset \PK}{}\xspace}

\def\Dbar    {{\ensuremath{\offsetoverline{\PD}}}\xspace}

\def\DorDbar {\kern \thebaroffset\optbar{\kern -\thebaroffset \PD}\xspace}

\def\Dzb     {{\ensuremath{\Dbar{}^0}}\xspace}

\def\B       {{\ensuremath{\PB}}\xspace}

\def\BorBbar {\kern \thebaroffset\optbar{\kern -\thebaroffset \PB}\xspace}

\def\Bd      {{\ensuremath{\B^0}}\xspace}

\def\BdorBdbar {\kern \thebaroffset\optbar{\kern -\thebaroffset \Bd}\xspace}
\def\Bu      {{\ensuremath{\B^+}}\xspace}

\def\Bs      {{\ensuremath{\B^0_\squark}}\xspace}

\def\BsorBsbar {\kern \thebaroffset\optbar{\kern -\thebaroffset \Bs}\xspace}
\def\Bc      {{\ensuremath{\B_\cquark^+}}\xspace}

\def\Y#1S{\ensuremath{\PUpsilon{(#1S)}}\xspace}

\def\LorLbar     {\kern \thebaroffset\optbar{\kern -\thebaroffset \PLambda}\xspace}

\def\ra                 {\ensuremath{\rightarrow}\xspace}
\def\to                 {\ensuremath{\rightarrow}\xspace}

\def\Vub  {{\ensuremath{V_{\uquark\bquark}}}\xspace}
\def\Vcb  {{\ensuremath{V_{\cquark\bquark}}}\xspace}

\def\AT#1     {\ensuremath{A_{\mathrm{T}}^{#1}}\xspace}           

\def\C#1      {\ensuremath{\mathcal{C}_{#1}}\xspace}                       
\def\Cp#1     {\ensuremath{\mathcal{C}_{#1}^{'}}\xspace}                    
\def\Ceff#1   {\ensuremath{\mathcal{C}_{#1}^{\mathrm{(eff)}}}\xspace}        
\def\Cpeff#1  {\ensuremath{\mathcal{C}_{#1}^{'\mathrm{(eff)}}}\xspace}       
\def\Ope#1    {\ensuremath{\mathcal{O}_{#1}}\xspace}                       
\def\Opep#1   {\ensuremath{\mathcal{O}_{#1}^{'}}\xspace}                    

\newcommand{\aunit}[1]{\ensuremath{\text{\,#1}}}       

\newcommand{\tev}{\aunit{Te\kern -0.1em V}\xspace}
\newcommand{\gev}{\aunit{Ge\kern -0.1em V}\xspace}
\newcommand{\mev}{\aunit{Me\kern -0.1em V}\xspace}
\newcommand{\kev}{\aunit{ke\kern -0.1em V}\xspace}
\newcommand{\ev}{\aunit{e\kern -0.1em V}\xspace}
\newcommand{\mevc}{\ensuremath{\aunit{Me\kern -0.1em V\!/}c}\xspace}
\newcommand{\gevc}{\ensuremath{\aunit{Ge\kern -0.1em V\!/}c}\xspace}
\newcommand{\mevcc}{\ensuremath{\aunit{Me\kern -0.1em V\!/}c^2}\xspace}
\newcommand{\gevcc}{\ensuremath{\aunit{Ge\kern -0.1em V\!/}c^2}\xspace}

\def\m    {\aunit{m}\xspace}

\def\gsim{{~\raise.15em\hbox{$>$}\kern-.85em
          \lower.35em\hbox{$\sim$}~}\xspace}
\def\lsim{{~\raise.15em\hbox{$<$}\kern-.85em
          \lower.35em\hbox{$\sim$}~}\xspace}

\def\evtgen     {\mbox{\textsc{EvtGen}}\xspace}

\def\pythia     {\mbox{\textsc{Pythia}}\xspace}

\def\xgboost    {\mbox{\textsc{XGBoost}\xspace}}

\def\tell1  {TELL1\xspace}
\def\ukl1   {UKL1\xspace}

\def\BcToTauNu{\Bc \ra \taup \neut }
\def\BuToTauNu{\Bu \ra \taup \neut }

\def \Vcb{\ensuremath{V_{cb}}\xspace}
\def \Vub{\ensuremath{V_{ub}}\xspace}

\def \fBu {\ensuremath f_{B^+}\xspace}
\def \fBc {\ensuremath f_{B_{c}}\xspace}

%% file: introduction.tex
\section{Introduction }
\label{sec:introduction}

While the Standard Model (SM) has been reigning uncontested since its formulation over 50 years ago~\cite{Glashow:1961tr,Weinberg:1967tq,Salam:1968rm}, our community has long been searching for evidence of physics beyond it (BSM). To this end, particularly useful probes are leptonic decays of heavy flavour mesons, such as $\BcToTauNu$ or $\BuToTauNu$\footnote{Charge conjugation is implied throughout this work, unless stated otherwise.}. 
The description of such decays in the SM is very simple, with the meson decay constants as the only hadronic inputs to the computation of their branching ratios.
The  decay constants for $B^+_c /\BuToTauNu$ decays have been computed with extreme precision, thanks to the help of simulations performed employing lattice QCD (LQCD)~\cite{FLAG:2021npn}, resulting in very clean and precise theoretical predictions. 

The study of these channels has been gathering further interest in recent years, especially for the case of $B_c^+\to\tau^+\nu_\tau$ decays~\cite{Alonso:2016oyd,Li:2016vvp,Blanke:2018yud,Blanke:2019qrx}, due to recently reported discrepancies in semileptonic $B$-meson decays mediated at the quark level by the same process, namely the $b\to c \ell \nu_\ell$ transitions. In particular, the lepton flavour universality ratio
\begin{equation}
{\cal R}(D^{(\ast)}) = \dfrac{\mathcal{B}(B\to D^{(\ast)}\tau \bar{\nu})}{\mathcal{B}(B\to D^{(\ast)} l\bar{\nu})}\Bigg{\vert}_{l=e,\mu}   
\end{equation}
have been measured at LHC and the $B$-factories~\cite{BaBar:2012obs,BaBar:2013mob,Aaij:2015yra,Belle:2015qfa,Belle:2016ure,Hirose:2016wfn,Aaij:2017uff,Hirose:2017dxl,Aaij:2017deq,Belle:2019rba,LHCb:2023zxo}, showing a combined $3.2~\sigma$ deviation between the experimental average and their SM prediction~\cite{HFLAV:2022pwe}. A similar deviation, albeit not as statistically significant and experimentally precise, has been observed in ${\cal R}(J/\psi)= \mathcal{B}(B_c^+ \to J/\psi \tau^+ \nu_\tau)/\mathcal{B}(B_c^+\to J/\psi \mu^+ \nu_\mu)$~\cite{Aaij:2017tyk}. It is worth mentioning that, lately, a first experimental measurement for the ratio ${\cal R}(\Lambda_c)=\mathcal{B}(\Lambda_b \to\Lambda_c \tau\bar\nu)/\mathcal{B}(\Lambda_b \to\Lambda_c \ell\bar\nu)$ has been performed by LHCb~\cite{LHCb:2022piu}.
Despite being described at the parton level by the same transition~\cite{Blanke:2018yud,Blanke:2019qrx}, this measurement demonstrated a different behaviour from the ${\cal R}(J/\psi)$ results.
However, further analyses are required for ${\cal R}(\Lambda_c)$, particularly due to how the normalisation employed in the experimental measurement could impact its final determination~\cite{Bernlochner:2022hyz,Fedele:2022iib}. 

It is also interesting to notice that, going potentially beyond the ${\cal R}(D^{(\ast)})$ anomalies, $\BcToTauNu$ and $\BuToTauNu$ decays are in general excellent probes of New Physics (NP) contributions due to an extended pseudoscalar sector, as predicted by BSM theories like Two-Higgs-Doublet Models (2HDM)~\cite{Branco:2011iw} or certain leptoquark models (LQ)~\cite{Buchmuller:1986zs,Dorsner:2016wpm}.

The measurement of $\BcToTauNu$ and $\BuToTauNu$ decays are of particular interest not only in the context of BSM searches but also as a test of an exquisitely SM feature, the extraction of Cabibbo-Kobayashi-Maskawa (CKM) elements. In the Standard Model, the branching ratio responsible for this class of decays is determined by 
\begin{equation}\label{eq:BRBqtaunu}
\mathcal{B}(B_q^+ \to \tau^+ \nu_\tau)^{\textrm {SM}} =\tau_{B_q^+} \frac{G_{F}^2 |V_{qb}|^{2} f_{B_q^+}^2 m_{B_q^+} m_\tau^2}{8\pi} \left( 1- \frac{m_\tau^{2}}{m_{B_q^+}^{2}} \right )^{2},  ~~q = u,c
\end{equation}
where $\tau_{B_q^+}$ denotes the  $B_q^+$ meson lifetime, $G_F$ is the Fermi constant, and $m_{B_q^+}$ and $m_\tau$ are the masses of the $B_q^+$ meson and the $\tau^+$ lepton, respectively. The $B_q^+$ meson decay constants are $f_{B_q^+}$: for the charmed meson, the latest result computed via LQCD~\cite{McNeile:2012qf,Colquhoun:2015oha} gives $\fBc=427(6)$~MeV, while the latest LQCD $N_f=2+1+1$ calculations for the $B^+$ decay constant~\cite{Dowdall:2013tga,ETM:2016nbo,Hughes:2017spc,Bazavov:2017lyh} yield to the averaged value $\fBu=190.0(1.3)$ MeV~\cite{FLAG:2021npn}. Finally, $\BcToTauNu$ and $\BuToTauNu$ are sensitive to $\vert\Vcb\vert$ and $\vert\Vub\vert$, respectively, whose precise determination has been however hindered by the long-standing inclusive vs. exclusive puzzle~\cite{UTfit:2022hsi,CKMfitter}. Several ways to extract these CKM elements have been indeed advocated in the past, differentiated by whether the study of a specific exclusive channel or rather an inclusive determination, would be used to this end. While these different approaches are expected yield to compatible values for the extracted CKM elements, it has not been the case for either $\vert\Vcb\vert$ or $\vert\Vub\vert$, for both of which the average of the exclusive determinations differs from the inclusive one at the $3.3~\sigma$ level~\cite{HFLAV:2022pwe}. Taking as an example the latest determinations for the CKM elements from the global SM CKM fit performed by the UTfit collaboration~\cite{UTfit:2022hsi}, namely $|V_{cb}|^\mathrm{excl.}=42.22(51)\times 10^{-3}$ and $|V_{ub}|^\mathrm{excl.}=3.70(11)\times 10^{-3}$, we obtain the SM predictions for the branching ratios
\begin{eqnarray}
\mathcal{B}(B_c^+\to \tau^+ \nu_\tau)^\mathrm{SM}&=&2.29(9)\times 10^{-2}\,,\\
\mathcal{B}(B^+\to \tau^+ \nu_\tau)^\mathrm{SM}&=&0.87(5)\times 10^{-4}\,,
\end{eqnarray}
where the main sources of uncertainty come from the CKM matrix elements and the meson decay constants. Therefore, the aforementioned cleanness of the leptonic $\BcToTauNu$ and $\BuToTauNu$ decays makes them a theoretically perfect choice for an independent extraction of these CKM elements, potentially capable to play a relevant role in understanding the inclusive vs. exclusive puzzle, provided a measurement with enough precision is achievable at colliders.

Unfortunately, the level of theoretical precision is not yet matched on the experimental side, particularly for the $B_c^+ \to \tau^+ \nu_\tau$ decay. 
In spite of its sizeable branching ratio in the SM ($\approx 2\%$), a search for the $\BcToTauNu$ decay is very challenging at current collider experiments.
The selection and profiling of such decay rely on the reconstruction of $\tau$ decay vertices as well as the measurement of the missing energy.
At hadron collider experiments, due to the high multiplicity of simultaneous interactions in each bunch crossing and the lack of knowledge of the centre-of-mass energy of the $b\bar{b}$ production process, it is practically impossible to separate $\BcToTauNu$ events from the overwhelming background processes.
The current $e^+e^-$ $B$-factories, in spite of their relatively clean environments and the full knowledge of the centre-of-mass energy, operate at energies (for example $\Upsilon(5S)$) lower than the production threshold of the $B_c$ meson.
In comparison, future $Z$-factories, with a large number of $Z\to b\bar{b}$ events and an exquisite reconstruction of event kinematics, would provide the ideal environment to study this decay.
Meanwhile, the measurement of $\BuToTauNu$, albeit having been conducted at $B$-factories with the current best precision of 20\%~\cite{HFLAV:2022pwe}, can also benefit significantly from the large dataset at $Z$-factories.

The Future Circular Collider (FCC) project~\cite{Gomez-Ceballos:2013zzn, Benedikt:2651299, Mangano:2651294} aims at a design of a multi-stage collider complex for $e^+e^-$, $pp$, and $ep$ collisions hosted in a 91~km tunnel near CERN, Geneva. 
The first stage, FCC-ee, is envisioned for $e^+e^-$ collisions at four different centre-of-mass energy windows: around 91~GeV for the $Z$ production, around 161~GeV for the $W^+W^-$ production, about 240~GeV for the $ZH$ production, and more than 350~GeV for the $t\bar{t}$ production.
The $Z$-pole operation offers an unprecedented dataset not only for precision measurements of electroweak parameters related to the properties of the $Z$~boson itself, but also the properties of all of its decay products, such as the $b$ and $c$~quarks, and the $\tau$ lepton.

The integrated luminosity of the FCC-ee $Z$-pole data is proposed to be 150~$\text{ab}^{-1}$ in the original conceptual design reports~\cite{Benedikt:2651299, Mangano:2651294}, corresponding to $N_Z \approx 5\times10^{12}$. 
Recent studies in the FCC-ee design have updated such expectations to 180~$\text{ab}^{-1}$ and $N_Z \approx 6\times10^{12}$. 
A dedicated study of $\BcToTauNu$ decay at FCC-ee based on the original design has been reported in Ref.~\cite{Amhis_2021}, and a similar study at the Circular Electron Position Collider (CEPC)~\cite{cepc_cdr1,cepc_cdr2} has been presented in Ref.~\cite{Zheng:2021xuq}. Both have demonstrated promising physics prospects in this channel.
In this work, we revise the analysis for $\BcToTauNu$ at FCC-ee and extend the work to a combined measurement of both $\BcToTauNu$ and $\BuToTauNu$ decays.
As done in Ref.~\cite{Amhis_2021}, we focus on the hadronic $\TauThreePi$ decay for both $\BcToTauNu$ and $\BuToTauNu$ cases.

It is worth mentioning at this point that, in the case of the $\BcToTauNu$ decay, the measurement of its branching ratio is hampered by the lack of knowledge of the $B_c^+$ meson hadronisation fraction, $f(B_c^\pm) \equiv f(b\to B_c^\pm)$~\cite{Mangano:1997md}. The strategy followed in this work is the same as the one adopted in Ref.~\cite{Amhis_2021}, which consists in using $B_c^+\to J/\psi \mu^+ \nu_\mu$ as a normalisation mode and therefore relies on its LQCD prediction~\cite{Harrison:2020gvo,Harrison:2020nrv} to extract a value for $\mathcal{B}(B_c^+ \to \tau^+ \nu_\tau)$. As will be reported in Sec.~\ref{sec:interpretation}, the relative precision of the LQCD prediction for this normalisation will be one of the main limiting factors in the determination of $\mathcal{B}(B_c^+ \to \tau^+ \nu_\tau)$. On the other hand, the $B^+$ meson hadronisation fraction, $f(B^\pm) \equiv f(b\to B^\pm)$, being known at the percent level~\cite{PDG}, does not pose an impediment for the measurement of $\mathcal{B}(B^+ \to \tau^+ \nu_\tau)$.

The remainder of this paper is organised as follows: Sec.~\ref{sec:analysis} summarises the experimental setup, describes the analysis workflow, and provides estimates of the range of signal precision in different scenarios; Sec.~\ref{sec:interpretation} illustrates the phenomenological impact of these measurements, both in the SM case and in selected relevant NP benchmarks.

The work in this paper has been conducted following the recommendation of the European Strategy Update for Particle Physics to investigate the technical and financial feasibility of a future hadron collider at CERN with a centre-of-mass energy of at least 100 TeV and with an electron-positron Higgs and electroweak factory as a possible first stage. This article provides a detailed description of the key ingredients for physics analyses at FCC-ee.

%% file: analysis.tex
 \section{Analysis}
\label{sec:analysis}

The procedure of this analysis is modified from the preceding work on the measurement of the $\BcTauNu$ at FCC-ee, reported in Ref.~\cite{Amhis_2021}. Major changes in the procedure are listed as follows:
\begin{itemize}
    \item Multi-classification, instead of binary classification, is deployed in the second-stage training, which achieves good separation between $\BTauNu$, $\BcTauNu$, and background processes and ensures orthogonality in later selections.
    \item The procedure for selection efficiency estimate is fully revised to minimise potential biases from the limited number of events that pass the final selection. Additional background samples are generated in the signal-enriched phase-space to further improve the robustness of the efficiency estimate.
    \item The method for the final fit is updated to allow for the simultaneous evaluation of $\BTauNu$ and $\BcTauNu$ yields. Additional validation studies are performed to make sure the selection procedure does not introduce shifts in the background shapes.
    \item Additional studies are performed to estimate the potential impact of background inflation and fluctuation to the signal precision in order to understand the range of expected sensitivity of these measurements.
\end{itemize}

This section provides a description of the full analysis.
The detector concepts, simulated samples, and the analysis setup are described in detail in Ref.~\cite{Amhis_2021} and not repeated here.
Additional simulated samples used in this work are summarised in Sec.~\ref{sec:sig_bkg_samples} and the updated analysis tools are publicly available in Ref.~\cite{perez_emmanuel_2023_8208701}.
The main kinematic features to characterise events in this analysis are discussed in Sec.~\ref{sec:thrust}.
In this analysis, events are selected with a series of rectangular selections on their kinematic properties, followed by two boosted decision tree (BDT) classifiers described in Secs.~\ref{sec:BDT1} and~\ref{sec:BDT2}.
The BDT-based final selection, detailed Sec.~\ref{sec:sel_opt}, separates the $\BcTauNu$ and $\BTauNu$ signals into two categories with high signal-to-background ratios~($S/B$) and little cross-contamination. 
A template fit is performed simultaneously for two signal modes to estimate their individual yields and precisions, as described in Sec.~\ref{sec:fit}. 
Considerations of potential systematic uncertainties and the expected experimental precision in scenarios with nonideal background modelling are also discussed in Secs.~\ref{sec:syst} and~\ref{sec:scenarios}. 
Finally, Secs~\ref{sec:less_data} discusses the projected precision in case of a smaller dataset.

\subsection{Signal and background samples}
\label{sec:sig_bkg_samples}

Event samples are generated with \pythia~8.303~\cite{Sjostrand:2014zea} and \evtgen~02.00.00~\cite{LANGE2001152} and the detector responses are simulated with \delphes~\cite{deFavereau:2013fsa}, following the configurations detailed in Ref.~\cite{Amhis_2021}.
This work uses the full set of samples described in Ref.~\cite{Amhis_2021}. 
A few additional samples, listed in App.~\ref{app:excl_samp}, are produced for exclusive background processes in order to improve the estimates of background efficiencies.

Orthogonal sets of samples are deployed for the BDT training and the statistical analysis to avoid potential biases from overtraining.
The training sample set comprises signal samples and inclusive $Z \to b\bar{b}$, $c\bar{c}$, and $q\bar{q}$ processes, where $q \in \{u,d,s\}$. 
The analysis sample set is composed of signal samples, inclusive background decays, and a selected list of exclusive background processes. 
The inclusive decay samples, each containing $10^9$ events, are used to estimate background yields and shapes. 
They become insufficient for estimating background yields towards final selections, where backgrounds are rejected at more than the $10^9$ level.
A collection of exclusive decay modes are chosen to provide a good representation of a subset of background processes, which share with signals more similarities of the kinematic profile.
The exclusive samples are used to estimate the background efficiency under high BDT requirements. 

The exclusive $b$-hadron decay modes in the $Z \to b\bar{b}$ process considered in this analysis are listed as follows:
 \begin{multicols}{3}
 \begin{itemize}
    \item $Z \to b\bar{b}$, $B \to D \tau^+ \nu_\tau$
    \item $Z \to b\bar{b}$, $B \to D^* \tau^+ \nu_\tau$
    \item $Z \to b\bar{b}$, $B \to D e^+ \nu_e$
    \item $Z \to b\bar{b}$, $B \to D^* e^+ \nu_e$
    \item $Z \to b\bar{b}$, $B \to D \mu^+ \nu_\mu$
    \item $Z \to b\bar{b}$, $B \to D^* \mu^+ \nu_\mu$
    \item $Z \to b\bar{b}$, $B \to D \pi^+ \pi^+ \pi^-$
    \item $Z \to b\bar{b}$,~$B \to D^*\pi^+\pi^+ \pi^-$
    \item $Z \to b\bar{b}$, $B \to D D_s^+$
    \item $Z \to b\bar{b}$, $B \to D^* D_s^+$
    \item $Z \to b\bar{b}$, $B \to D^* D_s^{*+}$
    \item[\vspace{\fill}]
\end{itemize}
\end{multicols}
The $B$ in the list stands for $\{B^0, B^+, B_s^0, \Lambda_b^0\}$ and the corresponding $D$ represents $\{D^-, \Dzb, D_s^-, \Lambda_c^-\}$. In each of the exclusive $b$-hadron samples, all of the $b$-hadron decay products are decayed further inclusively. The list of exclusive decays considered is not exhaustive and covers around \revision{24\%} of the decay width for each $b$-hadron. 
Hence, they are not used to estimate the absolute background yields, but only to evaluate the relative efficiency from the baseline selection to the final selections, as detailed in Sec.~\ref{sec:sel_opt}.

The exclusive $Z \to c\bar{c}$ decay modes considered in this analysis are:
\begin{multicols}{2}
\begin{itemize}
    \item $D^+ \to \tau^+\nu_\tau$      
    \item $D^+ \to K^0 \pi^+\pi^+\pi^-$ 
    \item $D_s^+ \to \tau^+\nu_\tau$     
    \item $D_s^+ \to \rho^+ \eta^\prime$ 
    \item $\Lambda_c^+ \to \Lambda^0 e^+\nu_e$    
    \item $\Lambda_c^+ \to \Lambda^0 \mu^+\nu_\mu$ 
    \item $\Lambda_c^+ \to \Lambda^0 \pi^+\pi^+\pi^-$ 
    \item $\Lambda_c^+ \to \Sigma^+ \pi^+\pi^-$  
\end{itemize}
\end{multicols}
All remaining unstable particles are decayed inclusively.
Among these samples, the $D^+ \to K^0 \pi^+\pi^+\pi^-$, $D_s^+ \to \rho^+ \eta^\prime$, $\Lambda_c^+ \to \Lambda^0 \pi^+\pi^+\pi^-$, and $\Lambda_c^+ \to \Sigma^+ \pi^+\pi^-$ modes are chosen to represent the high multiplicity hadronic decays.
Similar to the $Z \to b\bar{b}$ case, all exclusive samples are only used to evaluate the relative efficiency from the baseline selection to the final selection and not for the absolute yield.

\subsection{Thrust axis and event hemisphere definitions}
\label{sec:thrust}

In $Z \to b\bar{b}$ events at FCC-ee, the net momentum of the $Z$~boson is expected to be 0 and the two $b$ quarks are boosted in opposite directions along the $Z$~decay axis. 
The event can therefore be divided into two hemispheres, with one $b$~quark on each side.
The hemispheres are defined event-by-event, by a plane normal to the event thrust axis, which is reconstructed as the unit vector $\hat{\textbf{n}}$ to maximise
\begin{equation}
    T = \frac{\sum_{i}|\textbf{p}_i \cdot \hat{\textbf{n}}|}{\sum_i |\textbf{p}_i|},
\end{equation}
where $\textbf{p}_i$ is the momentum vector of the i$^\text{th}$ reconstructed particle.  
Reconstructed particles are assigned to either hemisphere based on the angle between their momentum vector and the thrust axis. 

In this analysis, the signal decays involve significant missing energy due to the presence of two neutrinos in the $B^+/\BcTauNu$ and $\TauThreePi$ decays, while a generic $b$-hadron decay usually have a higher fraction of reconstructed energy.
This leads to a sizeable energy imbalance between two hemispheres in events containing a signal decay on one side and a generic decay on the other side, 
in contrast to generic background events, where the energy distributions from two generic decays are well-balanced.
\rerevision{Furthermore, the significant missing energy in the signal hemisphere also skews the reconstructed thrust axis and biases the reconstruction of other kinematic properties of the event. Such interplay introduces further discrimination between the signals and backgrounds in variables that are not directly related to the signal decays.}
In the rest of this work, we designate the hemisphere with less energy as the signal (positive) hemisphere, and the other as the background (negative) hemisphere.

\subsection{First-stage BDT}\label{sec:BDT1}

The first-stage training (BDT1) is designed to loosely separate signals from generic backgrounds \rerevision{using event level kinematic features and not the specific decay processes.} The training follows the procedure described in Ref.~\cite{Amhis_2021}, the only change being the inclusion of the $\BTauNu$ process in the signal sample set. The BDT is trained with \xgboost~\cite{xgb} using the following features\revision{, which are chosen to provide the complete information of the overall event topology:}
\begin{itemize}
    \item Total reconstructed energy in each hemisphere;
    \item Total charged and neutral reconstructed energies in each hemisphere;
    \item Charged and neutral particle multiplicities in each hemisphere;
    \item Number of tracks in the reconstructed PV;
    \item Number of reconstructed $3\pi$ candidates in the event;
    \item Number of reconstructed vertices in each hemisphere;
    \item Minimum, maximum, and average radial distance of all decay vertices from the PV.
\end{itemize}
Distributions of all input features are summarised in Appendix~\ref{app:BDT_vars}. 

\begin{figure}[h!]
\centering
\includegraphics[width = 0.49\textwidth]{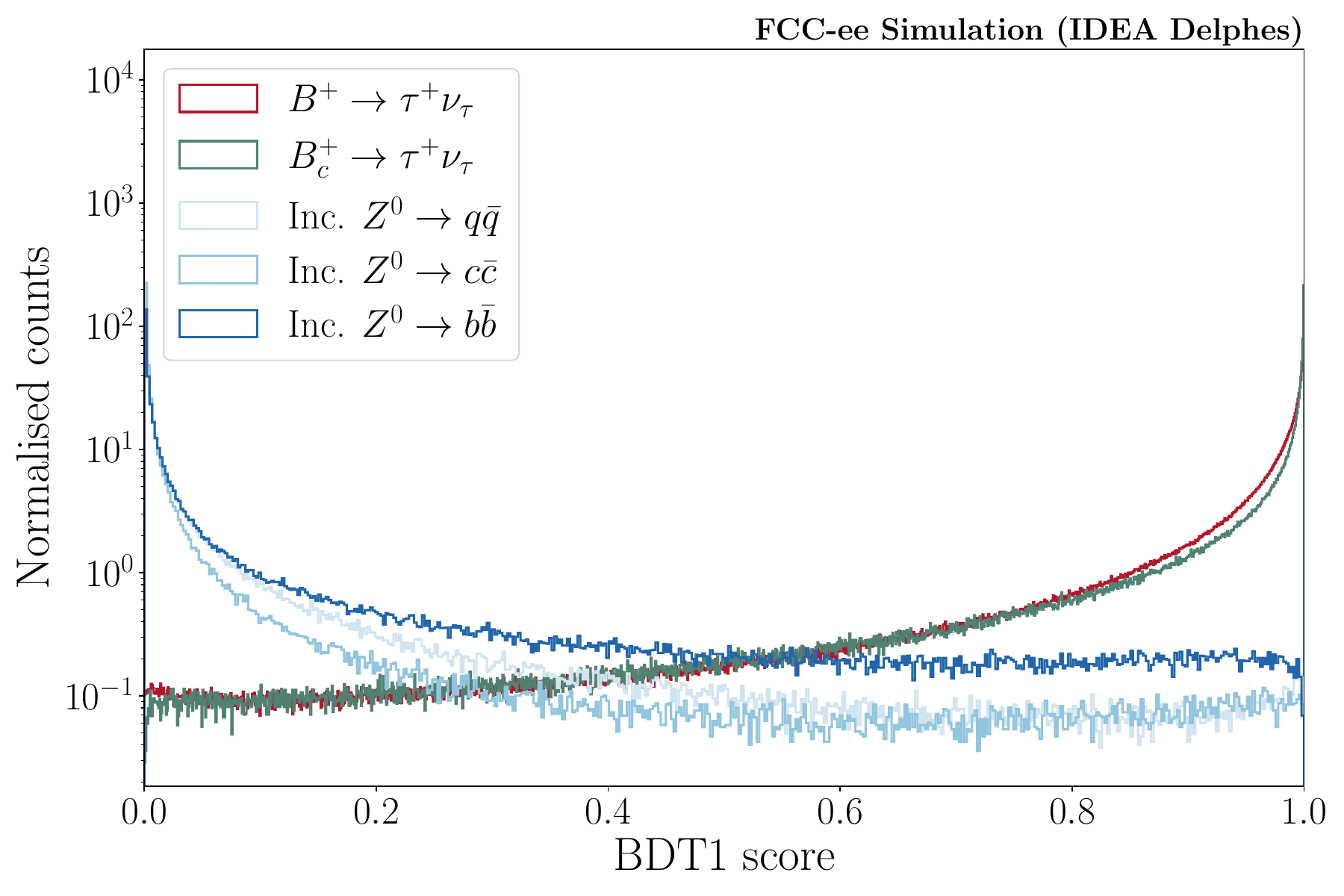} 
\includegraphics[width = 0.49\textwidth]{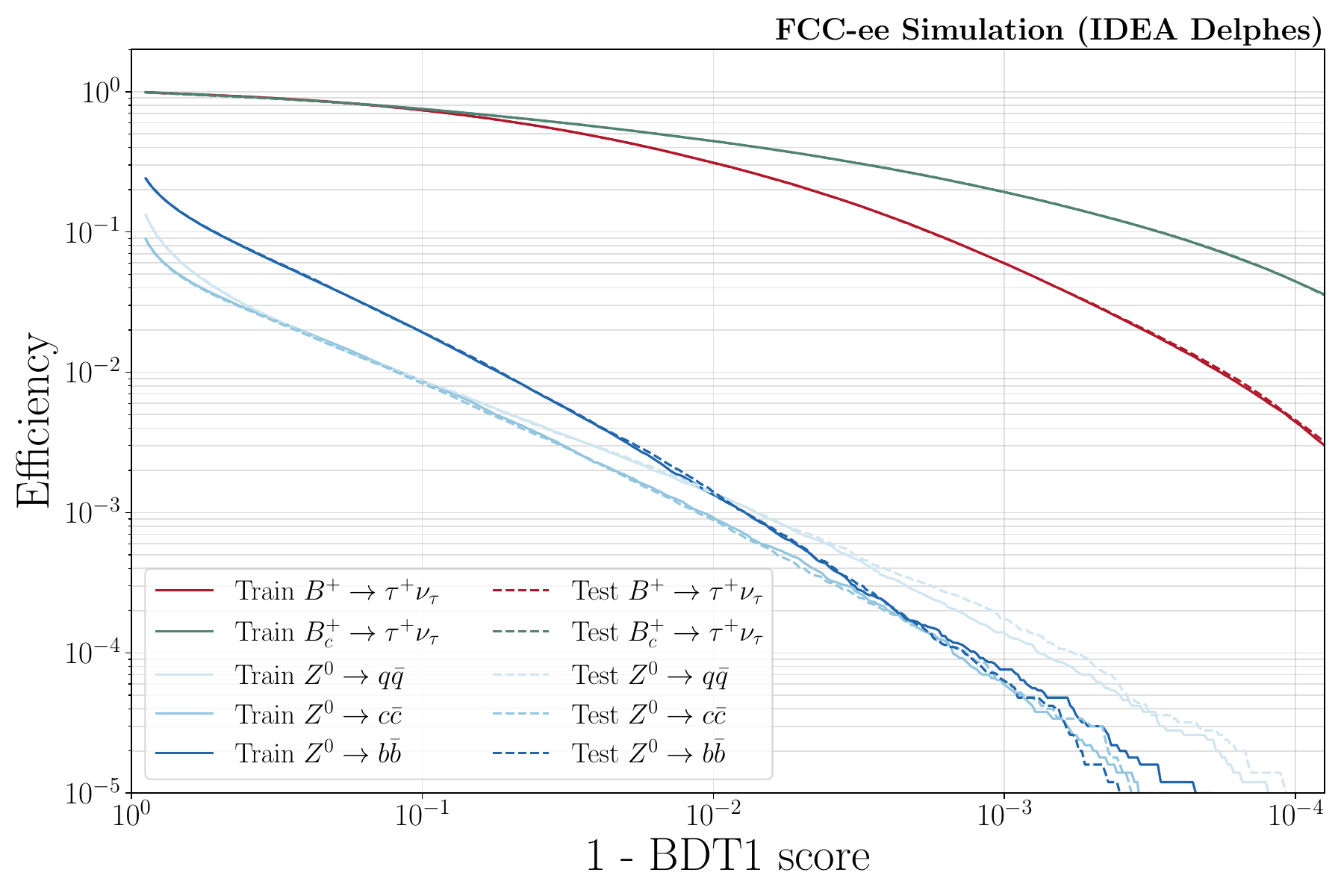}
\caption{(Left) First-stage BDT distribution for both signal modes and inclusive $Z$ background. (Right) Efficiencies in both training samples and testing samples for different processes, as a function of cut values on the first-stage BDT.}
\label{fig:BDT1}
\end{figure}

\revision{The optimal configuration of this is BDT found to feature 1000 trees, with a maximum number of cuts of 5 and a learning rate of 0.3.} 
It is found to have a receiver operating characteristic (ROC) curve area of 0.983, indicating an excellent separation between signals and backgrounds. The BDT performance is illustrated in Fig.~\ref{fig:BDT1}, where the BDT distributions in $\BcTauNu$, $\BTauNu$, and each type of inclusive $Z$ background are shown alongside their corresponding efficiency profiles.  
The signal efficiency for $\BTauNu$ is slightly worse than that of $\BcTauNu$ because its kinematic properties are more similar to generic $\Bu$ decays in $Z \to b\bar{b}$ backgrounds. 

A comparison of BDT performances on training and testing datasets is also provided in the right plot of Fig.~\ref{fig:BDT1}.
For this purpose, the testing dataset is a randomly sampled subset of the analysis samples, which are statistically independent of the training samples.
Each line in this plot is profiled with $5\times10^5$ events.
A good agreement in the efficiency profiles of training and testing samples is seen for efficiencies down to $10^{-3}$.
When background efficiencies go below $10^{-3}$, the efficiency profiles of training and testing samples start to mildly deviate from each other, due to the limited number of remaining events.

As will be shown in Sec.~\ref{sec:sel_opt}, very tight BDT selections are applied in the final statistical analysis, corresponding to BDT1~$\gtrsim 0.9999$ and BDT1 efficiencies of~O($10^{-5}$) for backgrounds. 
There are not enough events in the training samples to precisely evaluate background efficiencies to this level.
However, this should not inflict a concern about an over-optimistic efficiency estimate for the final analysis, 
because the efficiency estimates are always conducted with analysis samples, which are free from any statistical bias in the training.
Nonetheless, any potential discrepancy between the real data and the simulated samples in this analysis would lead to an inaccurate expectation of the BDT performance on the actual data.
To address the lack of knowledge of the data expected to be collected a few decades in the future, instead of quoting the BDT efficiencies as is for simulations, we evaluate the range of the final precision in a few scenarios in which the BDT performance on data is worse than its expectation.
The study is described in detail in Sec.~\ref{sec:scenarios}.

\FloatBarrier

\subsection{Second-stage BDT}\label{sec:BDT2}

Prior to the second-stage training, events are required to have the first-stage BDT~$>0.6$. This selection is over 80\% efficient for both signals and removes more than 90\% of each type of background. In addition, the energy difference between the background and signal hemispheres is required to exceed 10~GeV, in order to further remove background-like events with little energy imbalance. 
 
The second-stage BDT (BDT2) is designed to further refine signal selection using the full kinematic properties of the reconstructed $3\pi$ candidate for the $\tau$ decay as well as properties of other reconstructed decay vertices in the event.
The $\TauThreePi$ candidate is chosen as the $3\pi$ vertex with the smallest vertex fit $\chi^2$ in the signal hemisphere. 
It is further required to have an invariant mass below that of the $\tau$ lepton, and have at least one $m(\pi^+\pi^-)$ combination within the range 0.6--1.0~GeV.
These selections retain candidates consistent with the $a_1(1260)^+ \to (\rho^0 \to\pi^+\pi^-)\pi^+$ decay, via which all $\TauThreePi$ decays proceed.

The setup of the second-stage training is similar to that in Ref.~\cite{Amhis_2021}. Instead of using a binary classifier for the signal-background separation, the BDT2 in this work is a multi-classifier with three labels: Bc for $\BcTauNu$, Bu for $\BTauNu$, and Bkg for backgrounds. 
The full size of $\BcTauNu$ and $\BTauNu$ training samples after the aforementioned selections are used in the training, corresponding to about $1\times 10^6$ events for $\BcTauNu$ and $6\times 10^5$ for $\BTauNu$. 
The background training sample is a combination of $Z \to b\bar{b}$, $Z \to c\bar{c}$, and $Z \to q\bar{q}$ events, whose proportions are based on their branching ratios and their efficiencies after previous selections, leading to roughly $1 \times 10^4$~$Z \to q\bar{q}$, $1 \times 10^5$~$Z \to c\bar{c}$, and $1 \times 10^6$~$Z \to b\bar{b}$ events.

Input variables to the second-stage BDT include the ones used in Ref.~\cite{Amhis_2021}, \revision{which are designed to describe the full kinematic information of the $3\pi$ candidate as well as its correlation with other displaced vertices.} 
A few variables are added characterising extra $D$~meson decays in the event.
$D$~mesons are expected in the signal hemisphere of $\BcTauNu$ events because of the presence of charm quarks in the hadronisation of the $B_c^+$ meson.
Variables explicitly tagging $D$~meson candidates have shown to provide a small gain in selecting the $\BcTauNu$ signal.
The $D$~meson candidate is selected as the secondary vertex in the signal hemisphere with the invariant mass closest to 1.85~GeV.

A summary of all features used in the second-stage training is listed as follows:
\begin{itemize}
    \item $3\pi$ candidate mass, and masses of the two $\pi^+\pi^-$ combinations;
    \item Number of $3\pi$ candidates in the event;
    \item Radial distance of the $3\pi$ candidate from the PV;
    \item Vertex $\chi^2$ of the $3\pi$ candidate;
    \item Momentum magnitude, momentum components, and impact parameter (transverse and longitudinal) of the $3\pi$ candidate;
    \item Angle between the $3\pi$ candidate and the thrust axis;
    \item Minimum, maximum, and average impact parameter (longitudinal and transverse) of all other reconstructed decay vertices in the event;
    \item Mass of the PV;
    \item Nominal $B$ meson energy, defined as the $Z$ mass minus all reconstructed energy apart from the $3\pi$ candidate;
    \item The mass and vertex displacement of the $D$~meson candidate. 
\end{itemize}
Distributions of all input features are summarised in Appendix~\ref{app:BDT_vars}. 

\begin{figure}[h!]
\centering
\includegraphics[width = 0.6\textwidth]{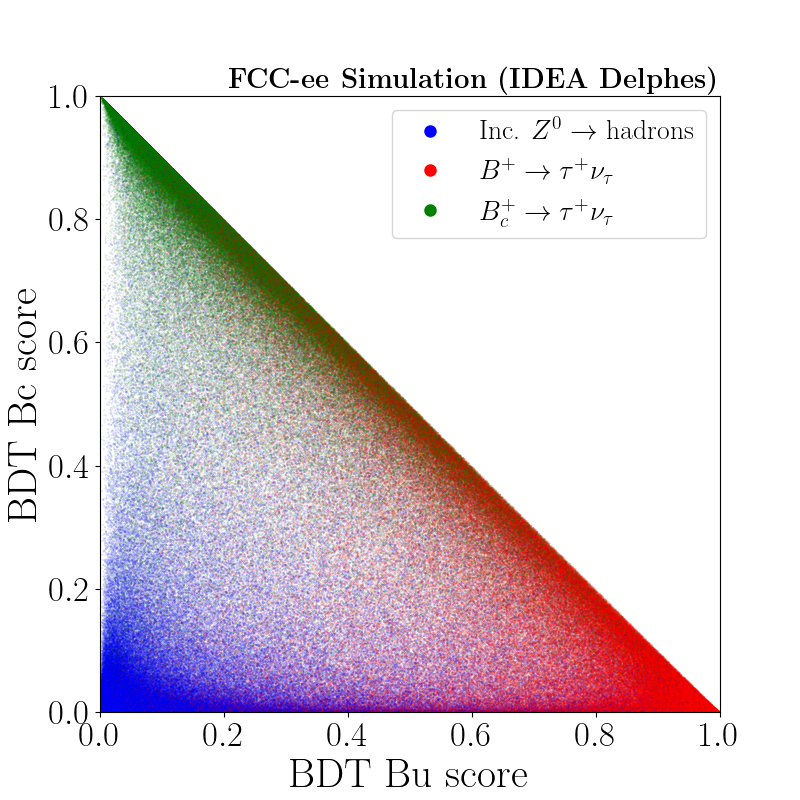}
\caption{Distribution of the second-stage BDT output in the Bc-Bu plane.}
\label{fig:BDT2_scatter}
\end{figure}

\revision{The optimal configuration of this is BDT found to feature 1000 trees, with a maximum number of cuts of 3 and a learning rate of 0.3.}
This BDT achieves a high level of separation between backgrounds and each signal, which is illustrated in the distribution of different processes in the Bc score vs Bu score plane in Fig.~\ref{fig:BDT2_scatter}.
The output of the multi-classifier is regulated such that the Bc score, Bu score and background score sum up to~1.
The closer to the origin of the Bc-Bu plane, the higher the background score.
The area under the Bc vs non-Bc ROC curve is found to be 0.921 and the area under the Bu vs non-Bu ROC curve to be 0.886.

\begin{figure}[h!]
\centering
\includegraphics[width = 0.49\textwidth]{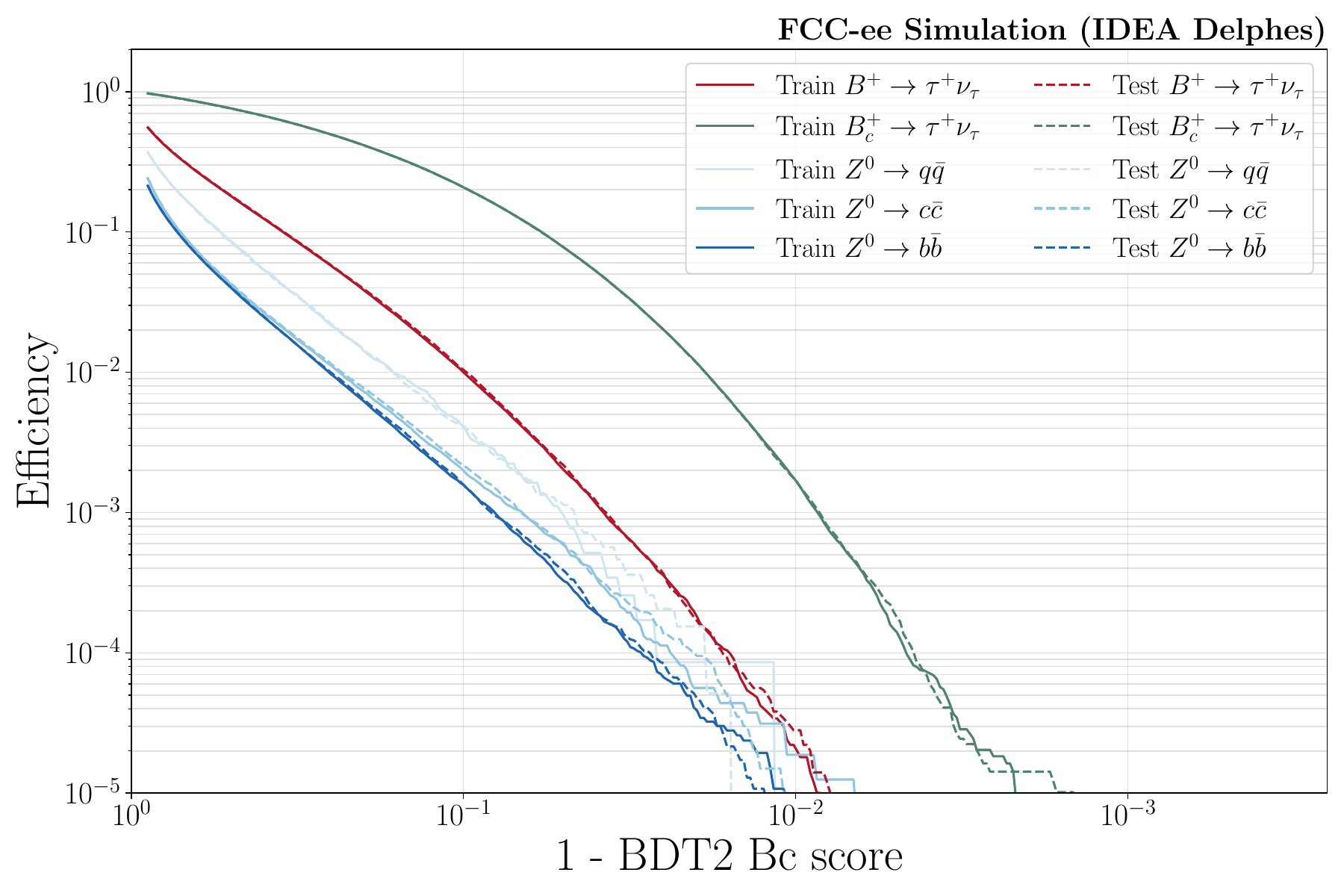}
\includegraphics[width = 0.49\textwidth]{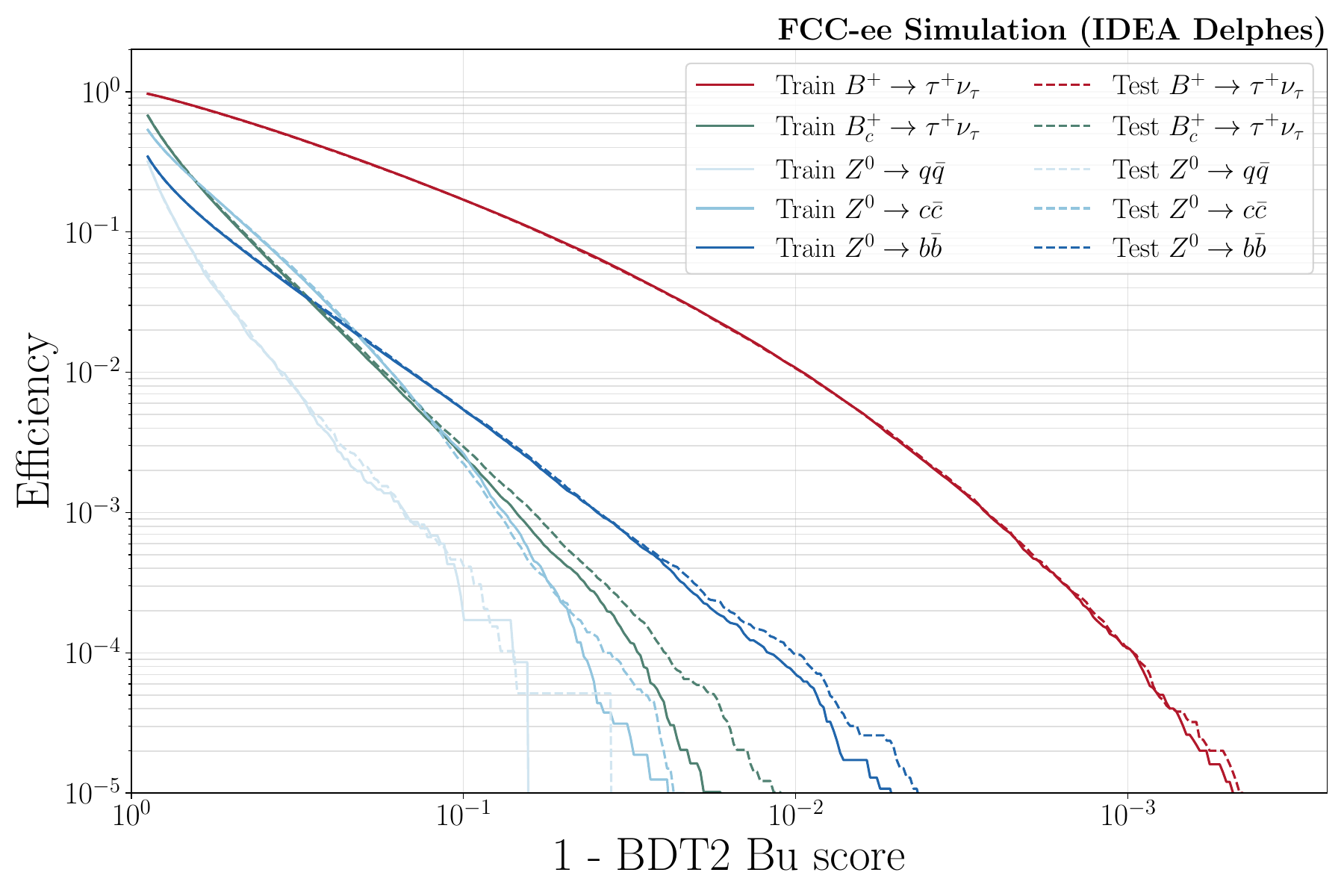}
\includegraphics[width = 0.49\textwidth]{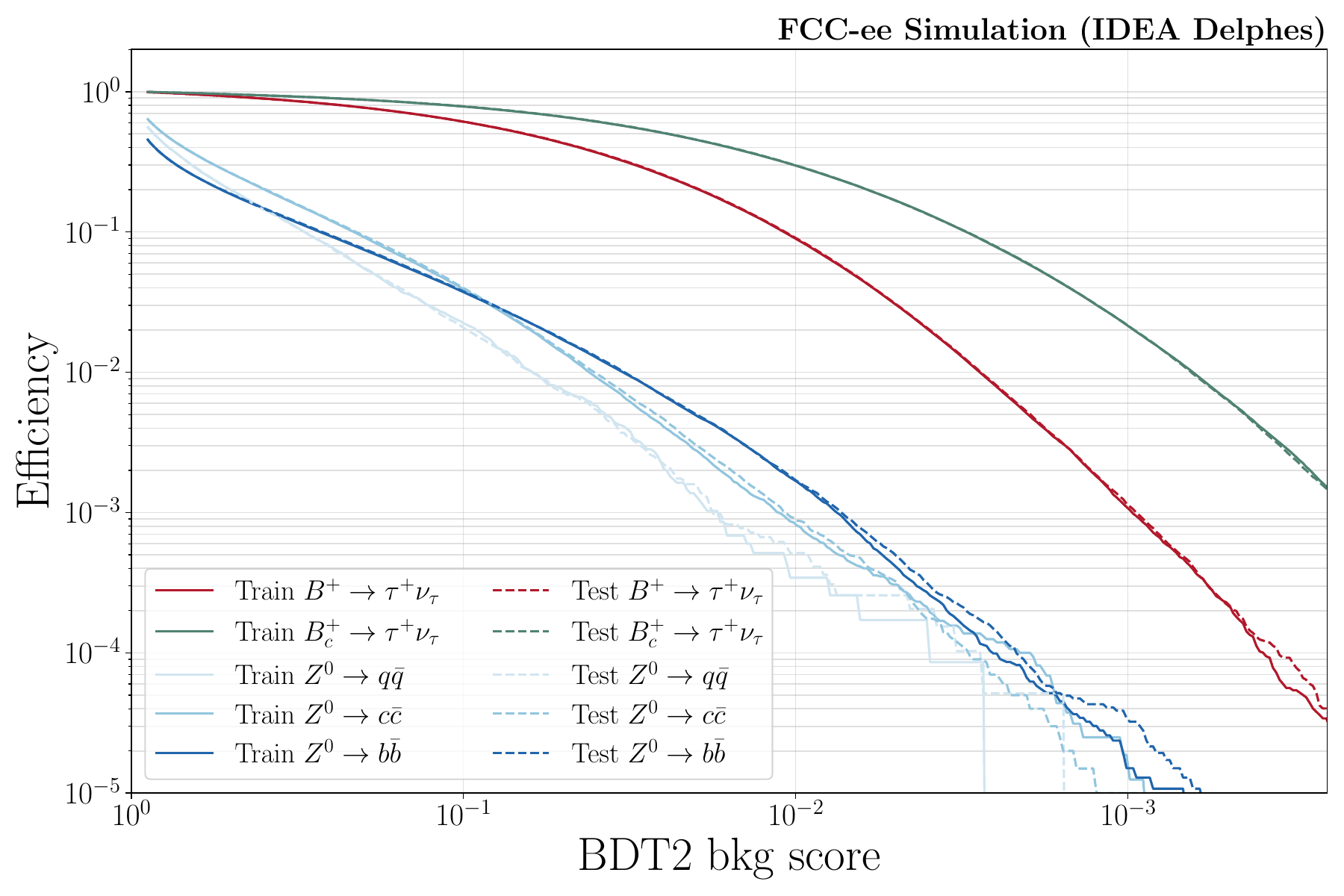}
\caption{Efficiencies in both training samples and testing samples for different processes, as a function of cut values on the Bc score (top left), Bu score (top right), and Bkg score (bottom) of the second-stage BDT.}
\label{fig:BDT2_eff}
\end{figure}

The second-stage BDT performance is shown in Fig.~\ref{fig:BDT2_eff}, along with a comparison between training samples and testing samples. 
The top left plot shows the selection efficiency of events with Bc score greater than different cut values, the top right plot shows the efficiency with regard to the Bu score and the bottom plot regarding the Bkg score.
Each line for $\BcTauNu$, $\BTauNu$, and $Z \to b\bar{b}$ processes is profiled with $5\times10^5$ events.
The $Z \to c\bar{c}$ and $Z \to q\bar{q}$ are already rejected with high rates by selections prior to the second-stage training and have fewer remaining events in the training samples.
As a results, $2\times 10^5$ events are used for each line of $Z \to c\bar{c}$ and $2\times 10^4$ for $Z \to q\bar{q}$.

The BDT2 performance is found to agree well between the training and testing samples in the Bc score plot and the Bkg score plot.  
In the Bu-enriched region, some deviations are observed between the training and testing samples for the $\BcTauNu$ and background processes. 
The largest difference is found to be about a factor of~5 in the efficiency of the $\BcTauNu$ process on the Bu score selection around BDT2 Bu~$= 0.95$.
The discrepancy is avoided in the final analysis, detailed in Sec.~\ref{sec:sel_opt}, as the final selection criterion on the Bu score is kept at~0.64.
Furthermore, similar to the case discussed in Sec.~\ref{sec:BDT1}, we do not assume the BDT performance on current simulations to be a precise depiction of what it will be on real data.  
The impact of non-ideal data modelling is addressed in Sec.~\ref{sec:scenarios}.

\subsection{Final selection}\label{sec:sel_opt}

To select each signal process with high purity, two orthogonal event categories are selected targeting the $\BcTauNu$ and $\BTauNu$ signals, respectively.
Each category is defined with a combined selection on the BDT1, BDT2 signal\footnote{The signal score here refers to the Bu score for the Bu category and Bc score for the Bc category. The same applies to the rest of the text.}, and BDT2 background scores.
The selection criteria are optimised by evaluating the yields for different processes under the selection combinations.

The yields for two signal modes are evaluated directly with simulated samples.
For backgrounds, an indirect measure is taken because the final background rejection is beyond~$10^9$ level, and the number of events in the simulated samples is insufficient for a direct estimate. 
The background yields are evaluated with a three-step procedure: 
\begin{itemize}
    \item The inclusive samples are used to estimate a yield with a baseline selection (BDT1~$> 0.9$, BDT2 signal~$> 0.6$, and BDT2 bkg~$< 0.1$).
    \item The inclusive samples become statistically limited after the baseline selection. The samples for exclusive decays listed in Sec.~\ref{sec:sig_bkg_samples} are used to estimate the efficiencies from this baseline selection to a tight selection (BDT1~$> 0.99$, BDT2 signal~$> 0.6$, and BDT2 bkg~$<0.05$). Each exclusive decay is weighted by its cross section multiplied by its efficiency regarding previous steps. The combination of all exclusive samples provides the best approximation to the inclusive phase-space.
    \item Beyond the tight selection, the exclusive samples also become limited in statistical precision. To estimate efficiencies with further selections, the first-stage and second-stage BDTs are assumed to be independent of each other and their efficiencies are evaluated separately with the exclusive samples. The efficiency profiles with regard to BDT1 are evaluated with a series of cut values and smoothed with splines. Similarly, the efficiency profiles with regard to BDT2 are evaluated on a 2D grid (BDT2 signal, BDT2 background) and smoothed with 2D splines. The efficiencies with regard to two BDTs are multiplied to give the total efficiency, as shown in Eq.~\ref{eq:final_eff}.
\end{itemize}

\begin{equation}
\begin{split}
    \epsilon_1(\alpha) &= \epsilon(\text{BDT1} > \alpha \hspace{0.1cm}|\hspace{0.1cm} \text{tight selection}), \\
    \epsilon_2(\beta, \gamma) &= \epsilon(\text{BDT2}_\text{sig} > \beta, \text{BDT2}_\text{bkg} < \gamma \hspace{0.1cm}|\hspace{0.1cm} \text{tight selection}), \\
    \epsilon_{tot}(\alpha, \beta, \gamma) &= \epsilon_1(\alpha) \times \epsilon_2(\beta, \gamma)
\end{split}
\label{eq:final_eff}
\end{equation}
The efficiency profiles of $Z \to b\bar{b}$ and $Z \to c\bar{c}$ processes are evaluated independently.
The light flavour $Z \to q\bar{q}$ decay is already rejected at~$10^9$ level with the baseline selection and is therefore not considered for further steps.

The final selection criteria are optimised separately for two categories by scanning through a three-dimensional grid of BDT1, BDT2 signal, and BDT2 background scores to find the best-expected signal purity, defined as $P = S/(S+B)$, where $S$ and $B$ are the expected signal and background yields.
The optimal selections are found to be BDT1~$>$~0.99993, BDT2 Bc~$>$~0.782, and BDT2 background~$<$~0.0088 for the Bc category, and BDT1~$>$~0.99946, BDT2 Bu~$>$~0.640, and BDT2 background~$<$~0.0123 for the Bu category. 
The efficiencies for $Z \to b\bar{b}$ and $Z \to c\bar{c}$ processes factorised for each step are listed in Tab.~\ref{tab:bkg_eff}. 
\revision{The signal efficiencies factorised for each step are listed in Tab.~\ref{tab:sig_eff}.}
The corresponding expected yields, in the scenario of $N_Z = 6 \times 10^{12}$, are listed in Tab.~\ref{tab:expected_yield}

\begin{table}[h!]
    \centering
    \begin{tabular}{c|c|c|c|c}
        \hline
        Selection stage & \multicolumn{2}{c|}{Bc category} & \multicolumn{2}{c}{Bu category} \\
        \hline 
        {}       & $Z \to b\bar{b}$ & $Z \to c\bar{c}$ & $Z \to b\bar{b}$ & $Z \to c\bar{c}$ \\ 
        \hline
        Baseline selection   & $1.99\times 10^{-5}$ & $4.63\times 10^{-6}$ & $2.76\times 10^{-5}$ & $5.79\times 10^{-6}$ \\
        Tight selection      & $1.39\times 10^{-1}$ & $1.97\times 10^{-1}$ & $7.39\times 10^{-2}$ & $7.63\times 10^{-2}$ \\
        Final selection BDT1 & $1.55\times 10^{-3}$ & $1.93\times 10^{-3}$ & $1.80\times 10^{-2}$ & $2.24\times 10^{-2}$ \\
        Final selection BDT2 & $1.13\times 10^{-1}$ & $7.91\times 10^{-2}$ & $1.00\times 10^{-1}$ & $4.79\times 10^{-3}$ \\
        \hline
        Total                & ~$4.82\times 10^{-10}$ & ~$1.39\times 10^{-10}$  & ~$3.68\times 10^{-9}$ & ~$4.74\times 10^{-11}$ \\
        \hline
    \end{tabular}
    \caption{Summary of background efficiencies in each step with regard to previous steps, for both $Z \to b\bar{b}$ and $Z \to c\bar{c}$ processes in Bc and Bu categories.}
    \label{tab:bkg_eff}

    \revision{
    \begin{tabular}{c|c|c|c|c}
        \hline
        Selection stage & \multicolumn{2}{c|}{Bc category} & \multicolumn{2}{c}{Bu category} \\
        \hline 
        {}       & $\BcTauNu$ & $\BTauNu$ & $\BcTauNu$ & $\BTauNu$ \\ 
        \hline
        Baseline selection   & $0.246$ & $0.049$ & $0.022$ & $0.173$ \\
        Tight selection      & $0.659$ & $0.512$ & $0.409$ & $0.358$ \\
        Final selection BDT1 & $0.099$ & $0.023$ & $0.216$ & $0.094$ \\
        Final selection BDT2 & $0.568$ & $0.301$ & $0.224$ & $0.258$ \\
        \hline
        Total                & $0.0090$ & $0.00017$  & $0.00042$ & $0.0015$ \\
        \hline
    \end{tabular}
    \caption{\revision{Summary of signal efficiencies in each step with regard to previous steps, for both $\BcTauNu$ and $\BTauNu$ signals in Bc and Bu categories.}}
    \label{tab:sig_eff}
    }
\end{table}

\begin{table}[h!]
    \centering
    \begin{tabular}{l|c|c}
        Process             & Bc category & Bu category \\
        \hline 
        $N(\BcTauNu)$       & 12000.8 & 569.0 \\
        $N(\BTauNu)$        & 1373.6  & 12011.4 \\
        $N(Z \to b\bar{b})$ & 437.4   & 3334.6 \\
        $N(Z \to c\bar{c})$ & 100.7   & 34.2 \\
    \end{tabular}
    \caption{Expected yield of different processes after the final selection in Bc and Bu categories, assuming a total of $6 \times 10^{12}$ $Z$~bosons.}
    \label{tab:expected_yield}
\end{table}

\subsection{Fit to measure the signal yield}\label{sec:fit} 

The signal yield is measured by fitting the distributions of certain kinematic variables, which discriminate signals from backgrounds.
\revision{The kinematic distributions are modelled directly for signals after the final selection.
For the background, however, there is not enough simulated events after final selections, so its fit variable shapes need to be extrapolated from looser selections. 
Such extrapolation relies on the assumption that there is no significant variation in the background shapes between loose and tight BDT selections.}
\rerevision{Since the two BDTs utilise the full kinematic description of the event, all kinematic variables are, to different extents, correlated with the BDT scores. However, as will be discussed below, some variable shapes are found to be more sculpted by very loose BDT selections and become stable against tighter cuts.}

\rerevision{The first-stage BDT imposes strong selection on the decay processes in both hemispheres of background events. Semileptonic decays, which include large missing energy and have more kinematic resemblance to signals, are found to be significantly preferred in both hemispheres. This effect leads to intrinsic differences between the signals and backgrounds in the non-signal hemisphere, which is otherwise supposed to be random decays.
The second-stage BDT focuses on the signal-like hemisphere, in which the signal decays feature more missing energy and larger angle to the thrust axis than backgrounds. As mentioned in Sec.~\ref{sec:thrust}, this indirectly changes the shape of the non-signal hemisphere energy, therefore the energy variable still exhibits discriminating power after BDT selections.
Most of the background process filtering and the kinematic variable morphing described above is attained at low BDT regions, and the background composition become rather stable towards tight BDT selections.  
On the other hand, the discriminating power in the energy of the signal hemisphere is preserved after BDT selections because very few background processes contain as much missing energy as the signals.
}

Therefore the sum of energy in the signal hemisphere (minimum hemisphere energy) and that in the non-signal hemisphere (maximum hemisphere energy) are chosen as candidates for the final fit, as described in Ref.~\cite{Amhis_2021}.
\revision{The candidate variables are test against the assumption of stable background shapes, with details provided in App.~\ref{app:fit_var}. 
From App.~\ref{app:fit_var}, no significant variation is found for the shape of maximum hemisphere energy under further BDT selections, while the minimum hemisphere energy is clearly sculpted.}
Therefore only the maximum hemisphere energy is considered in the final fit.

The fit is constructed as a combined maximum likelihood fit across two event categories, each with four binned distribution templates for the four processes: $\BcTauNu$,~$\BTauNu$ signals, and~$Z \to b\bar{b}$,~$Z \to c\bar{c}$ backgrounds.
The shape templates for signals are directly modelled by simulations.
Whereas for backgrounds, as mentioned in the previous section, there are not enough events in simulations after the final selection to make a direct model of their shapes. 
The background shape templates are taken as the shapes of the inclusive samples after baseline BDT selections (BDT1~$> 0.9$, BDT2 signal~$> 0.6$, and BDT2 bkg~$<0.1$) and are modelled separately for $Z \to b\bar{b}$ and $Z \to c\bar{c}$ processes.
Since the fit variable is independent of BDT selections, we do not expect the template shapes to change from the baseline to the final selection.

\begin{figure}[h!]
    \centering
    \includegraphics[width=0.45\textwidth]{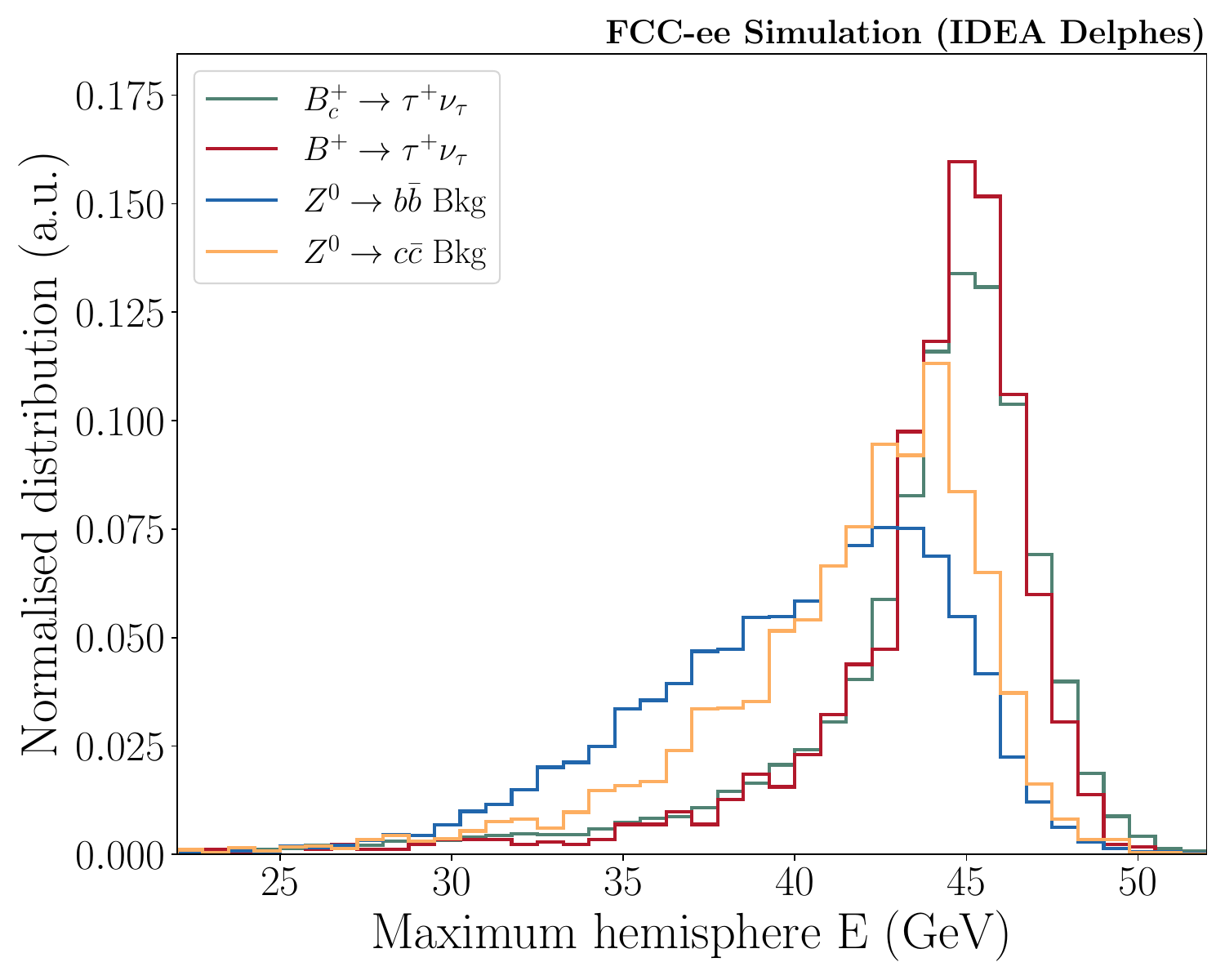}
    \includegraphics[width=0.45\textwidth]{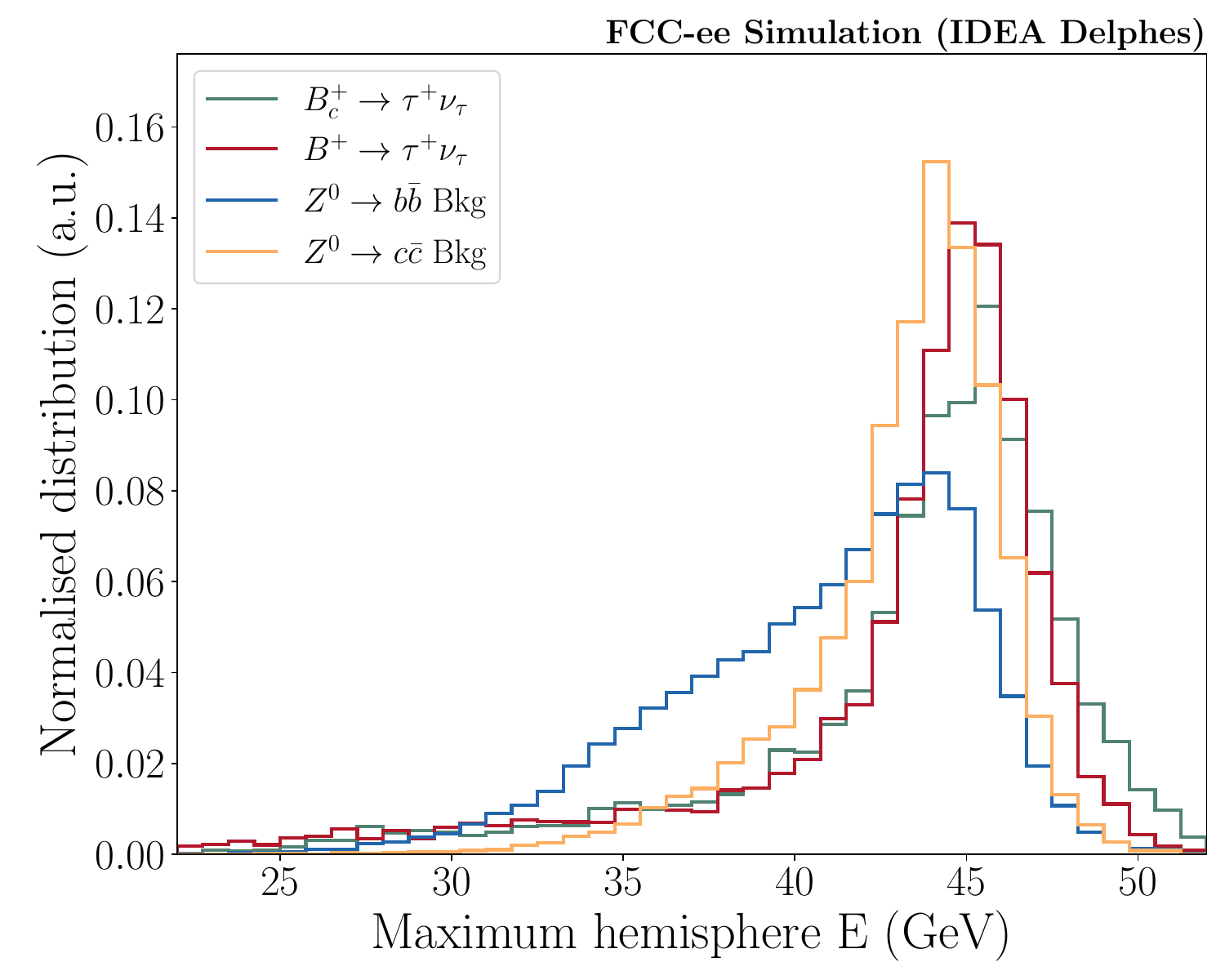}
    \caption{Shape templates of different processes for the final fit in the Bc category (left) and Bu category (right).}
    \label{fig:template_shapes}
\end{figure}

The template shapes in both categories are shown in Fig.~\ref{fig:template_shapes}. 
Signal distributions are in general more centred towards 45~GeV than the background ones.
The two signal modes have similar shapes and their individual contributions are mainly determined by their contrasting expected yields in different categories.
The background shapes, featuring a further spread over low-energy regions, can in general be distinguished from signals in the fit.
The only exception is the $Z \to c\bar{c}$ background in the Bu category, shown as the orange line in the right plot of Fig.~\ref{fig:template_shapes}, which resembles signals in shape.
Given the very small anticipated~$Z \to c\bar{c}$ yield in this category, as provided in Tab.~\ref{tab:expected_yield}, we do not expect it to lead to a visible bias in the signal estimates as long as the best-fit value of the $Z \to c\bar{c}$ events stays in a reasonable range.
This is ensured by posing a very loose constraint on the~$Z \to c\bar{c}$ yield in the Bu category in the likelihood fit, 
which is detailed in the following part of this section as well as Sec.~\ref{sec:syst}.

\begin{figure}[h!]
    \centering
    \includegraphics[width=0.45\textwidth]{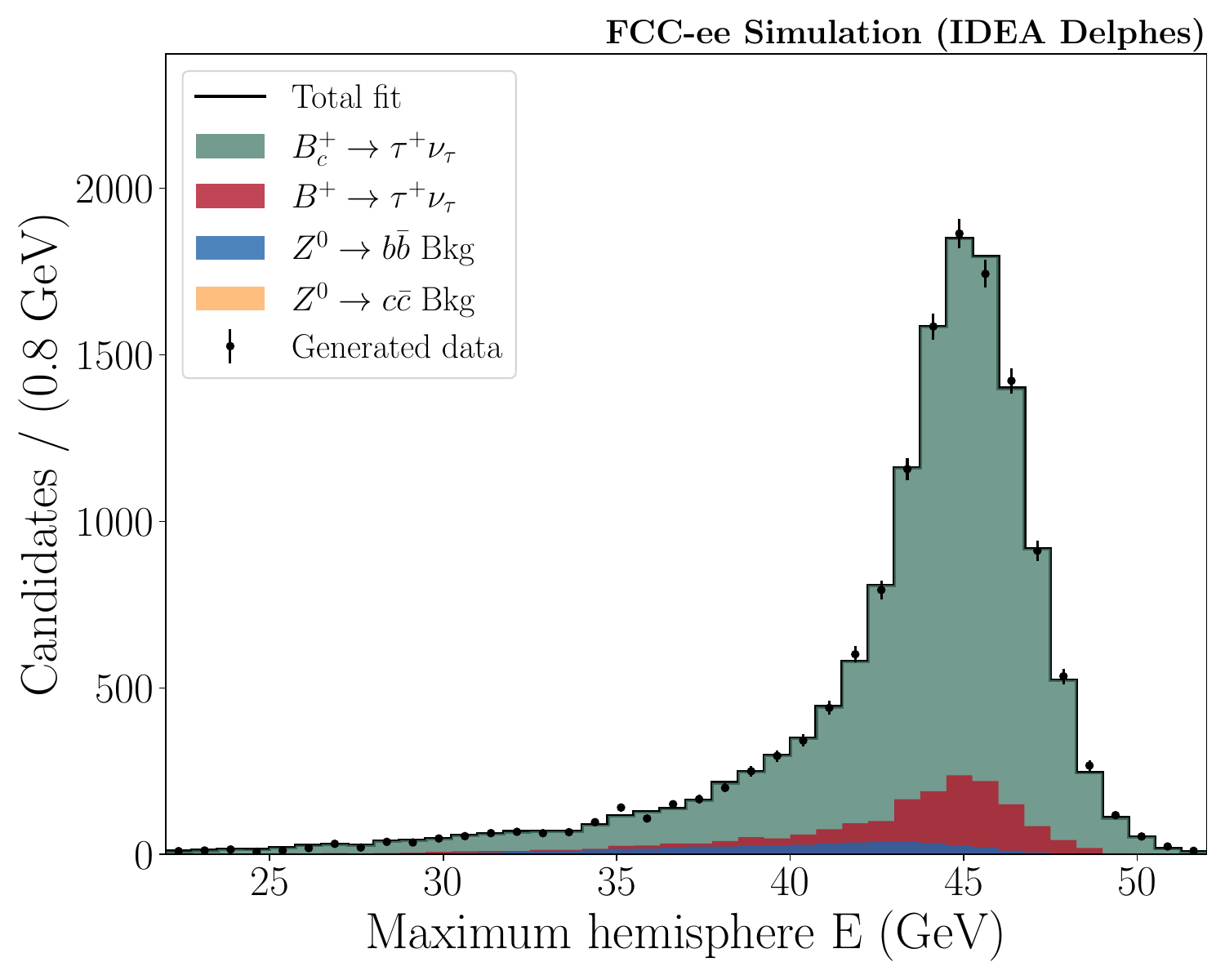}
    \includegraphics[width=0.45\textwidth]{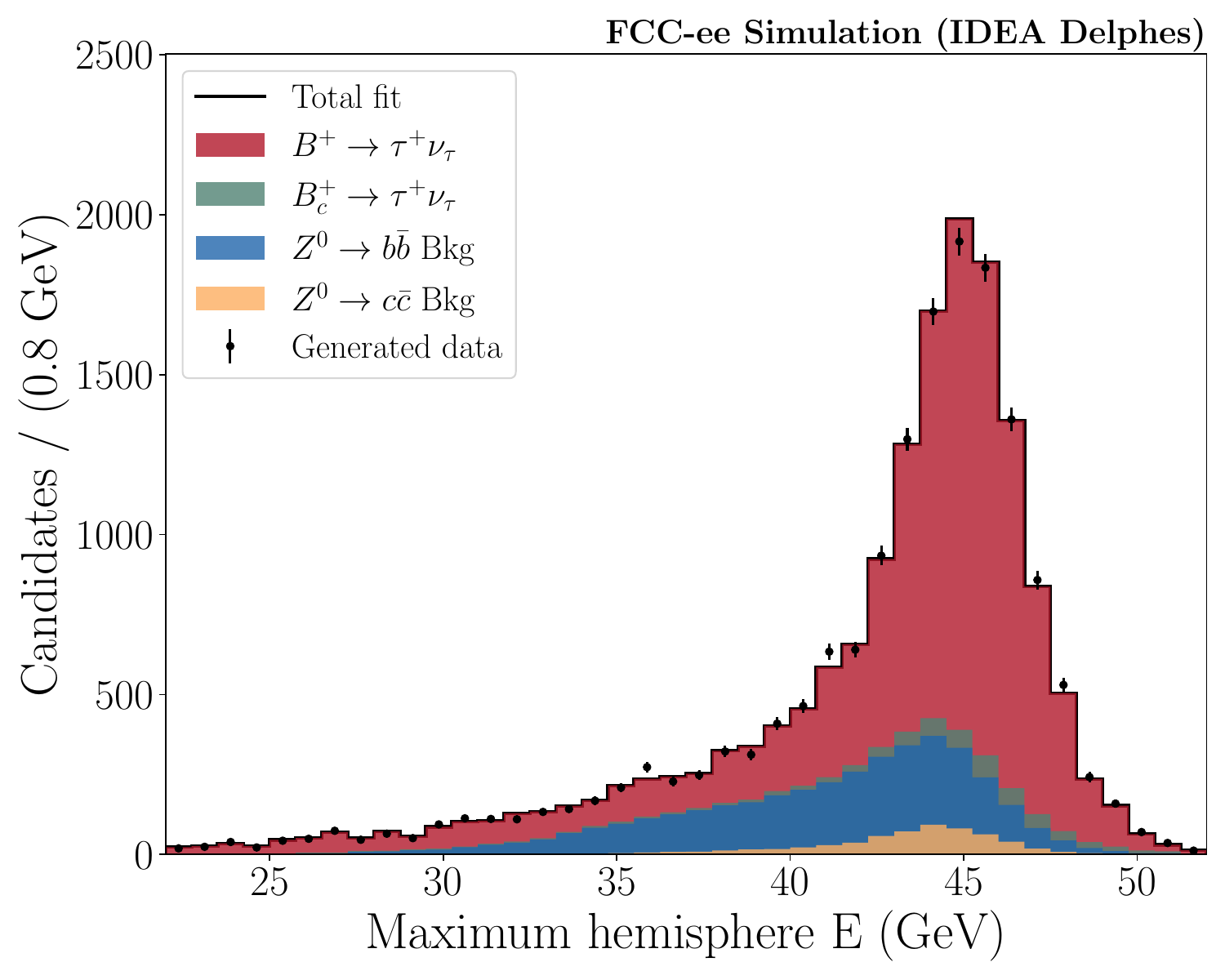}
    \caption{An example of the template fit performed on a pseudo-experiment. Post-fit distributions of signal and background processes are shown alongside the pseudo-data distribution in the Bc category (left) and Bu category (right). }
    \label{fig:fit_example}
\end{figure}

All templates are normalised by their expected yields listed in Tab.~\ref{tab:expected_yield}.
In total, 6 parameters are considered in the fit:
\begin{itemize}
    \item $\mu(\BcTauNu)$ and $\mu(\BTauNu)$: The signal strength modifiers relative to their expected yields. They are correlated across two categories and are fully floating.
    \item $\mu_\text{Bc}(Z \to b\bar{b})$, $\mu_\text{Bc}(Z \to c\bar{c})$, $\mu_\text{Bu}(Z \to b\bar{b})$, $\mu_\text{Bu}(Z \to c\bar{c})$: The background strength modifiers relative to their expected yields in each category. They only appear in their designated category and are uncorrelated from each other. 
    The modifiers $\mu_\text{Bc}(Z \to b\bar{b})$, $\mu_\text{Bc}(Z \to c\bar{c})$, and $\mu_\text{Bu}(Z \to b\bar{b})$ are fully floating.
    The modifier $\mu_\text{Bu}(Z \to c\bar{c})$ is allowed to float without boundary but with a penalty in the log-likelihood of the fit, which corresponds to a lognormal uncertainty equal to 10 times its expected yields. 
\end{itemize}
Figure~\ref{fig:fit_example} shows an example of the fit performed on a pseudo-dataset.
Pseudo-datasets are generated from the sum of the expected templates of all processes, with a bin-by-bin random Poisson fluctuation based on the event count in each bin.

\begin{figure}[h!]
\centering
\includegraphics[width = 0.45\textwidth]{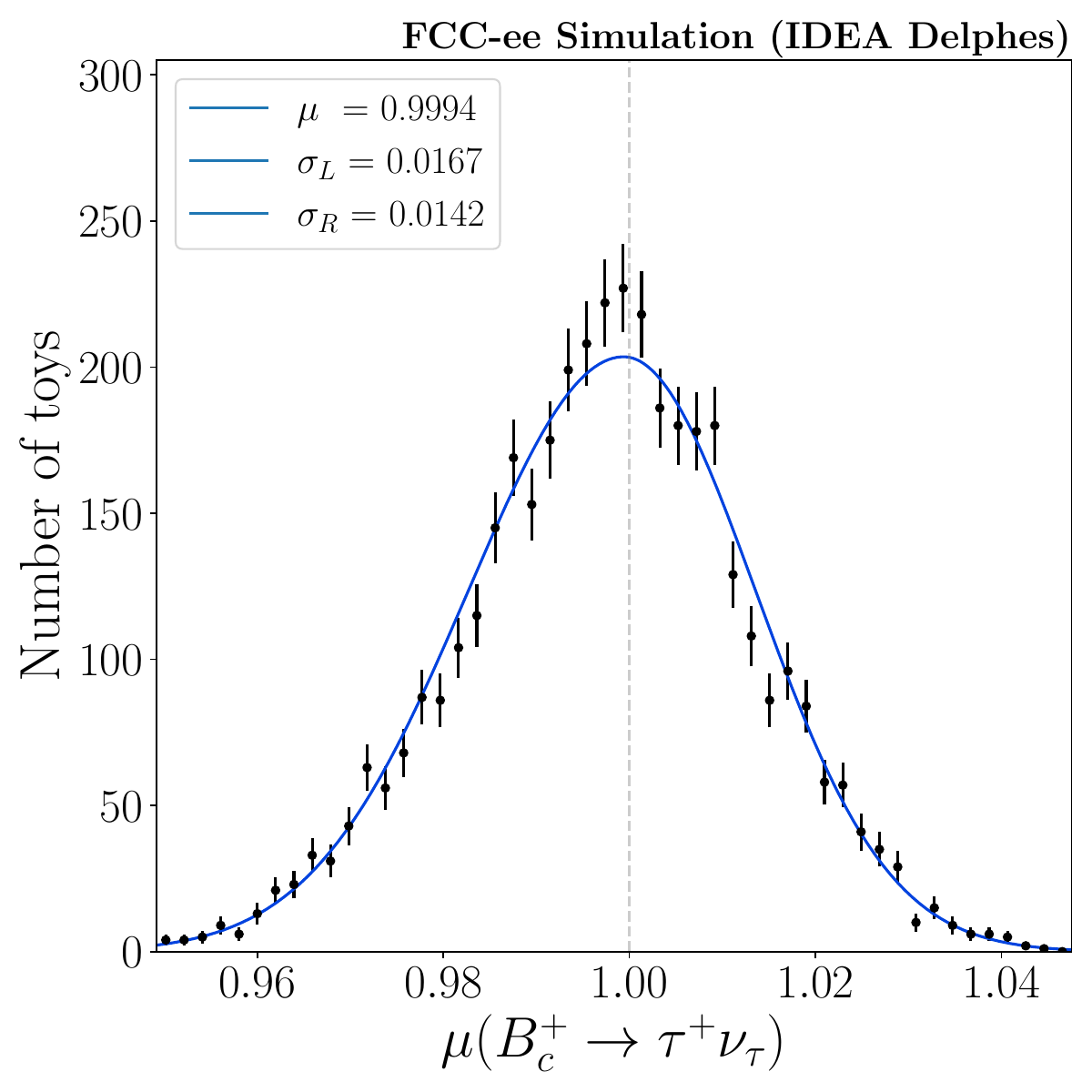}
\includegraphics[width = 0.45\textwidth]{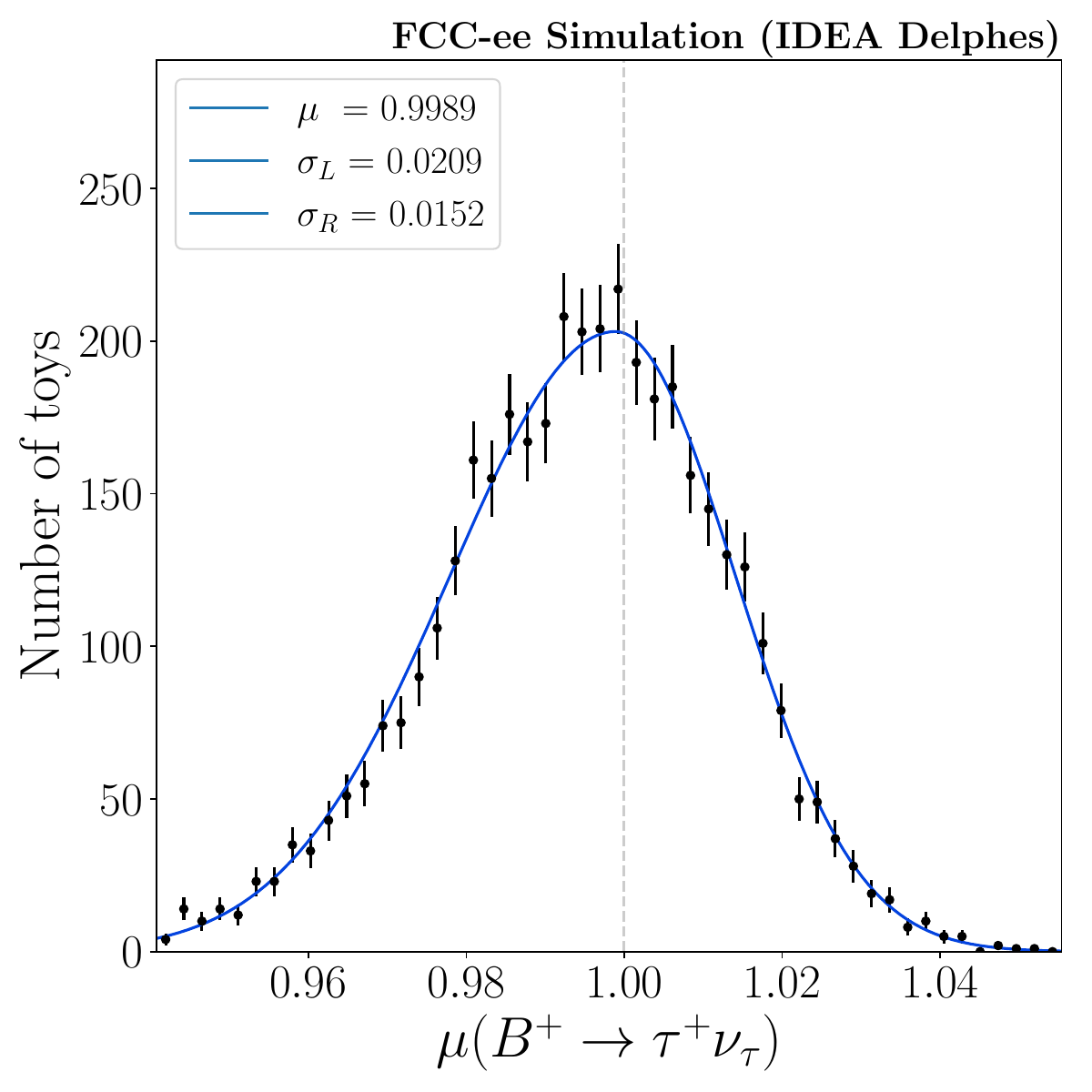}
\caption{Distribution of signal strengths measured in 4000 pseudo-experiments, for $\BcTauNu$ signal (left) and $\BTauNu$ signal (right). A double-sided Gaussian function is used to fit the distributions, shown as the solid blue lines.}
\label{fig:toy_stat}
\end{figure}

To evaluate statistical uncertainties and potential biases from the fit, 4000 pseudo-datasets are generated and the corresponding fit results are summarised for statistical tests.
Figure~\ref{fig:toy_stat} shows the distributions of the best-fit signal strengths from these 4000 pseudo-experiments.
This distribution is fit with a double-sided Gaussian function to examine its central value and uncertainties.
The mean values for both signal strengths have roughly 0.1\% level deviations from 1.0, which are 10~times smaller than the corresponding signal uncertainties.
This indicates that there is no visible bias in the fit results from this approach.
The relative uncertainties on signal strengths are found to be ${}^{+1.4\%}_{-1.7\%}$ for~$\BcTauNu$ signal and ${}^{+1.5\%}_{-2.1\%}$ for~$\BTauNu$ signal.
Uncertainties extracted at this stage are purely statistical. 
Considerations of systematic uncertainties are discussed in the next section.

\subsection{Considerations for systematic uncertainties}\label{sec:syst}

In this section, we discuss the a few main types of systematic uncertainties and their potential impacts on the results.
Note that such considerations are base on the current knowledge of FCC, which is largely speculative at this early stage.
We aim to inspect whether there are major ineluctable uncertainties that would impair the feasibility of these measurements.
Detailed investigations on detector responses and data conditions will be required to provide accurate estimates of systematic uncertainties in the future.

\begin{figure}[!t]
    \centering
    \includegraphics[width = 0.45\textwidth]{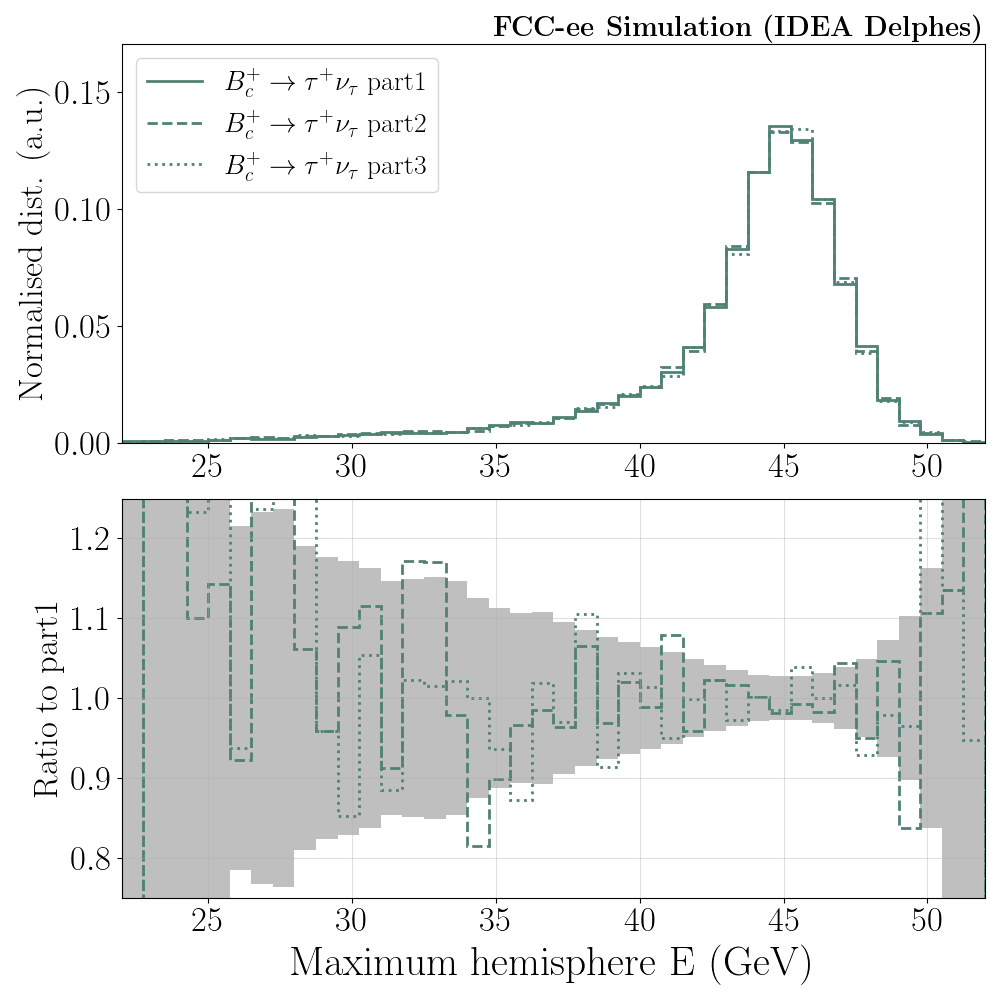}
    \includegraphics[width = 0.45\textwidth]{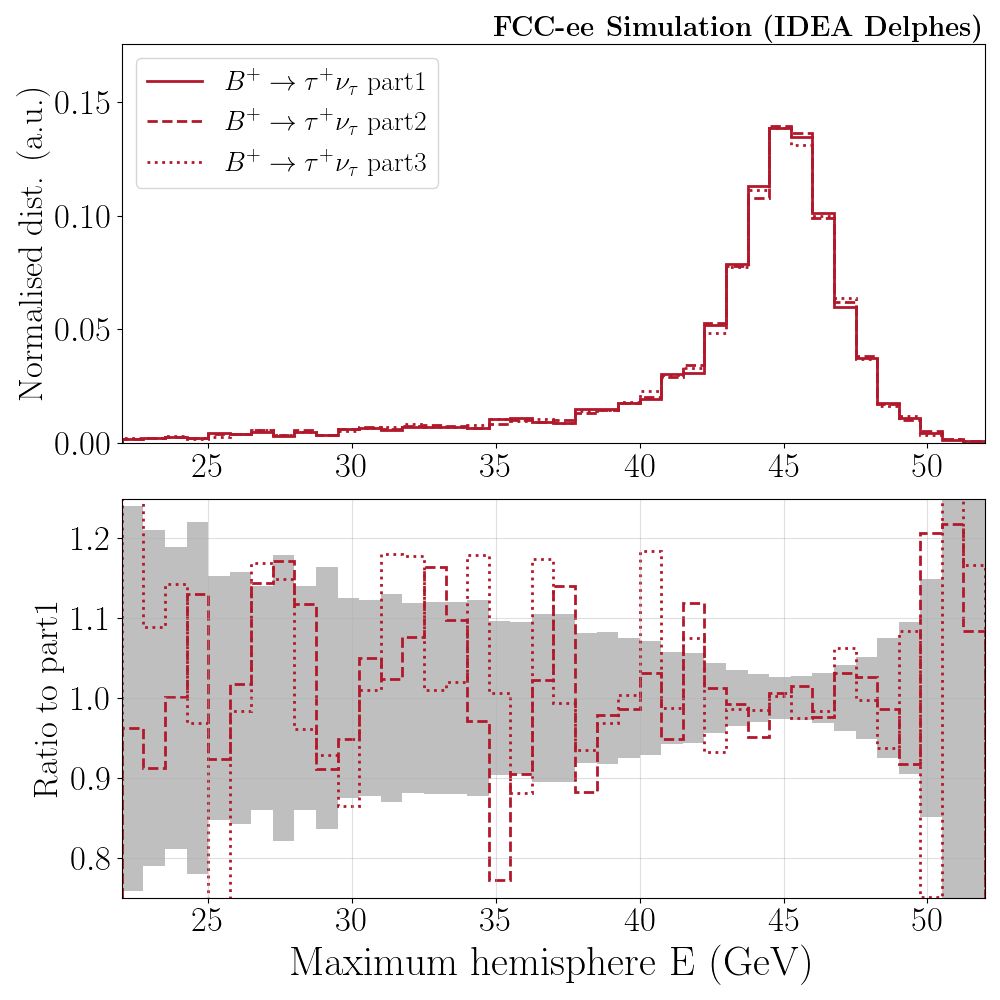}
    \caption{Statistical variation of the signal shape for the $\BcTauNu$ (left) and $\BTauNu$ (right) processes.
             The upper panels show the distributions of the energy in the maximum hemisphere in three equal subsets of the signal simulation.
             The lower panels compare the ratio of the distributions in the second and third subsets (dashed and dotted lines) with regard to the distribution in the first subset. 
             The Poisson uncertainties based on the expected yields in data are also shown as the grey band in the lower panel. }
    \label{fig:shape_syst}
\end{figure}

\begin{itemize}
    \item Particle identification and vertex reconstruction efficiency: Signal events feature one displaced vertex in the signal hemisphere with 3 charged pions. The successful reconstruction of signal events relies on the identification of pions and reconstruction of secondary vertices. In this analysis, particle identification is assumed to be perfect. A prospective study~\cite{Wilkinson:2778909} has shown that pions and kaons can be distinguished at above $3 \sigma$ level in the full kinematic phase-space relevant to this analysis. Therefore we expect the signal uncertainty from particle identification to be at per mille level or smaller. In this analysis, the vertex reconstruction is seeded with true decay vertices from simulation. With real data, we do not expect there to be a significant change of signal vertexing efficiency, as the signal $\TauThreePi$ vertex is very displaced from other hadron activities, but we do expect the imperfect vertexing to lead to a non-negligible amount of combinatorial backgrounds. Impacts of additional backgrounds is discussed in Sec.~\ref{sec:scenarios}.
    \item Signal shape uncertainty: The stability of signal shapes after the final selection is tested by splitting the signal samples into three equal subsets and comparing the signal shape in each subset. Figure~\ref{fig:shape_syst} illustrates the level of signal shape variation in different subsets as well as the bin-by-bin Poisson uncertainty based on the expected event yields. The shapes in the three subsets are found to be consistent within the statistical fluctuation. To evaluate the impact of such variations, a test is performed including these shape variations as a systematic uncertainty on the signal templates with 4000 pseudo-experiments following the procedure described in Sec.~\ref{sec:fit}. The combined uncertainties are found to be ${}^{+1.4\%}_{-1.7\%}$ for~$\BcTauNu$ and ${}^{+1.5\%}_{-2.1\%}$ for~$\BTauNu$, in which the exact numbers differ from those in Sec.~\ref{sec:fit} only in the third digit and are omitted after rounding.
    At the current stage, there is very limited information to speculate how the shape uncertainties would be in an analysis with real data. 
    This test example provides a rough estimate of the robustness of the signal precision against shape uncertainties. We conclude that signal shape variations do not introduce a significant source of uncertainty to the best of our knowledge.
    \item \revision{Potential background shape sculpting from BDT}: As explained in App.~\ref{app:fit_var}, the background shape is very stable against tight BDT selections. Although there is not enough simulated events to provide a quantitative estimate of background shape variations in the final selection, we expect such variations to be very small and not to lead to a visible change in the fit results.
    \item Background efficiency: The background efficiency estimate relies on many assumptions: the stability of BDT performance at high BDT values as discussed in Sec.~\ref{sec:BDT1} and~\ref{sec:BDT2}, the agreement between inclusive background efficiencies and exclusive background efficiencies, and other potential background sources that are not taken into consideration, for example, the combinatorial backgrounds mentioned above.
    In the fit approach described in Sec.~\ref{sec:fit}, the background yields are freely floating. \footnote{$\mu_\text{Bu}(Z \to c\bar{c})$ is loosely constrained but effectively freely floating in the range of interest.} Since the fit does not constrain background yields to their expected values in simulations, any bias in the background efficiency should not affect the fit as a systematic uncertainty.  
    \item \revision{Uncertainty of branching ratios of $B \to DD$ components: as described in Sec.~\ref{sec:sig_bkg_samples}, several $B\to DD$ decay samples are considered as the exclusive decays, which are used for background efficiency estimates in Sec.~\ref{sec:sel_opt}.
    The branching ratios of these decay are not yet known to high precision, and such uncertainties should be propagated into the efficiency estimates and background shape variations.
    We consider a 30\% uncertainty on the branching ratios of all $B \to DD$ processes and check for the impact on the signal sensitivity. The impact turns out to be negligible. This study is summarised in App.~\ref{app:BDD}.}
\end{itemize}

Overall, we do not expect any significant systematic uncertainty from known sources.
However, to make a prediction on the experimental sensitivity, as performed in Sec.~\ref{sec:fit}, pseudo-datasets are generated based on the expected background yields in simulations.
We evaluate the lack of knowledge in the exact background yields in a series of nonideal scenarios 
and provide a range of expected experimental sensitivity in Sec.~\ref{sec:scenarios}.

\subsection{Fit performance for nonideal scenarios}\label{sec:scenarios}

In this analysis, we consider two treatments on the background yields to provide sensitivity estimates in several nominal scenarios, spanning optimistic and very pessimistic cases.
\begin{itemize}
    \item Inflation factor: The expected background yields, as reported in Sec.~\ref{sec:sel_opt}, are multiplied by the inflation factor. In each pseudo-dataset, the same inflation factor is applied to both $Z \to b\bar{b}$ and $Z \to c\bar{c}$ backgrounds in both the Bc and Bu categories. Scenarios of $N_\text{bkg}=$ [1,2,5,10]$\times N_\text{bkg}^\text{exp}$ are considered.
    \item Random fluctuation: When generating pseudo-datasets, in addition to the Poisson fluctuation based on the event yields, a random factor following a lognormal distribution is applied to each background process. The lognormal distribution has a central value (value at 50\% quantile) at 1.0 and a 68\% coverage band of $[1/\BkgFluc, \BkgFluc]$, where $\BkgFluc$ is the relative background yield fluctuation. In each pseudo-dataset, the fluctuation parameter of each background process in each category is independently sampled from the same lognormal distribution. Scenarios of $\BkgFluc=$[1,2,5,10] are considered\footnote{By construction, $\BkgFluc=1$ refers to a scenario without lognormal fluctuation, in which the 68\% coverage band is $[1, 1]$.}.
\end{itemize}

\begin{figure}[h!]
\centering
\includegraphics[width = 0.45\textwidth]{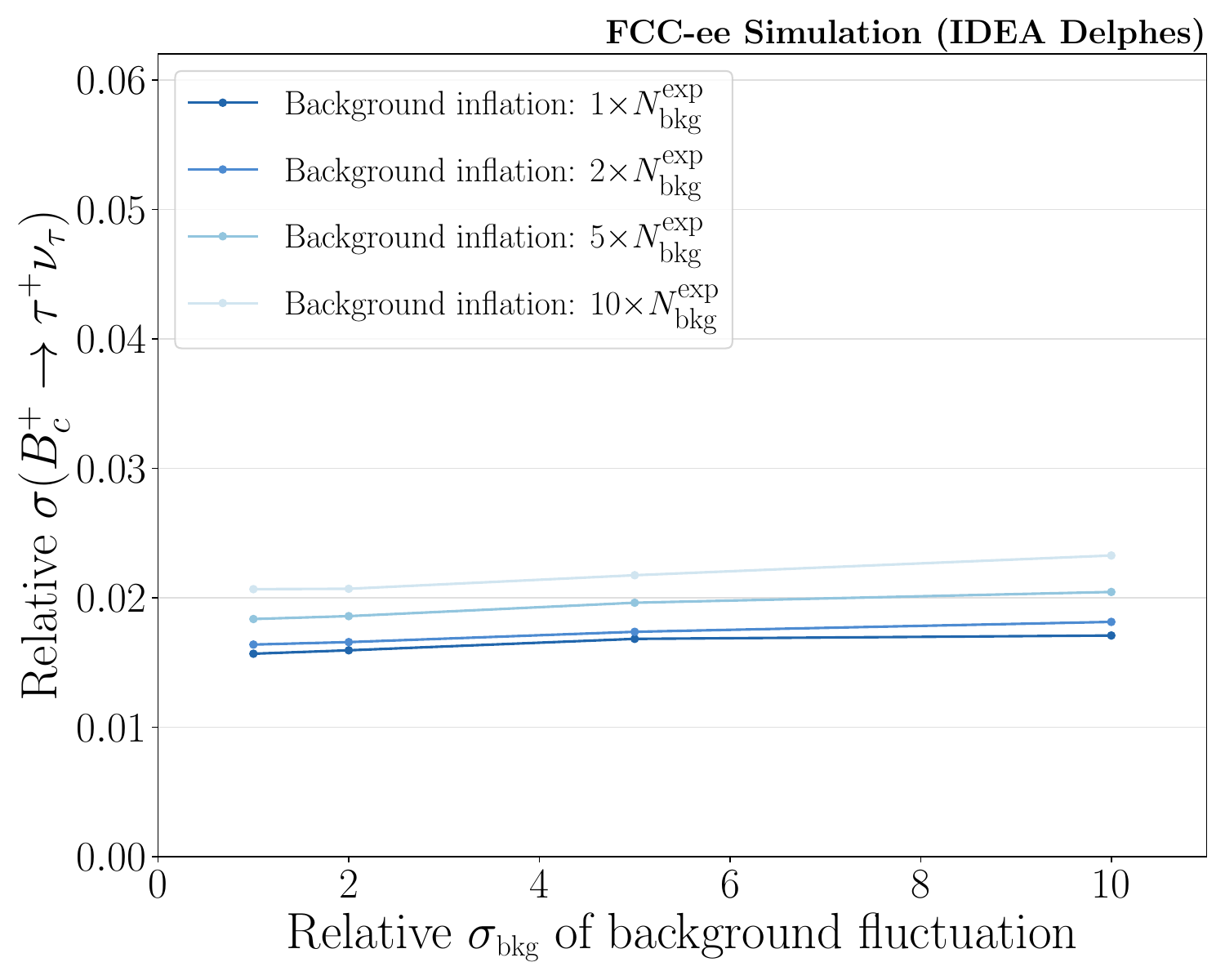}
\includegraphics[width = 0.45\textwidth]{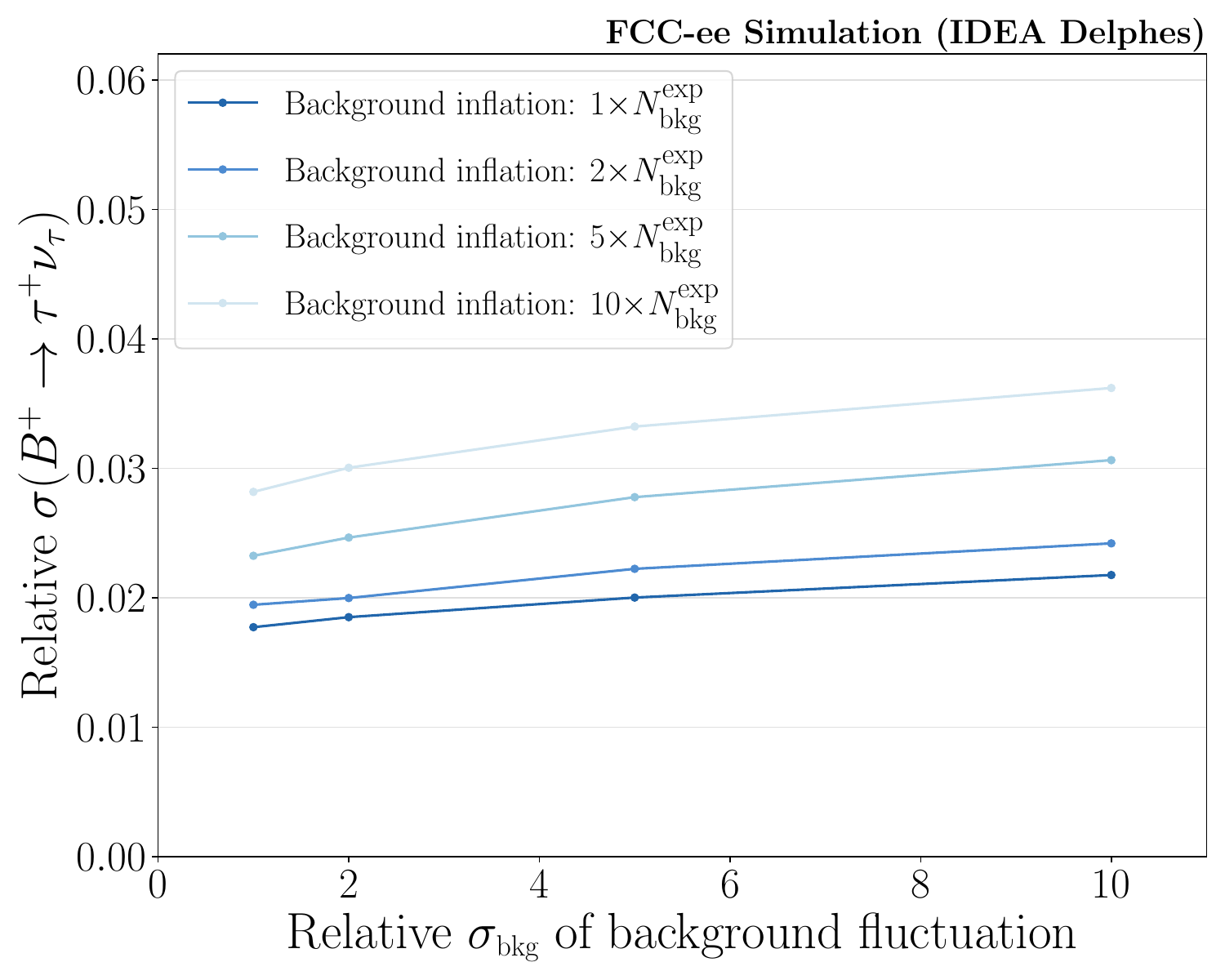}
\caption{Symmetrised relative precision on $\BcTauNu$ (left) and $\BTauNu$ (right) in different scenarios of background inflation and fluctuation. Each point is extracted from 4000 pseudo-experiment fits.}
\label{fig:bkg_scene}
\end{figure}

The signal sensitivity is evaluated with 4000 pseudo-experiments in each scenario of background inflation and fluctuation. 
Figure~\ref{fig:bkg_scene} summarises the expected sensitivity for both $\BcTauNu$ (left) and $\BTauNu$ (right) signals. 
The relative uncertainties shown in Fig.~\ref{fig:bkg_scene} are symmetrised from their up and down uncertainties.
The scenario with $\BkgFluc=1$ and $N_\text{bkg}=N_\text{bkg}^\text{exp}$ is the idealistic case as demonstrated in Sec.~\ref{sec:fit}.

The experimental sensitivity is shown to be reasonably stable against background inflation and fluctuations.
From the idealistic case to the most pessimistic case, the~$\BcTauNu$ precision deteriorates from~$1.6\%$ to~$2.3\%$ and the~$\BTauNu$ precision from~$1.8\%$ to~$3.6\%$.
This high level of resilience against exaggerated backgrounds is a result of the high signal-to-background ratios in both categories. 
We expect the background inflation and fluctuation in pessimistic cases to be large enough to account for most of the uncertainties discussed in Sec.~\ref{sec:syst} 
and such results to provide a reasonable expectation of the achievable precision range at FCC-ee.

\subsection{Expected performance with different data sizes}
\label{sec:less_data}

All experimental predictions shown in previous sections are based on an expected nominal FCC-ee $Z$-pole dataset of 180~ab$^{-1}$, containing~$6\times10^{12}$ $Z$~bosons. 
Note that the nominal luminosity is updated since the previous work on the $\BcTauNu$ decay~\cite{Amhis_2021}. 
This change of expected luminosity leads to a marginal improvement of the experimental precision, while the main improvement of the $\BcTauNu$ precision comparing to the previous work comes from the updated categorisation strategy.
In this paper we only provide results based on $6\times10^{12}$ $Z$~bosons, while similar predictions on experimental sensitivities with different sample sizes, 
in case of further changes of the FCC design or operation schedule, can be inferred with simple calculations.
The uncertainties evaluated in Sec.~\ref{sec:scenarios} are purely statistical uncertainties, which scale inversely with the square root of the sample size, $\propto 1/\sqrt{N_Z}$.
Therefore predictions on the experimental precision, assuming the same analysis strategy, can be calculated by scaling the numbers provided by Sec.~\ref{sec:scenarios}.

\FloatBarrier

%% file: interpretation.tex
\section{Phenomenological Implications in the Standard Model and Beyond}
\label{sec:interpretation}

In this section, we inspect the phenomenological implications that the measurements of $\mathcal{B}(B^+\to \tau^+ \nu_\tau)$ and $\mathcal{B}(B_c^+\to \tau^+ \nu_\tau)$ would have in the Standard Model and Beyond. Our predictions are based on the projected precision with which these modes will be measured at FCC-ee at 180~ab$^{-1}$, as studied in the previous section and summarised in Fig.~\ref{fig:bkg_scene}. For $N(B^+\to \tau^+ \nu_\tau)$, we will consider two different benchmarks corresponding to precisions of $\approx 2\%$ and $\approx 4\%$ in the idealistic and pessimistic case respectively. Whereas for $N(B_c^+\to \tau^+ \nu_\tau)$, as will be discussed in this section, theoretical uncertainties from other sources are non-negligible, and the difference between the idealistic ($1.6\%$) and pessimistic ($2.3\%$) experimental precisions becomes insignificant. Therefore we adopt a single benchmark corresponding to a $\approx 2\%$ precision. As we stated in the introduction it is worth to mention that, while for the $B^+$ channel its signal yield can be easily translated into a branching ratio measurement and employed to infer phenomenological implications, this is not the case for the $B_c^+$ one due to the poor knowledge of its hadronisation fraction $f(B_c^\pm)$. 

In order to circumvent this issue, we choose to factor out $f(B_c^\pm)$ by normalising the $\BcTauNu$ decay to the semileptonic decay mode $B_c^+ \to J/\psi \mu^+\nu_\mu$.
As explained in our previous work~\cite{Amhis_2021}, the normalisation ratio 
\begin{equation}
    \mathcal{R} = \frac{N(B_c^+\to \tau^+ \nu_\tau)}{N(B_c^+ \to J/\psi \mu^+\nu_\mu)} 
                = \frac{\mathcal{B}(B_c^+\to\tau^+ \nu_\tau)} {\mathcal{B}(B_c^+ \to J/\psi \mu^+\nu_\mu)}
                = \frac{\Gamma(B_c^+\to\tau^+ \nu_\tau)}{\Gamma(B_c^+ \to J/\psi \mu^+\nu_\mu)}
\end{equation}
is measured purely from the experiment, in which the experimental uncertainty on $N(B_c^+ \to J/\psi \mu^+\nu_\mu)$ is expected to be significantly smaller than that of $N(B_c^+\to \tau^+ \nu_\tau)$.
In this work, we assume that the normalisation channel is not affected by NP, which is a well justified choice since we will consider scenarios where NP couples only to taus, and not to light leptons. 
Here we place a $\approx 1.5\%$ uncertainty for $N(B_c^+ \to J/\psi \mu^+\nu_\mu)$, which leads to a $\approx 3\%$ uncertainty for $\mathcal{R}$.
In this manner, we obtain the relative precision of $\mathcal{B}(B^+\to \tau^+ \nu_\tau)$ from $\mathcal{R}$ as well as a theoretical external input $\Gamma_{\rm theo}(B_c^+ \to J/\psi \mu^+\nu_\mu)/|V_{cb}|^2$.
For this input the current prediction obtained employing LQCD yields a relative uncertainty of $\approx 7 \%$~\cite{Harrison:2020gvo}. However, current estimates consider achievable to cut down this uncertainty to $\approx 2 \%$ already in the next decade~\cite{CDavies_private}, hence it is conceivable to assume such level of precision for our study. Our final expected precision on $\Gamma(B_c^+ \to J/\psi \mu^+\nu_\mu)/|V_{cb}|^2$ therefore corresponds to $\approx 3.5 \%$. It is interesting to notice that, if LQCD will be able to improve the precision on $\Gamma_{\rm theo}(B_c^+ \to J/\psi \mu^+\nu_\mu)/|V_{cb}|^2$ beyond the current estimates, this will translate in an increase of the expected precision for $\Gamma(B_c^+ \to J/\psi \mu^+\nu_\mu)/|V_{cb}|^2$ as well.

\subsection{Extraction of \texorpdfstring{$\vert V_{ub} \vert$}{|Vub|}}

The measurement of $\mathcal{B}(B^+\to \tau^+ \nu_\tau)$ allows us to perform a direct determination of $\vert V_{ub} \vert$, which we define as $\vert V_{ub} \vert^{\rm lep}$, by inverting Eq.~\eqref{eq:BRBqtaunu}. Observing that the $B^+$ meson lifetime is currently determined with a per mille accuracy~\cite{PDG}, the only other source of the uncertainty in determining $\vert V_{ub} \vert^{\rm lep}$ would come from the theory prediction of the decay constant $\fBu$. As stated in the introduction, the average of its latest LQCD determinations is equal to $\fBu=190.0(1.3)$ MeV~\cite{FLAG:2021npn}, with a $\approx 0.6\%$ relative error. However, according to present estimates this uncertainty could be halved within the next 10 years~\cite{CDavies_private}, well before the FCC timeline. Therefore, the precision of the future extraction of $\vert V_{ub} \vert^{\rm lep}$ will be fully determined by the experimental error.

\begin{figure}[!t!]
\centering
\includegraphics[width = 0.8\textwidth]{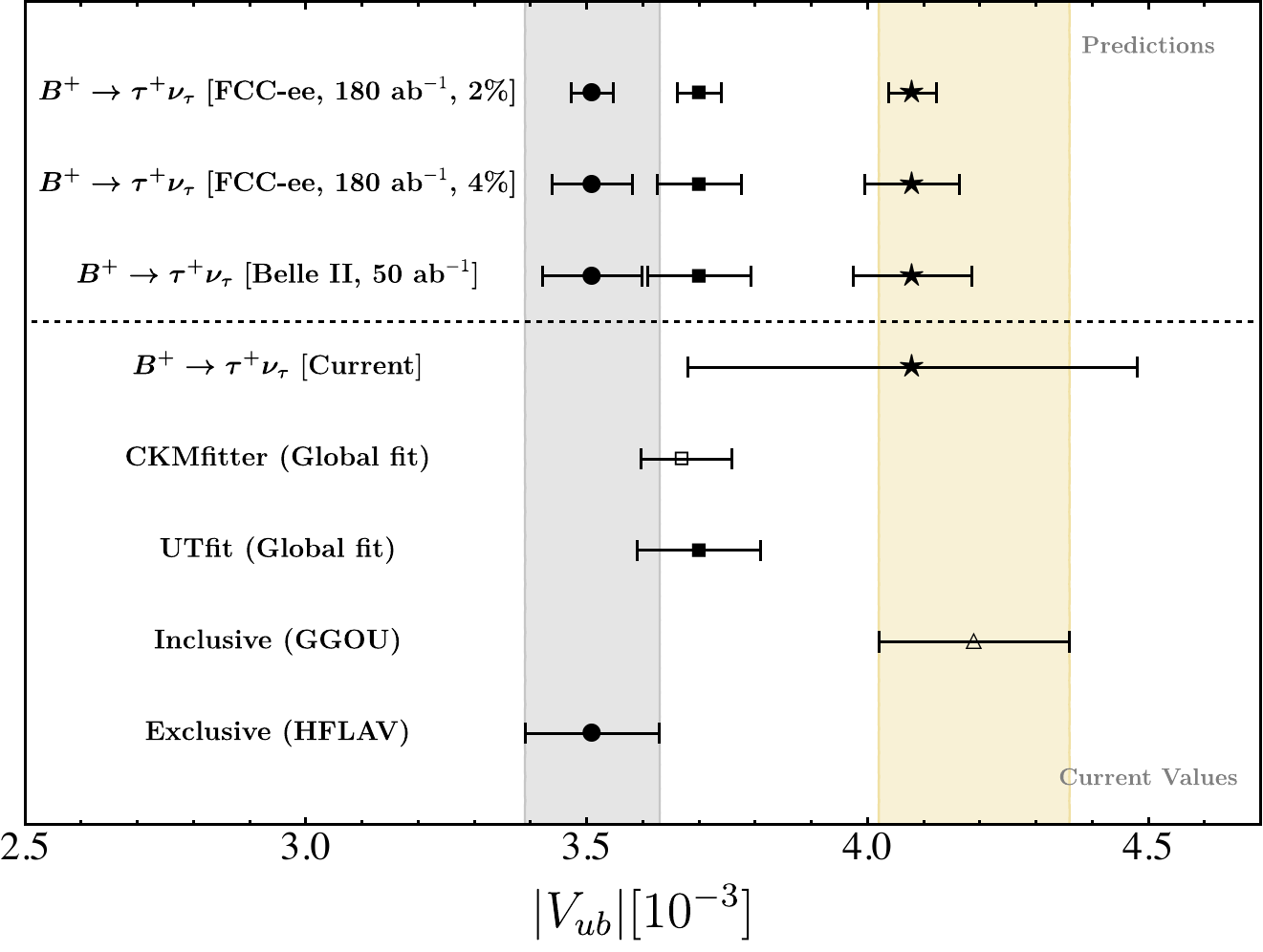}
\caption{Current determinations of $\vert V_{ub} \vert$ from various sources, and future predictions. In the lower part of the plot, we report from bottom to top the current values extracted by a fit to exclusive channels~\cite{HFLAV:2022pwe}, a determination from the inclusive channel~\cite{HFLAV:2022pwe} based on the method originally developed in Ref.~\cite{Gambino:2007rp}, the values obtained by the UTfit and CKMfitter global fits~\cite{UTfit:2022hsi,CKMfitter}, and the determination extracted by the average of current measurements of $B^+\to \tau^+ \nu_\tau$~\cite{HFLAV:2022pwe}. In the upper part of the plot, we give the predicted extractions by future measurements of $B^+\to \tau^+ \nu_\tau$ both at Belle II and at FCC-ee: for the former, we assume a precision of $5\%$ in the determination of the branching ratio~\cite{BelleII}, while for the latter we give both the pessimistic and idealistic cases of precision, namely $4\%$ and $2\%$. We give for each of these scenarios three different predictions according to the choice employed for the central value of $\vert V_{ub} \vert$, namely the current exclusive determination, the current UTfit result or the current $B^+\to \tau^+ \nu_\tau$ determination, identified by a circle, a square or a star marker, respectively. To facilitate a comparison, the grey and ochre vertical bands represent the current exclusive and inclusive determinations of $\vert V_{ub} \vert$, respectively.}
\label{fig:Vub}
\end{figure}

We report our findings in Fig.~\ref{fig:Vub}. In the lower part of this plot, we show the current status of the determination of $\vert V_{ub} \vert$ giving the values presently extracted by exclusive and inclusive~\cite{HFLAV:2022pwe} channel determinations, employing the method originally developed in Ref.~\cite{Gambino:2007rp} for the latter, respectively dubbed below as $\vert V_{ub} \vert^{\rm excl.}$ and $\vert V_{ub} \vert^{\rm incl.}$, together with the results from global fits performed by the UTfit~\cite{UTfit:2022hsi} and the CKMfitter~\cite{CKMfitter} collaborations, $\vert V_{ub} \vert^{\rm fit}$. Moreover, we also show the  determination extracted by the average of current measurements of $\mathcal{B}(B^+\to \tau^+ \nu_\tau)$~\cite{HFLAV:2022pwe}. The current value extracted for $\vert V_{ub} \vert^{\rm lep}$ is compatible with the present inclusive determination, but with a large error due the current precision of the $\mathcal{B}(B^+\to \tau^+ \nu_\tau)$ determination, equal to $\approx 20\%$. An incompatibility with the exclusive determination of the CKM element can therefore be hardly claimed. Hence, this channel currently plays no relevant role neither in the determination of $\vert V_{ub} \vert^{\rm excl.}$, nor in the attempt to understand and reconciling the long-standing $\vert V_{ub} \vert^{\rm incl.}$ vs. $\vert V_{ub} \vert^{\rm excl.}$ puzzle.

This situation will change in the future. Belle II expected precision on the determination of $\mathcal{B}(B^+\to \tau^+ \nu_\tau)$ is equal to $\approx 5\%$~\cite{BelleII}, and as we shown in the previous section, FCC-ee will bring it down potentially to $\approx 2\%$. Given the accuracy of the theoretical inputs described above, this will directly reflect into a significant increase in the precision of the extraction of $\vert V_{ub} \vert^{\rm lep}$, as the reader can see from the upper part of Fig.~\ref{fig:Vub}.\footnote{It is worth to mention that, with this predicted level of accuracy, it would be warranted to include radiative QED corrections to Eq.~\eqref{eq:BRBqtaunu}, presently not known given the current lack of motivation due to its experimental status. This kind of correction is indeed capable to halve the precision of the extraction of CKM elements in decays better measured, like $K$ or $\pi$ leptonic decays~\cite{DiCarlo:2019thl}.} In this section of the plot, we give predictions for Belle II and both the pessimistic and the idealistic scenarios at FCC-ee. Given the obtained level of accuracy, we present each of these predictions in three different benchmarks, differentiated by the choice adopted for the central value of $\vert V_{ub} \vert^{\rm lep}$: for the first benchmark, we assume that $\mathcal{B}(B^+\to \tau^+ \nu_\tau)$ will align to the other exclusive modes, and hence employ the current exclusive average for the central value of $\vert V_{ub} \vert^{\rm lep}$; in the second benchmark, we forecast that the channel to converge to what is currently suggested from global CKM fits, and assume a central value for the CKM element in according to the current UTfit determination; for the last benchmark, we instead keep the central value as the same of the current average of $\mathcal{B}(B^+\to \tau^+ \nu_\tau)$ measurements. These benchmarks are identified in the plot by a circle, a square and a star marker, respectively.

Several considerations are now in order. First and foremost, the level of accuracy with which $\vert V_{ub} \vert^{\rm lep}$ could be extracted at FCC-ee in the idealistic case is less than a third of the current precision stemming from exclusive average, while about a half in the pessimistic case. This implies that, contrary to the current situation, $\mathcal{B}(B^+\to \tau^+ \nu_\tau)$ will start to play a role in the extraction of the determination of $\vert V_{ub} \vert^{\rm excl.}$ in the future. Moreover, if the central value will be similar to what we assumed for our third benchmark, we would obtain a theoretically clean exclusive determination in accordance with the current inclusive one. This could be a hint to the reason behind the long-standing exclusive vs. inclusive discrepancy: on the one hand, the semileptonic decays currently employed for the $\vert V_{ub} \vert^{\rm excl.}$ are prone to theory errors due to the presence of form factors; on the other hand, the theoretical prediction of the fully inclusive rate used for the extraction of $\vert V_{ub} \vert^{\rm incl.}$ is cleaner, but also hindered by the need to perform phase space cuts due to experimental limitations. A measurement of $\vert V_{ub} \vert^{\rm lep}$ compatible with $\vert V_{ub} \vert^{\rm incl.}$ could therefore point of a better control of the latter theoretical errors, rather than the former ones. If we will instead observe a result in accordance to the first or second benchmark, this would strengthen the case of the exclusive determination of $\vert V_{ub} \vert^{\rm excl.}$, which is currently the preferred one given the better accordance with results coming from global CKM fits.

In conclusion, the determination of $\vert V_{ub} \vert^{\rm lep}$ through the measurement of $\mathcal{B}(B^+\to \tau^+ \nu_\tau)$ at FCC-ee will be of extreme relevance, potentially playing a key role in the resolution of the long-standing $\vert V_{ub} \vert^{\rm incl.}$ vs. $\vert V_{ub} \vert^{\rm excl.}$ puzzle. 

\subsection{New Physics Implications}

We proceed now to study the implications that the measurements of $\BFBTauNu$ and $\BFBcTauNu$ at FCC-ee would have on some specific NP models. In order to perform this kind of analysis, we start by introducing the most general dimension-six effective Hamiltonian for $b\to q\tau\nu$ transitions, with $q=u,c$, namely
\begin{align}\label{eq:left}
\mathcal{H}_\mathrm{eff} &= 2\sqrt{2}G_F \sum_{q=u,c} V_{qb}\Big{[}\left(1+C^q_{V_L}\right)\,\big{(}\bar{q}_{L}\gamma_\mu b_{L} \big{)}\big{(}\bar{\tau}_L \gamma^\mu\nu_{L}\big{)}+C^q_{V_R}\,\big{(}\bar{q}_{R}\gamma_\mu b_{R} \big{)}\big{(}\bar{\tau}_L \gamma^\mu\nu_{L}\big{)}\\[0.38em]
&+C^q_{S_L}\,\big{(}\bar{q}_{R} b_{L} \big{)}\big{(}\bar{\tau}_R \nu_{L}\big{)}+C^q_{S_R}\,\big{(}\bar{q}_{L} b_{R} \big{)}\big{(}\bar{\tau}_R\nu_{L}\big{)}+C^q_{T}\,\big{(}\bar{q}_{R}\sigma_{\mu\nu} b_{L} \big{)}\big{(}\bar{\tau}_R \sigma^{\mu\nu}\nu_{L}\big{)}\,\Big{]}+\mathrm{h.c.}\,,\nonumber
\end{align}
where the effective coefficients $C^q_{\alpha}\equiv C^q_{\alpha}(\mu)$ with $\alpha\in \lbrace V_{L(R)},S_{L(R)},T\rbrace$ are evaluated at the renormalisation scale $\mu=m_b$. The normalisation is chosen in such a way that the SM can be recovered by setting $C^q_{\alpha}=0$ for all coefficients. Note also that here and below we are assuming that NP couples only to taus, and not to light leptons.

Starting from this Hamiltonian, the NP contribution to the leptonic $B_q^+$ decays can be written as
\begin{align}
\label{eq:Bctaunu}
\mathcal{B}(B_q^+ \to \tau^+ \nu_\tau)= \mathcal{B}(B_q^+ \to \tau^+ \nu_\tau)^\mathrm{SM}\times\left|1-\left(C^q_{V_R} - C^q_{V_L}\right)+\left(C^q_{S_R} - C^q_{S_L}\right)\dfrac{m_{B_q}^2}{m_\tau(m_b+m_q)}\right|^2\,,
\end{align}
which can also be expressed in terms of common combinations $C^q_{A} \equiv C^q_{V_R}- C^q_{V_L}$ and $C^q_{P} \equiv C^q_{S_R}- C^q_{S_L}$. It is worth mentioning that, due to chiral enhancement, $\mathcal{B}(B_q^+ \to \tau^+ \nu_\tau)$ is particularly sensitive to contributions from the scalar coefficient $C^q_{P}$. The combinations $C^q_{V} \equiv C^q_{V_R}+ C^q_{V_L}$ and $C^q_{S} \equiv C^q_{S_R}+ C^q_{S_L}$, or the coupling $C^q_T$, cannot be probed by leptonic $B^+_q$ decays, but are on the other hand accessible in decays involving a pseudoscalar meson in the final state like, e.g., $B^0 \to D^-(\pi^-) \tau^+ \nu_\tau$ decays, for $q=c(u)$ respectively. Moreover, complementary information can be obtained by decays with a vector in the final state, such as $B^0 \to D^{* -} \tau^+ \nu_\tau$ and $B_c^+ \to J/\psi \tau^+ \nu_\tau$, or $B^0 \to \rho^- \tau^+ \nu_\tau$ and $B^+ \to \omega \tau^+ \nu_\tau$, respectively for the $q=c$ and $q=u$ channels. For a recent global analysis in $b\to c$ and $b\to u$ transitions, see e.g. Refs.~\cite{Iguro:2022yzr} and \cite{Bhatta:2020yvb}, respectively.

The effective couplings introduced in Eq.~\eqref{eq:left} can be used to parametrize at the low scale NP effects stemming from new heavy fields, once they are integrated out, and hence constrain such models. In this section, we focus particularly on two specific extensions of the Standard Model, namely the generic Two-Higgs-doublet model (G2HDM)~\cite{Branco:2011iw} and specific models containing scalar and vector LQs~\cite{Buchmuller:1986zs,Dorsner:2016wpm}. This choice is justified by the fact that these models predict a non-negligible value for the $C^q_{P}$ coefficients, particularly in light of a potential explanation of the anomalies observed in $b\to c$ transitions. Details on the results obtained for such models as discussed in the following sections.

\subsubsection{G2HDM}

\begin{figure}[!t!]
\centering
\includegraphics[width = 0.28\textwidth]{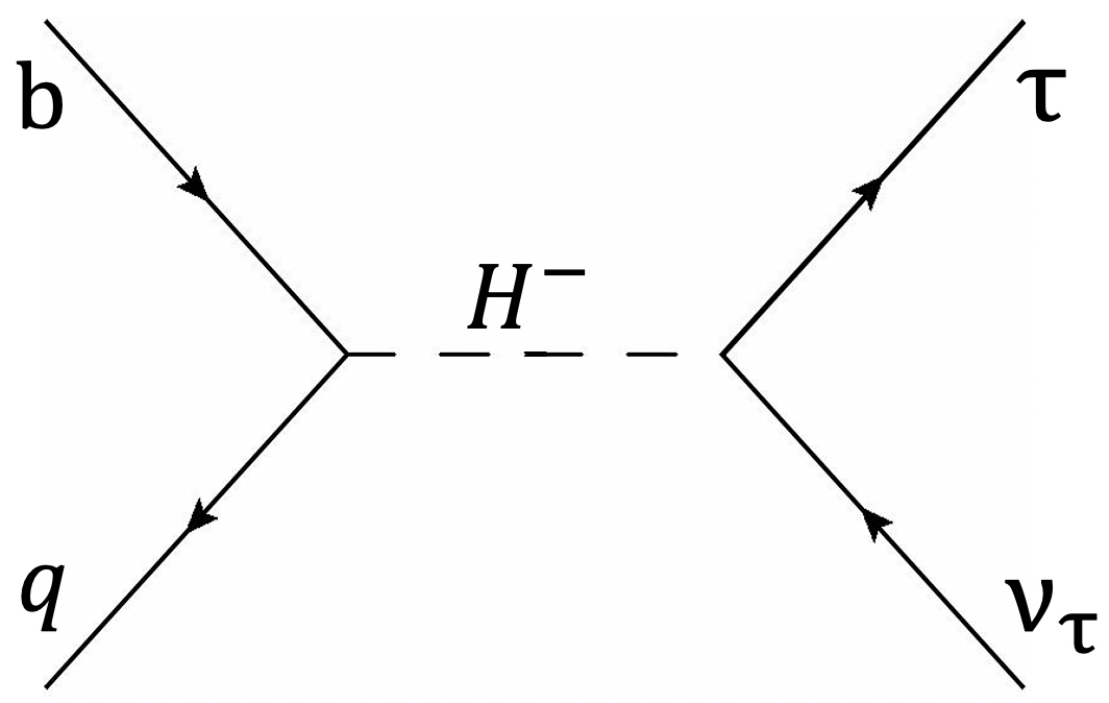}\hspace{3em}
\includegraphics[width = 0.28\textwidth]{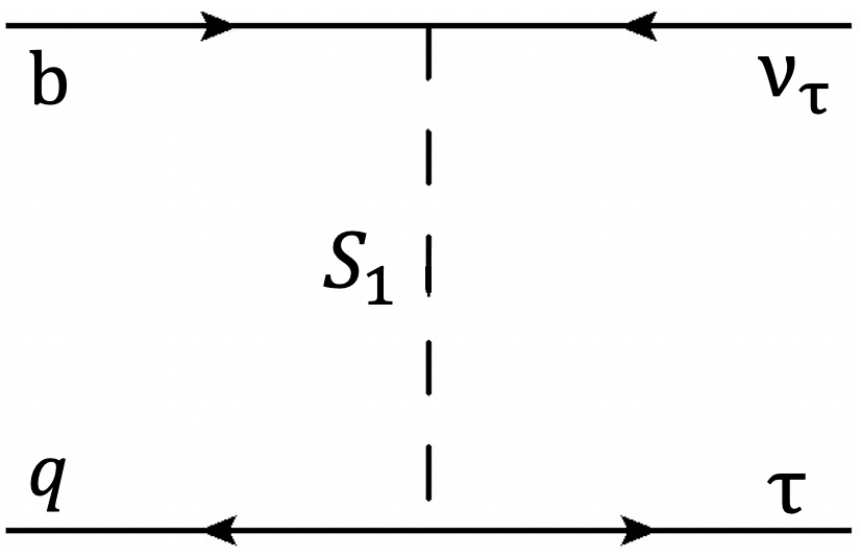}\hspace{3em}
\includegraphics[width = 0.28\textwidth]{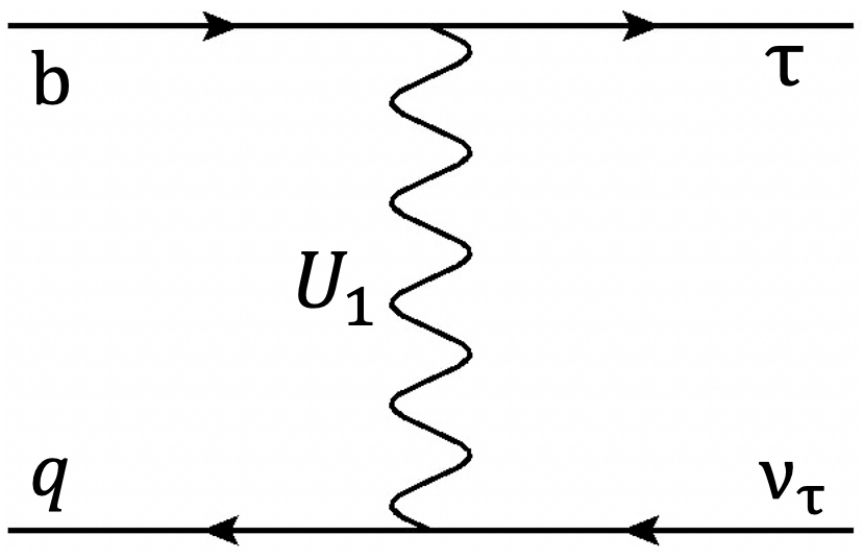}
\caption{Left diagram: additional tree-level contribution to $b\to q \tau \nu_\tau$ transitions due to the presence of a charged Higgs, $H^-$. Central diagram: additional tree-level contribution to $b\to q \tau \nu_\tau$ transitions due to the presence of the scalar leptoquark $S_1$. Right diagram: additional tree-level contribution to $b\to q \tau \nu_\tau$ transitions due to the presence of the vector leptoquark $U_1$.}
\label{fig:NP_diags}
\end{figure}

The first NP model that we consider here is obtained by extending the Higgs sector with an additional Higgs doublet, carrying the same quantum numbers under the SM gauge group~\cite{Branco:2011iw}. This addition will enrich the scalar sector of the SM with 4 new states: a second $C\!P$-even scalar, a neutral $C\!P$-odd one, and a charged Higgs. This last particle could mediate the same kind of transitions which occurs in the SM via the exchange of a $W$ boson, and hence would contribute to $B_q^+\to \tau^+ \nu_\tau$ decays as well, as depicted at the partonic level in the left diagram of Fig.~\ref{fig:NP_diags}. Allowing both Higgs doublets to couple to all fermions leads however to tree level flavour changing neutral currents, which are heavily suppressed in the SM. For this reason, often one imposes some kind of discrete symmetry on the fermions and scalar fields, such that only fermions of a specific chirality and hypercharge should couple to a single Higgs doublet~\cite{Glashow:1976nt}. However, it was observed that a G2HDM where such a symmetry is not imposed can be used to address the $b\to c$ anomalies, given it allows to evade bounds such the ones discussed in Ref.~\cite{Faroughy:2016osc}, and is currently the only realisation of 2HDM still capable to explain such measurements, see e.g. Ref.~\cite{Iguro:2022uzz,Blanke:2022pjy} and references therein.

In particular, defining the simplified Lagrangian
\begin{equation}\label{eq:G2HDM_lag}
\mathcal{L}_{\rm G2HDM} \supset y^q_Q H^-(\bar{b}P_R q) - y_\tau H^-(\bar\tau P_L \nu_\tau) + \textrm{h.c.}\,,
\end{equation}
where we introduced the complex coefficients $y^q_Q$ and $y_\tau$ which are mediating the interaction between the charged Higgs boson and the SM fermions, we can relate such a Lagrangian to the effective coefficients appearing at Eq.~\eqref{eq:left} once integrating out the heavy charged Higgs. In particular, it is possible to write
\begin{equation}
C_{S_L}^q= \frac{1}{2\sqrt{2}G_FV_{qb}}\frac{y_Q^{q*}y_\tau}{m_{H^-}^2}\,,
\end{equation}
 where the coefficients are related to the couplings introduced in Eq.~\eqref{eq:G2HDM_lag} and the mass of the charged Higgs mass $m_{H^-}$.

\begin{figure}[!t!]
\centering
\includegraphics[width = 0.49\textwidth]{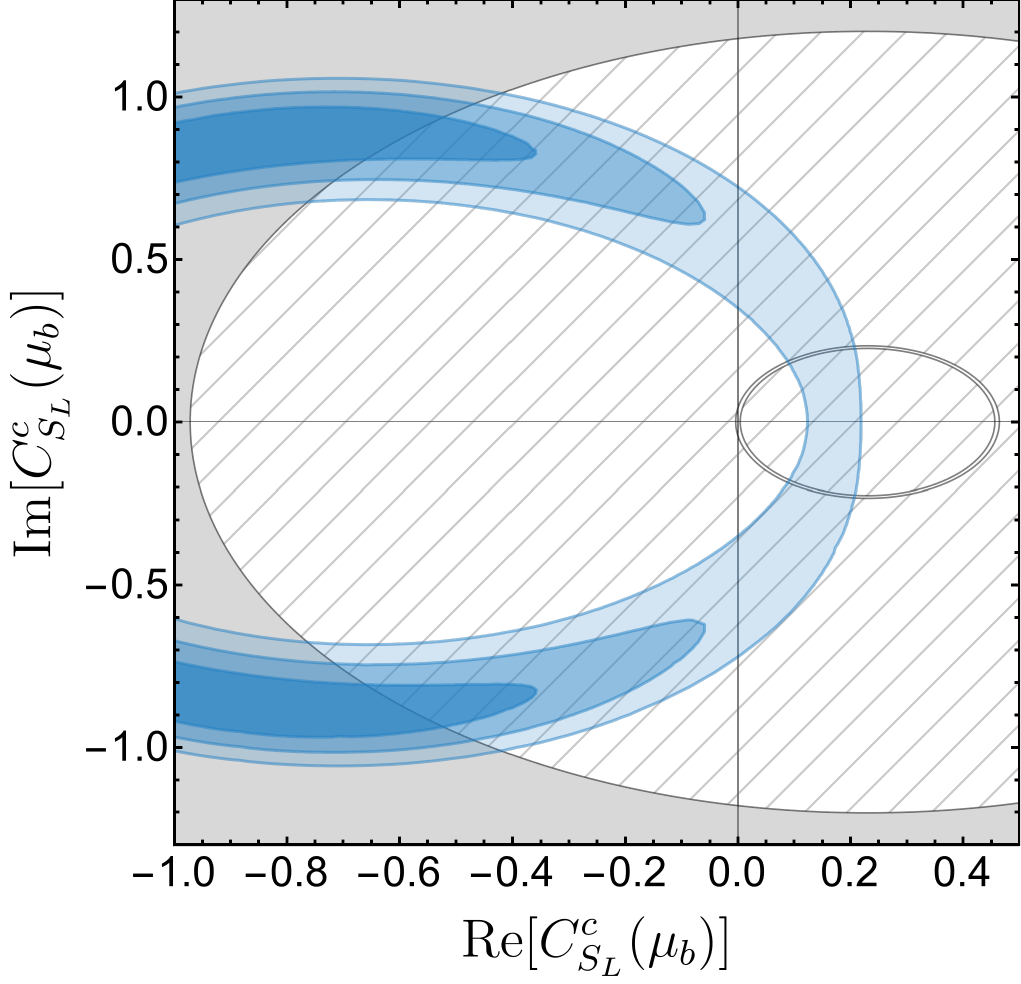}\hspace{0.5em}
\includegraphics[width = 0.49\textwidth]{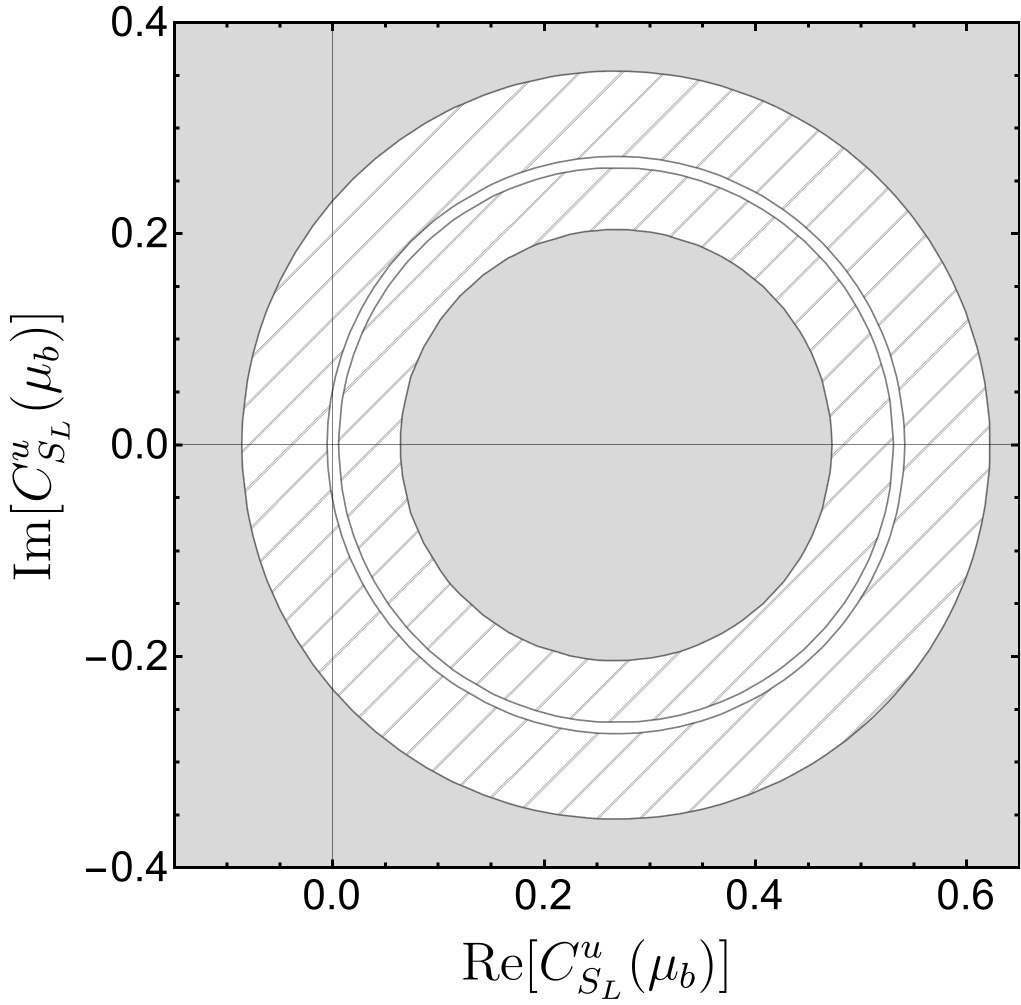}
\caption{Left panel: allowed region for the G2HDM induced complex $C_{S_L}^c$ coefficient. The grey region represents the constraint coming from current constraint on $\mathcal{B}(B_c^+\to \tau^+ \nu_\tau)$, while the hashed region corresponds to the future constraint induced by FCC-ee. In blue are shown the 1$\sigma$, 2$\sigma$ and 3$\sigma$ regions (darker to lighter) inferred on the couplings from $b\to c$ anomalies. Right panel: analogous to the left panel, but for the $B^+\to \tau^+ \nu_\tau$ channel.}
\label{fig:G2HDM}
\end{figure}

The expected FCC-ee constraints on the plane $\textrm{Re}(C_{S_L}^q)$ vs.~$\textrm{Im}(C_{S_L}^q)$ are shown in Fig.~\ref{fig:G2HDM}, where we report on the left panel the bounds for the $c$ sector and on the right panel the ones for the $u$ one. In this plots, we show both current and future constraints coming from $B_q^+\to \tau^+ \nu_\tau$ measurements. In particular, for the current constraint on the still-to-be observed $\Bc$ tauonic decay we employed the conservative estimate $\BFBcTauNu \leq 60\%$~\cite{Blanke:2018yud,Blanke:2019qrx}, which is the reason why the unconstrained region appears as a disk and not as an annulus, while for the $\Bu$ measurement we can rely on the average $\BFBTauNu = (1.094 \pm 0.208 \pm 0.043) \times 10^{-4}$~\cite{HFLAV:2022pwe}. The future constraints are inferred employing for their central values the current SM prediction in the $\Bc$ channel and the current experimental average in the $\Bu$ one, and from a measurement with a relative precision of $3.5\%$ and $4\%$, respectively. Here and below, we refrain ourselves to show the results obtained assuming a $2\%$ precision in the $\Bu$ system as well, in order not to impede the readability of the plots by the reader. As can be observed from Fig.~\ref{fig:G2HDM}, the future measurements at FCC-ee in these channels will strongly constrain the currently allowed parameter space for the G2HDM. Of particular relevance in the $\Bc$ sector, due to its relation to the aforementioned $b\to c$ anomalies. To this end, we show in the plot also the current 1$\sigma$, 2$\sigma$ and 3$\sigma$ regions required in order to address such anomalies: while the current limit on $\BFBcTauNu$ already imposes a stringent constraint on the allowed parameter space, FCC-ee will be able to fully probe the 1$\sigma$ and 2$\sigma$ regions, leaving only a marginal part of the 3$\sigma$ one untested. The measurement of this channel at FCC-ee will be therefore of extreme relevance in the context of the current $B$-meson charged current anomalies, capable of playing a major, independent role in confirming or rebutting the presence of NP in such channels.
It is worth mentioning that, while for $m_{H^-} \ge 400\,$GeV stringent constraints are imposed by $\tau\nu$ resonance searches, this is not the case for $m_{H^-} \le 400\,$GeV where the lack of constraints still allows the explanation of the $b\to c$ anomalies~\cite{Iguro:2018fni}. We therefore assume a mass for the charged Higgs below this value, hence evading current constraints.

\subsubsection{Leptoquarks}

The second class of NP models considered in our analysis are obtained adding to the SM field contents a new class of particles, named leptoquarks, which couple directly to leptons and quarks in the same vertices. Several different realisation of LQs exists, given their spin and their quantum number under the SM gauge group~\cite{Buchmuller:1986zs}. In our analysis we will focus on two specific models, which have been recently object of extensive study due to their ability to satisfactorily address the aforementioned $b\to c$ anomalies, see e.g. Ref.~\cite{Iguro:2022yzr} and references therein. We classify these models by the LQ quantum numbers $(SU(3)_c,SU(2)_L,U(1)_Y)$, where $Q=Y+T_3$ is the electric charge, $Y$ denotes the hypercharge and $T_3$ is the third component of weak isospin.

The first LQ that we study here is the scalar Leptoquark $S_1$, characterised by the quantum numbers $(\mathbf{\bar{3}},\mathbf{1},1/3)$. $S_1$ is therefore an $SU(2)_L$ singlet, capable to mediate at tree-level the transition $b\to q\tau\nu$ being this kind of interaction induced by the Lagrangian
\begin{equation}
\mathcal{L}_{S_1} = y_L^{ij}\overline{Q_i^C}i\tau_2L_jS_1 + y_R^{ij}\overline{u_{Ri}^C}l_{Rj}S_1 + \textrm{h.c.}\,,
\end{equation}
where $\tau_2$ is the second Pauli matrix and $y_{L,R}^{ij}$ are generic Yukawa complex couplings. For a diagrammatic depiction of such a contribution, see the central diagram of Fig.~\ref{fig:NP_diags}. Once $S_1$ is integrated out of the theory, its low energy footprints in $b\to q \tau\nu$ transitions are mediated by the couplings
\begin{equation}\label{eq:S1_WC}
C_{V_L}^q(\mu_{\rm LQ})= \frac{1}{4\sqrt{2}G_FV_{qb}}\frac{y_L^{b\nu}\left(Vy_L^*\right)^{q\tau}}{m_{S_1}^2}\,, \quad\,
C_{S_L}^q(\mu_{\rm LQ})= - 4 C_T^q(\mu_{\rm LQ}) = -\frac{1}{4\sqrt{2}G_FV_{qb}}\frac{y_L^{b\nu}\left(y_R^*\right)^{q\tau}}{m_{S_1}^2}\,,
\end{equation}
where the matching is performed at the scale of the $S_1$ mass, $\mu_{\rm LQ}=\m_{S_1}$.  Both the vectorial and the combined scalar/tensor couplings are independently capable to address the $b\to c$ anomalies. However, since the former would also induce a tree-level contribution to $b\to s\nu\nu$ due to the $SU(2)_L$ symmetry implying additional constraint, we will focus below only on the latter.

\begin{figure}[!t!]
\centering
\includegraphics[width = 0.49\textwidth]{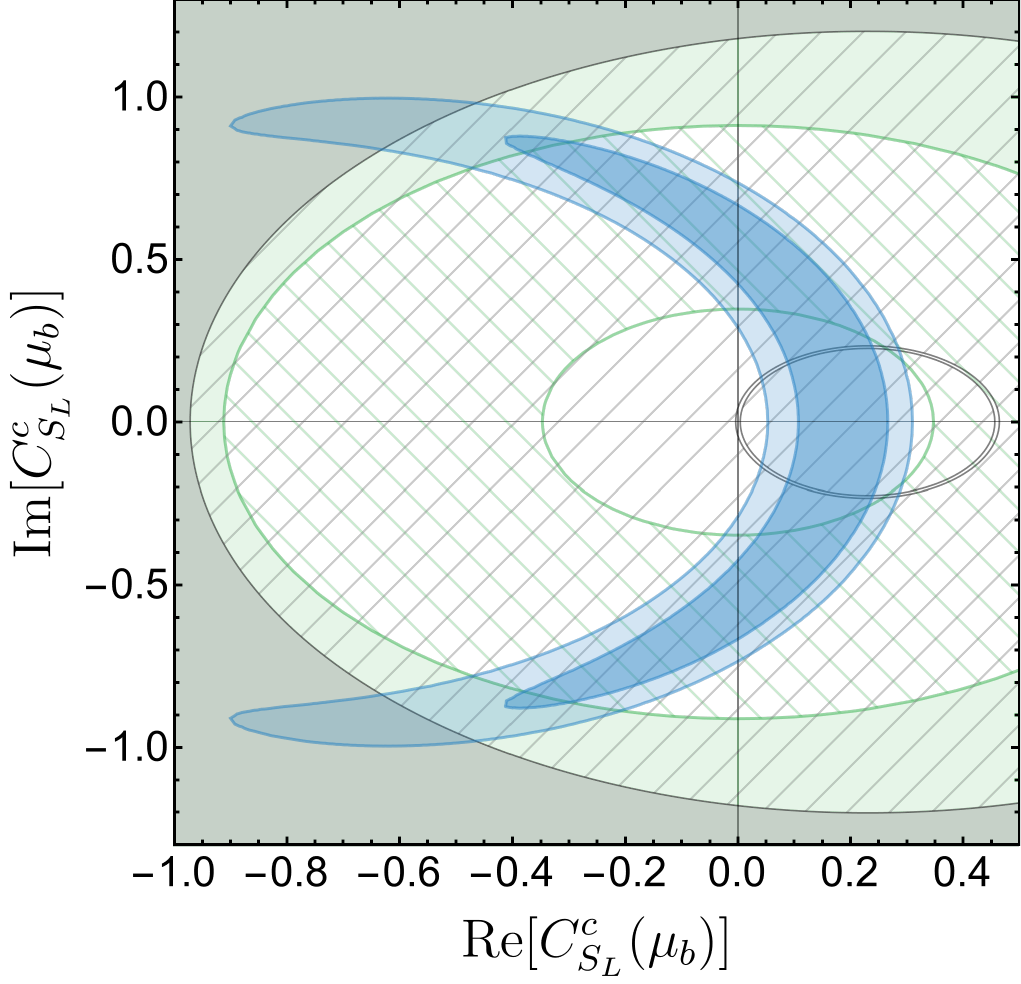}\hspace{0.5em}
\includegraphics[width = 0.49\textwidth]{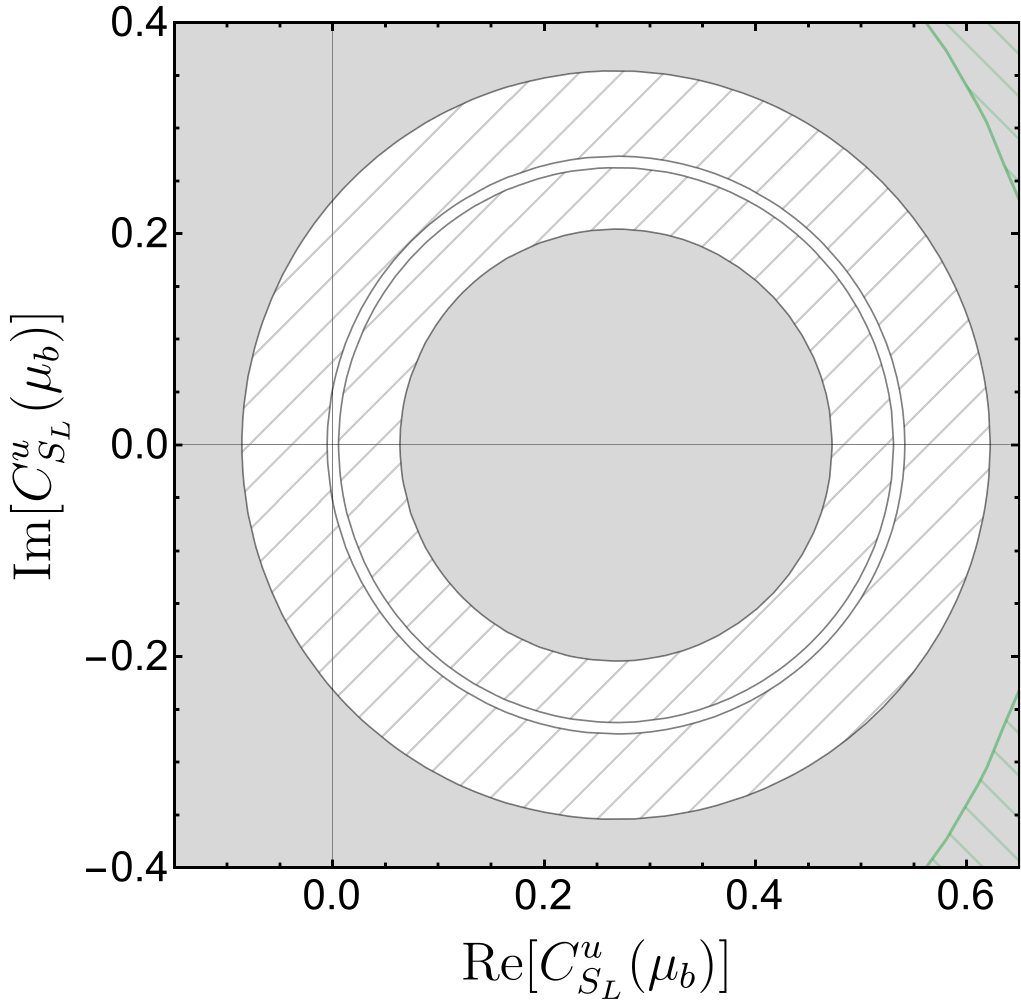}
\caption{Left panel: allowed region for the $S_1$ induced complex $C_{S_L}^c$ coefficient. The grey region represents the constraint coming from current constraint on $\mathcal{B}(B_c^+\to \tau^+ \nu_\tau)$, while the  grey hashed region corresponds to the future constraint induced by FCC-ee. The green region shows the constraint coming from current direct searches at LHC, with future limits from HL-LHC given in hashed green. In blue are shown the 2$\sigma$ and 3$\sigma$ regions (darker to lighter) inferred on the couplings from $b\to c$ anomalies. Right panel: analogous to the left panel, but for the $B^+\to \tau^+ \nu_\tau$ channel.
}
\label{fig:S1}
\end{figure}

We report in Fig.~\ref{fig:S1} the implications of future measurements of $B_q^+\to \tau^+ \nu_\tau$. In our analysis we assumed $\m_{S_1}=2$ TeV, which modifies the relation between the scalar and the tensor coefficients to $C_{S_L}^q= - 8.9 C_T^q$ due to renormalisation group evolution (RGE), once the coefficients are evolved down to $\mu_b=m_b$~\cite{Gonzalez-Alonso:2017iyc}. Following the same procedure employed in the previous section, we show both present and future exclusion coming from the measurement of the leptonic decays branching ratios. On top of those, we also overlay the collider limits coming from LQ mediated high $P_T$ mono-$\tau$ search at the LHC~\cite{Iguro:2020keo}, together with the projected sensitivity for HL-LHC~\cite{Iguro:2020keo,Endo:2021lhi}. Finally, we report the current 2$\sigma$ and 3$\sigma$ regions required for the coefficients in order to address the charged current anomalies. Different considerations are in order for the two channels here analysed. For the $B_c^+\to \tau^+ \nu_\tau$, given the present lack of a measurement for its branching ratio, the current main constraints come from direct searches at LHC. However this picture is going to change in the future, where FCC-ee and HL-LHC will play complementary roles in constraining the currently available parameter space, strongly reducing the possibility of an $S_1$ explanation of the anomalies but not able to completely rule it out. On the other hand, having already a measurement for $\mathcal{B}(B^+\to \tau^+ \nu_\tau)$ makes current LHC constraint non competitive. Moreover, the suppression coming from a different CKM normalisation in the coefficients of Eq.~\eqref{eq:S1_WC} compared to the $B^+_c$ channel is capable to overcome the PDF enhancement of this channel. Therefore, even after the advent of HL-LHC this picture will not be altered, making the measurement of $\mathcal{B}(B^+\to \tau^+ \nu_\tau)$ at FCC-ee of particular interest.

\begin{figure}[!t!]
\centering
\includegraphics[width = 0.496\textwidth]{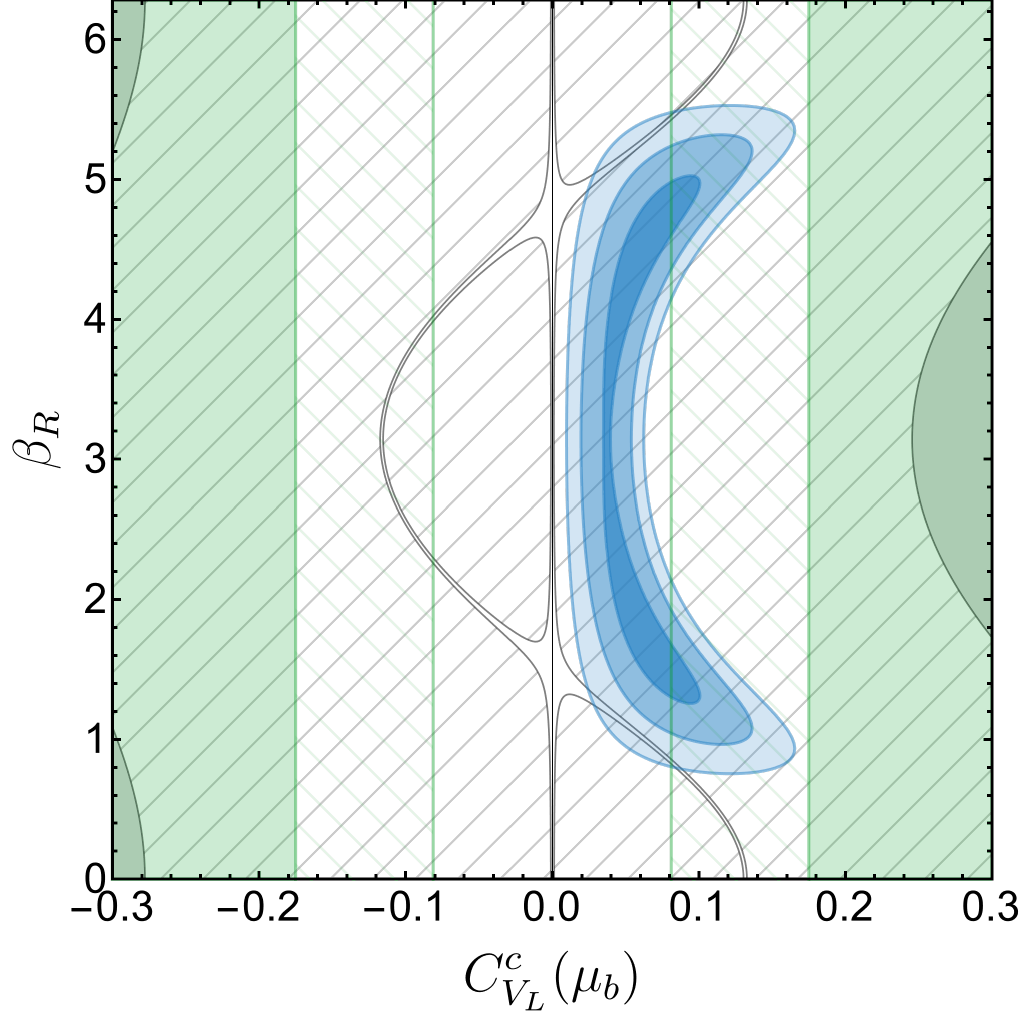}\hspace{0.5em}
\includegraphics[width = 0.484\textwidth]{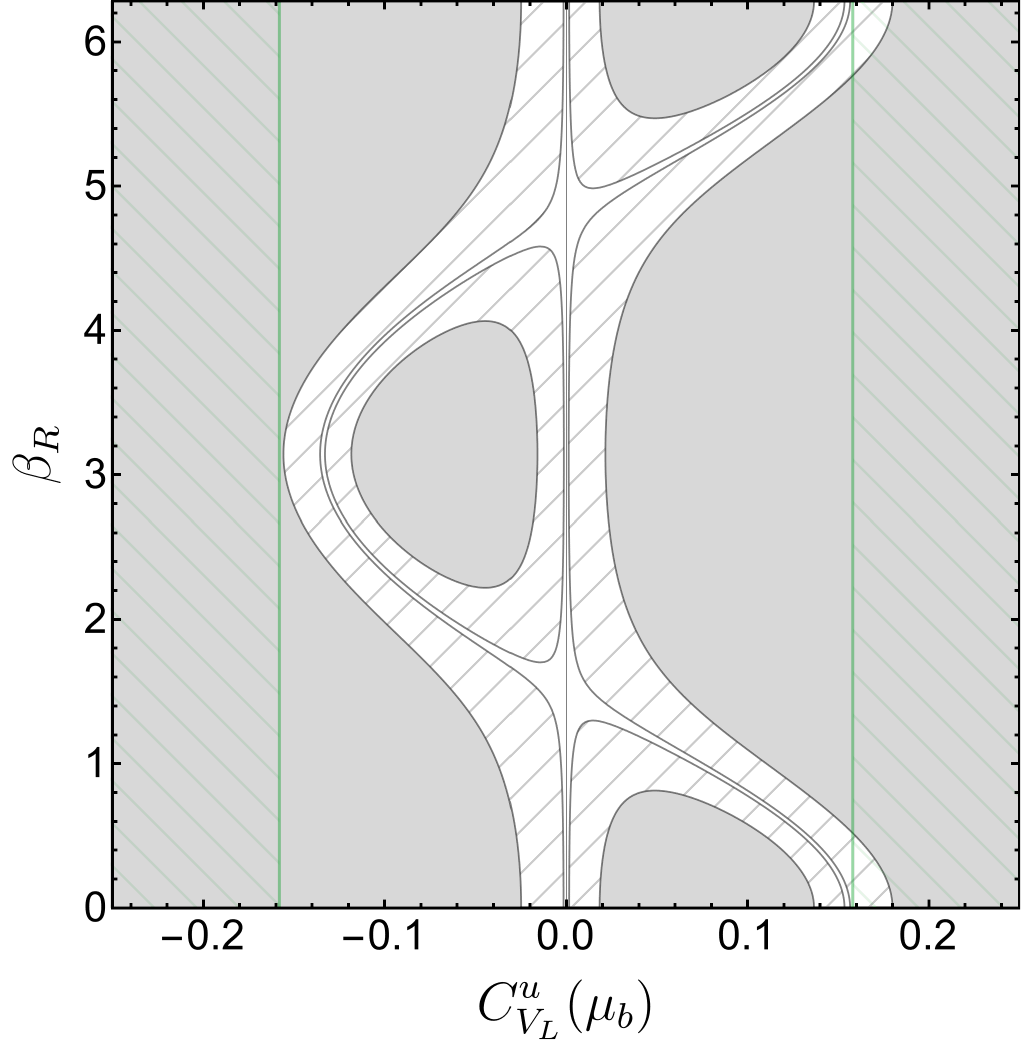}
\caption{Left panel: allowed region for the $U_1$ induced $C_{V_L}^c$ real coefficient and $\beta_R^q$ complex phase, see text for details. The colour scheme is the same of Fig.~\ref{fig:S1}. Right panel: analogous to the left panel, but for the $B^+\to \tau^+ \nu_\tau$ channel.}
\label{fig:U1}
\end{figure}

The second LQ model that we inspect is the singlet vector Leptoquark $U_1=(\mathbf{3},\mathbf{1},2/3)$. As depicted at in the right diagram of Fig.~\ref{fig:NP_diags}, also this field can mediate at tree-level the transition $b\to q\tau\nu$, being described by the Lagrangian
\begin{equation}
\mathcal{L}_{U_1} = \hat z_L^{ij}\overline{Q_i}\gamma_\mu L_jU^\mu_1 + \hat z_R^{ij}\overline{d}_{Ri}\gamma_\mu l_{Rj}U^\mu_1 + \textrm{h.c.}\,,
\end{equation}
where the generic complex couplings $\hat z_{L,R}^{ij}$ are defined in the interaction basis. Integrating out now the vector LQ and going to the mass basis, one obtains
\begin{equation}
C_{V_L}^q(\mu_{\rm LQ})= \frac{1}{2\sqrt{2}G_FV_{qb}}\frac{\left(Vz_L\right)^{q\tau}\left(z_L^*\right)^{b\tau}}{m_{U_1}^2}\,, \quad\quad
C_{S_R}^q(\mu_{\rm LQ})= - \frac{1}{\sqrt{2}G_FV_{qb}}\frac{\left(Vz_L\right)^{q\tau}\left(z_R^*\right)^{b\tau}}{m_{U_1}^2}\,,
\end{equation}
with the matching performed at the $U_1$ mass scale, $\mu_{\rm LQ}=\m_{U_1}$. Being the $U_1$ LQ a massive vector state, it requires some sort of UV completion to explain the origin of its mass: a common realisation consists into considering it as a gauge boson of a new $U(2)$ symmetry~\cite{Fuentes-Martin:2019mun}, and assuming that in the interaction basis it couples only to SM fields of the third generation, i.e., $\hat z_L^{b\tau}=\hat z_R^{b\tau}\neq 0$, and vanishing otherwise. Rotating now to the mass basis will induce two consequences: first, the appearance of couplings like $z_L^{c\tau}$ or $z_L^{u\tau}$; second, the mis-alignment of left- and right-handed couplings due to the physical phase $\beta_R$, coming from the rotation matrices and implying $z_R^{b\tau}= e^{-i\beta_R} z_L^{b\tau}$. We therefore obtain, in a UV completion of the $U_1$ LQ by means of a $U(2)$ symmetry, the following relation:
\begin{equation}
C_{S_R}^q(\mu_b)= -3.7 e^{i\beta_R} C_{V_L}^q(\mu_b)\,,
\end{equation}
where we have taken into account the RGE effects~\cite{Gonzalez-Alonso:2017iyc}. The implications of future measurements of $B_q^+\to \tau^+ \nu_\tau$ on this model are reported in Fig.~\ref{fig:U1}. Analogously to what done for the previous cases, we show present and future bounds on the parameter space coming both from $\mathcal{B}(B_q^+\to \tau^+ \nu_\tau)$ and high $P_T$ mono-$\tau$ searches at LHC. Similarly to the $S_1$ scenario, it is interesting to notice that in the $B_c$ channel FCC-ee and HL-LHC will play complementary roles in constraining this model. Of particular interest is the capability of FCC-ee to constrain the CP-violating phase $\beta_R$, which is instead not probed at HL-LHC. This will fully enable the probe of the 1$\sigma$ and 2$\sigma$ regions identified by the $b\to c$ anomalies, leaving only a minimal part of the 3$\sigma$ parameter space not probed. Once again, the high precision with which FCC-ee will be capable to measure $\mathcal{B}(B_c^+\to \tau^+ \nu_\tau)$ will allow it to play a prominent role into either confirming or refuting these anomalies. Regarding the $B^+$ channel, due to the different CKM normalisation present also in this scenario, HL-LHC will not have a meaningful impact in constraining the model, contrarily to the strong probing power of FCC-ee.


%% file: conclusion.tex
\section{Conclusion}
\label{sec:conclusion}

In this paper, a study on the prospects for precise measurements of $\BcTauNu$ and $\BTauNu$ decays at FCC-ee is presented.
This work is based on common FCC software tools from sample generation, event reconstruction, to statistical analysis.
Events are first selected with displaced vertices containing three charged pions for the $\TauThreePi$ decay and are further refined with two stages of BDT discriminators against the background processes of generic $Z$ to hadron decays.
High signal purities are achieved in the individual selections of $\BcTauNu$ and $\BTauNu$ events, and their experimental precisions are evaluated.
For the first time, it is demonstrated that $\BTauNu$ processes can be selected from both $\BcTauNu$ processes and generic $Z\to b\bar{b}$ backgrounds and measured with high precision. 
The impacts of many factors on the experimental results are studied, and a range of expected signal precision is provided from idealistic cases to pessimistic cases: 1.6\% to 2.3\% for $N(\BcTauNu)$ and 1.8\% to 3.6\% for $N(\BTauNu)$, assuming an FCC-ee $Z$-pole dataset of 180 ab$^{-1}$.
These results translates to a measurement of $\mathcal{B}(\BTauNu)$ with a 2\% to 4\% relative precision, and a measurement of the ratio $\mathcal{R}=\mathcal{B}(B_c^+\to \tau^+ \nu_\tau)/\mathcal{B}(B_c^+\to J/\psi \mu^+ \nu_\mu)$ with a 2\% to 3\% relative precision.

The phenomenological implications of measurements of $B_c^+\to\tau^+ \nu_\tau$ and $B^+\to\tau^+ \nu_\tau$ have also been discussed here, both in the SM and beyond. In particular, this work demonstrates how a precise measurement of $\mathcal{B}(B^+\to \tau^+ \nu_\tau)$ could lead to a very precise extraction of $\vert\Vub\vert$, which could play a prominent role in the resolution of the long-standing $\vert\Vub\vert_{\rm incl.}$ vs. $\vert\Vub\vert_{\rm excl.}$ puzzle. In the context of NP searches, this work illustrates how a measurement of both $\mathcal{B}(B_c^+\to \tau^+ \nu_\tau)$ and  $\mathcal{B}(B^+\to \tau^+ \nu_\tau)$ at FCC-ee could play a prominent role in constraining the allowed parameter space in G2HDM and specific leptoquark scenarios, complementary to what HL-LHC could do. These BSM models are selected since they are currently advocated as the best explanation of the $R_D$ and $R_{D^\ast}$ anomalies and their presently allowed parameter space would be probed almost entirely with an FCC-ee measurement of such channels.

\section*{Acknowledgements}
The authors would like to thank C. Davies, S. Monteil, E. Perez and O. Sumensari for the useful discussions and their input. The work of M. F. and S. I. is supported by the Deutsche Forschungsgemeinschaft (DFG, German Research Foundation) under grant  396021762 - TRR 257, ``Particle Physics Phenomenology after the Higgs Discovery''. This project has received support from the European Union’s Horizon 2020 research and innovation programme under grant agreement No. 951754.

%% file: internal.tex
\clearpage
\appendix

\section{Additional exclusive samples}\label{app:excl_samp}

All samples listed in Tab. 2 of~\cite{Amhis_2021} are used in this work. 
Additional samples are listed in Tab.~\ref{tab:excl_bb} for $Z\ to b\bar{b}$ process and Tab.~\ref{tab:excl_cc} for $Z \to c\bar{c}$ process.
These decays modes are selected by studying the decay chains in inclusive events that pass baseline BDT selections. 
Decays modes with the highest occurrence in those events are selected for exclusive sample generation. 

\begin{table}[h!]
    \centering
    \begin{tabular}{l|c}
        Decay & Number of events \\
        \hline
        $B^0 \to D^- e^+\nu_{e}$         & $1\times10^8$ \\
        $B^0 \to D^{*-} e^+\nu_{e}$      & $1\times10^8$ \\
        $B^0 \to D^- \mu^+\nu_{\mu}$     & $1\times10^8$ \\
        $B^0 \to D^{*-} \mu^+\nu_{\mu}$  & $1\times10^8$ \\
        \hline
        $B^+ \to \bar{D}^0 e^+\nu_{e}$         & $1\times10^8$ \\
        $B^+ \to \bar{D}^{*0} e^+\nu_{e}$      & $1\times10^8$ \\
        $B^+ \to \bar{D}^0 \mu^+\nu_{\mu}$     & $1\times10^8$ \\
        $B^+ \to \bar{D}^{*0} \mu^+\nu_{\mu}$  & $1\times10^8$ \\
        \hline
        $B^0_s \to D^-_s e^+\nu_{e}$         & $1\times10^8$ \\
        $B^0_s \to D^{*-}_s e^+\nu_{e}$      & $1\times10^8$ \\
        $B^0_s \to D^-_s \mu^+\nu_{\mu}$     & $1\times10^8$ \\
        $B^0_s \to D^{*-}_s \mu^+\nu_{\mu}$  & $1\times10^8$ \\
        \hline
        $\Lambda_b^0 \to \Lambda_c^- e^+\nu_{e}$         & $1\times10^8$ \\
        $\Lambda_b^0 \to \Lambda_c^{*-} e^+\nu_{e}$      & $1\times10^8$ \\
        $\Lambda_b^0 \to \Lambda_c^- \mu^+\nu_{\mu}$     & $1\times10^8$ \\
        $\Lambda_b^0 \to \Lambda_c^{*-} \mu^+\nu_{\mu}$  & $1\times10^8$ \\
        \hline
    \end{tabular}
    \caption{Exclusive $Z \to b\bar{b}$ samples additional to the previous work~\cite{Amhis_2021}.}
    \label{tab:excl_bb}
\vspace{1cm}
    \begin{tabular}{l|c}
        Decay & Number of events \\
        \hline
        $D^+ \to \tau^+\nu_\tau$      & $1\times10^8$ \\
        $D^+ \to K^0 \pi^+\pi^+\pi^-$ & $1\times10^8$ \\
        \hline
        $D_s^+ \to \tau^+\nu_\tau$      & $1\times10^8$ \\
        $D_s^+ \to \rho^+ \eta^\prime$  & $1\times10^8$ \\
        \hline
        $\Lambda_c^+ \to \Lambda^0 e^+\nu_e$      & $1\times10^8$ \\
        $\Lambda_c^+ \to \Lambda^0 \mu^+\nu_\mu$  & $1\times10^8$ \\
        $\Lambda_c^+ \to \Lambda^0 \pi^+\pi^+\pi^-$  & $1\times10^8$ \\
        $\Lambda_c^+ \to \Sigma^+ \pi^+\pi^-$     & $1\times10^8$ \\
        \hline
    \end{tabular}
    \caption{Exclusive $Z \to c\bar{c}$ samples additional to the previous work~\cite{Amhis_2021}.}    
    \label{tab:excl_cc}
\end{table}

\section{BDT input variables}\label{app:BDT_vars}

\begin{figure}[t!]
    \centering
    \includegraphics[width=0.3\textwidth]{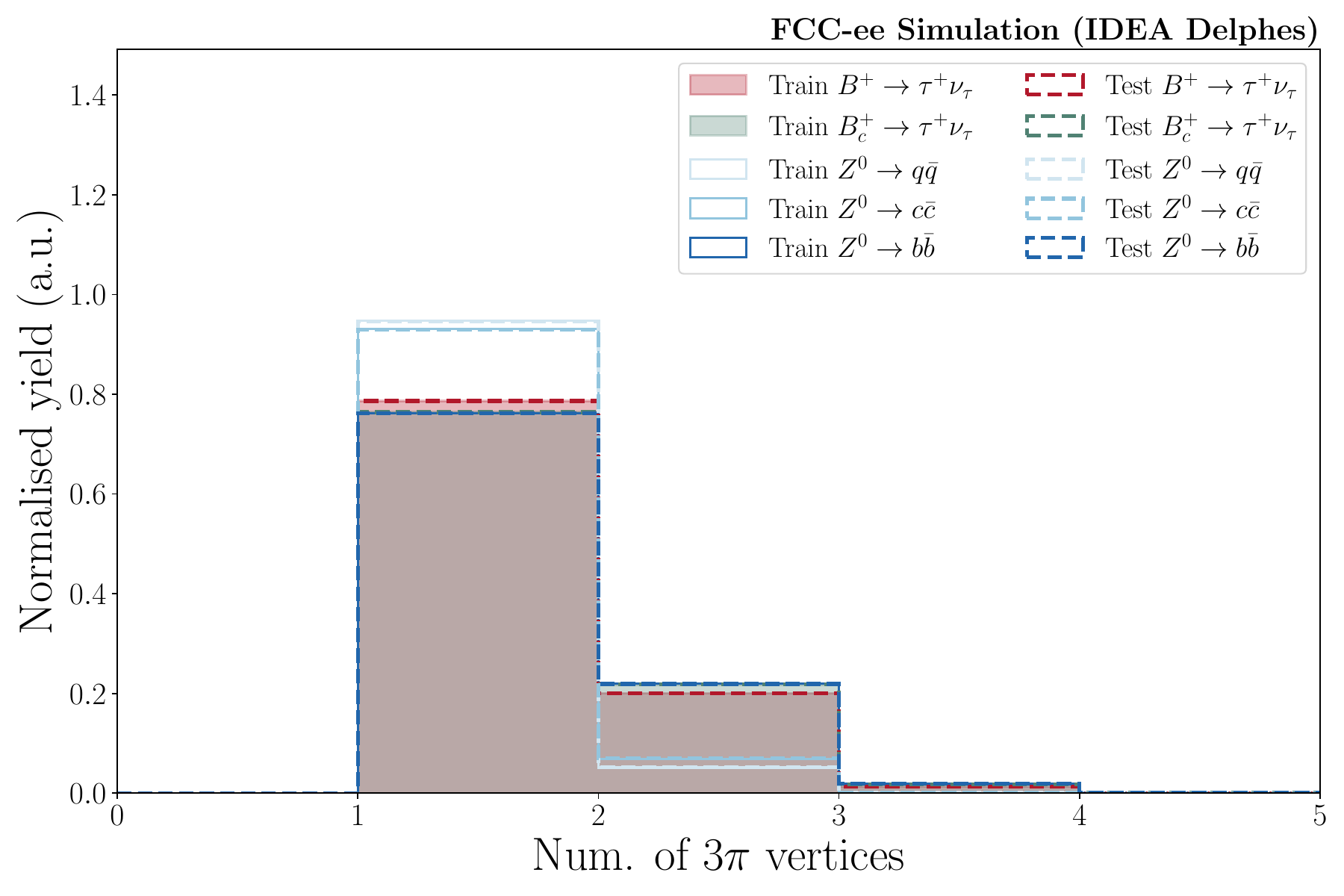}
    \includegraphics[width=0.3\textwidth]{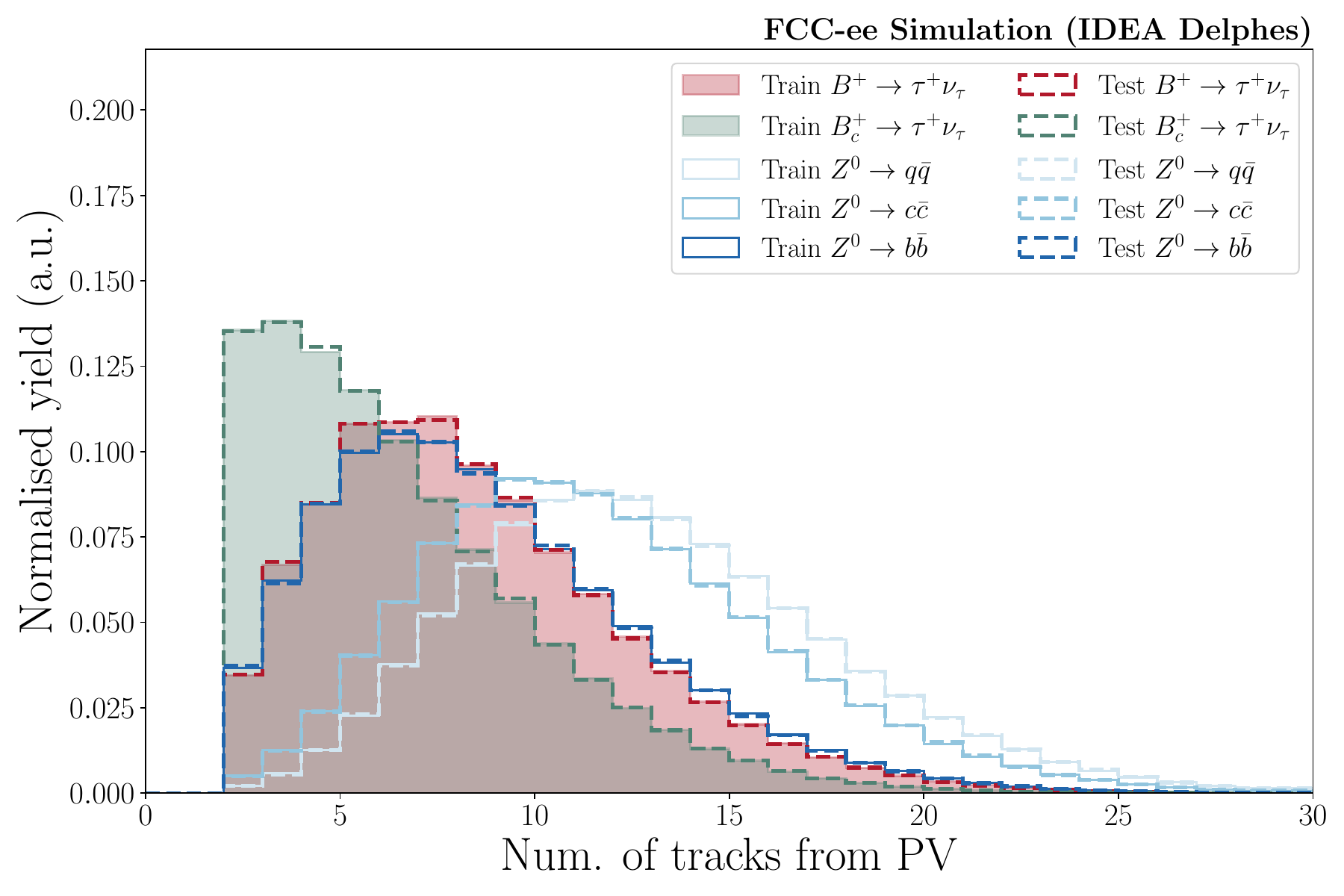}
    \includegraphics[width=0.3\textwidth]{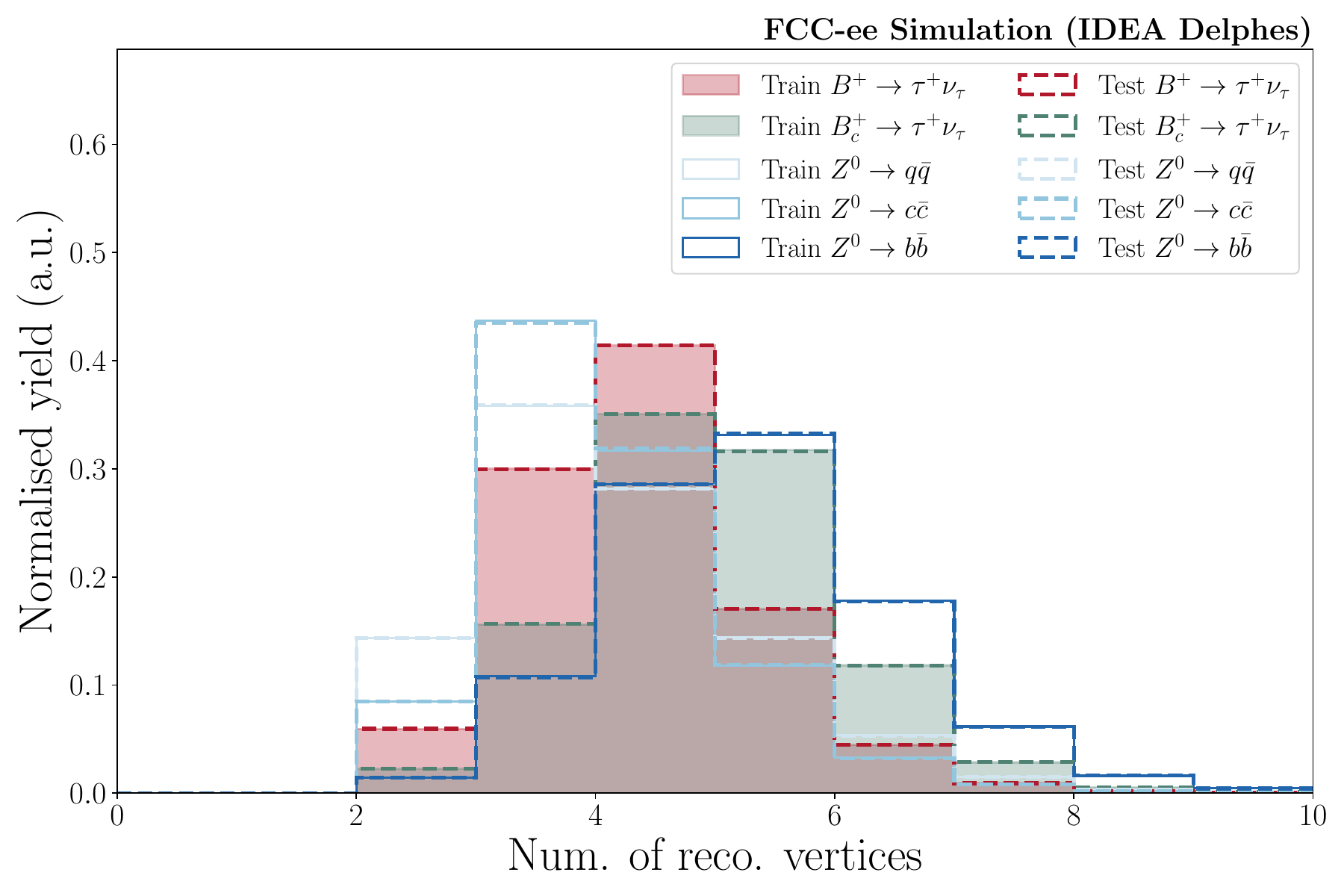} \\
    \includegraphics[width=0.3\textwidth]{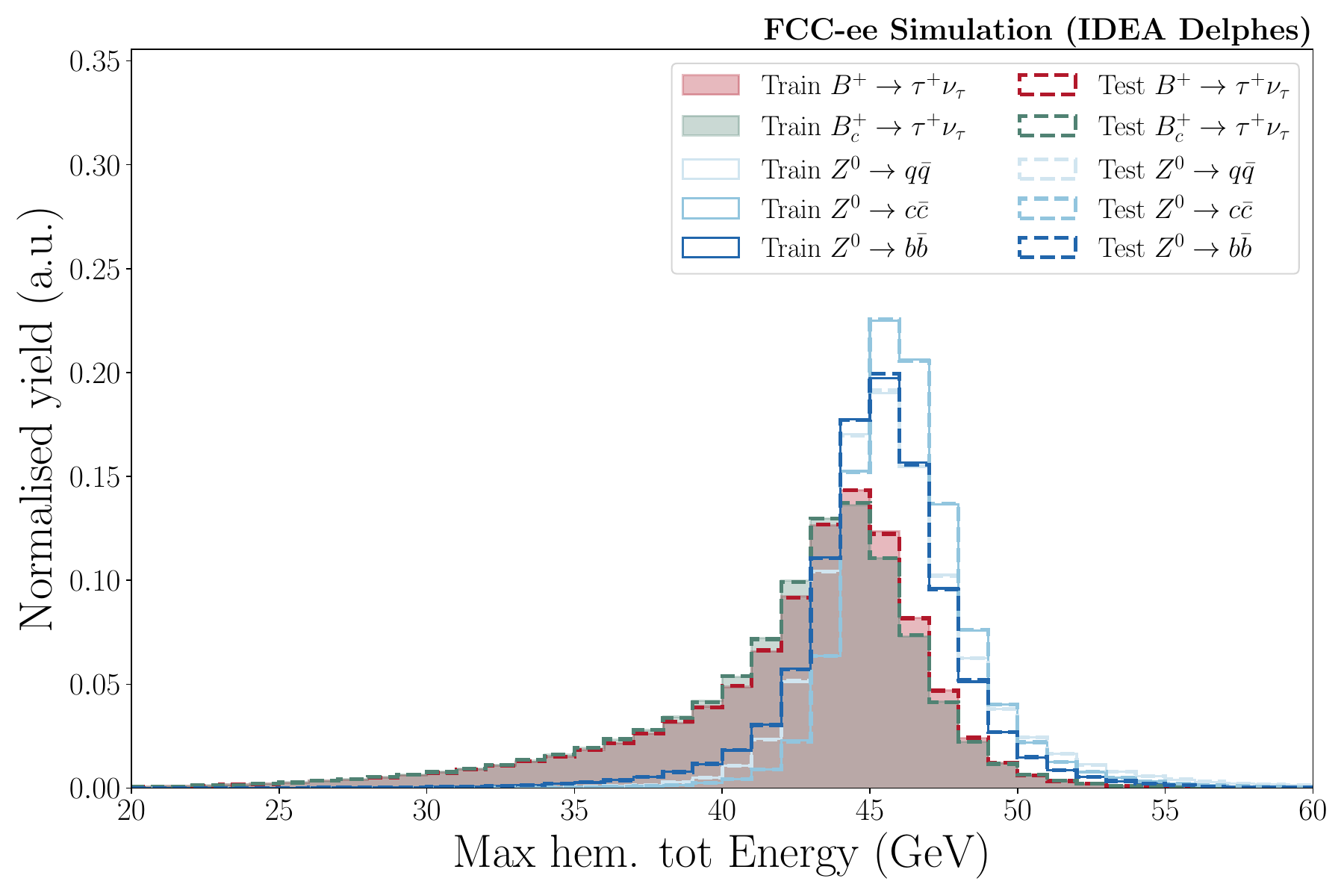}
    \includegraphics[width=0.3\textwidth]{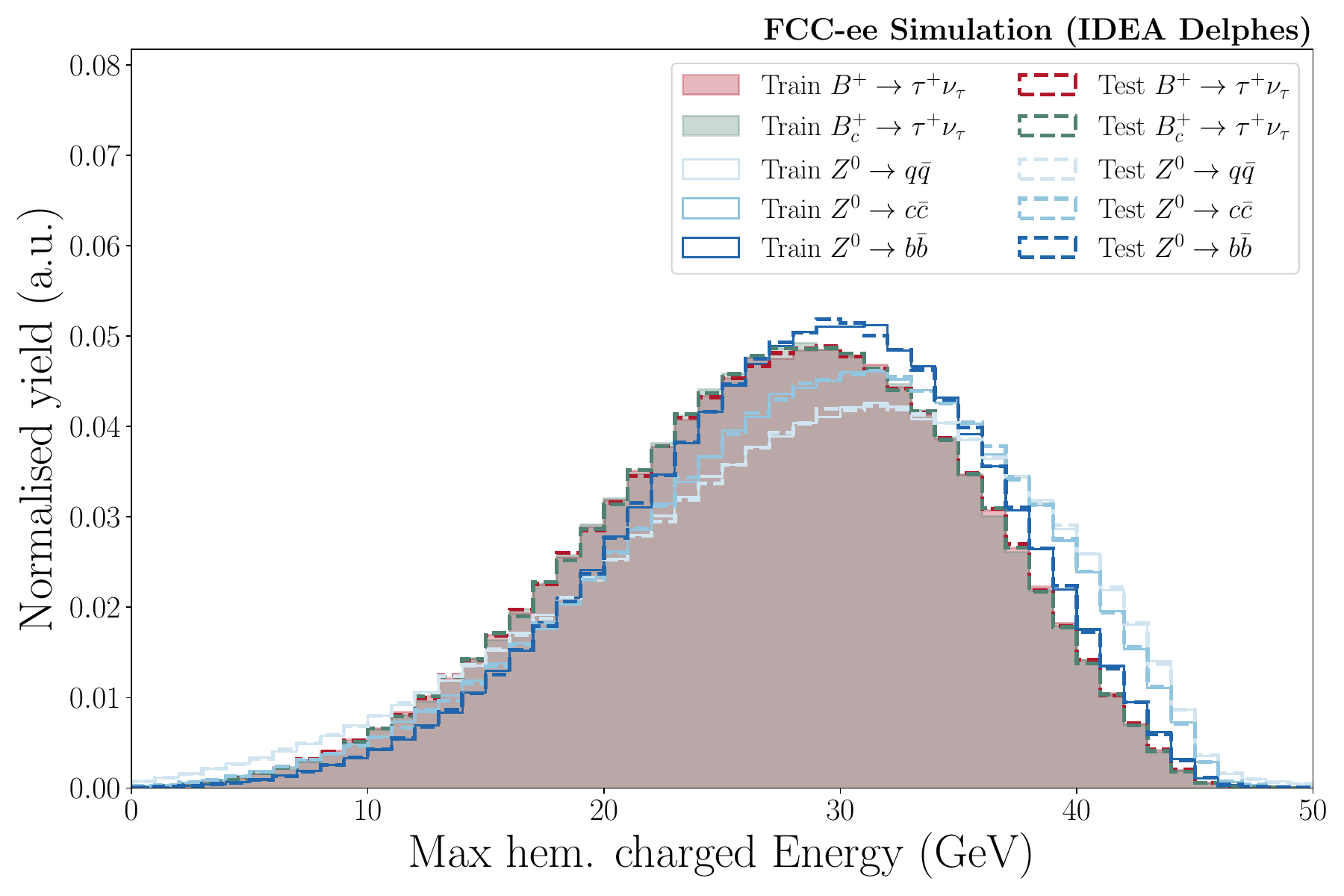}
    \includegraphics[width=0.3\textwidth]{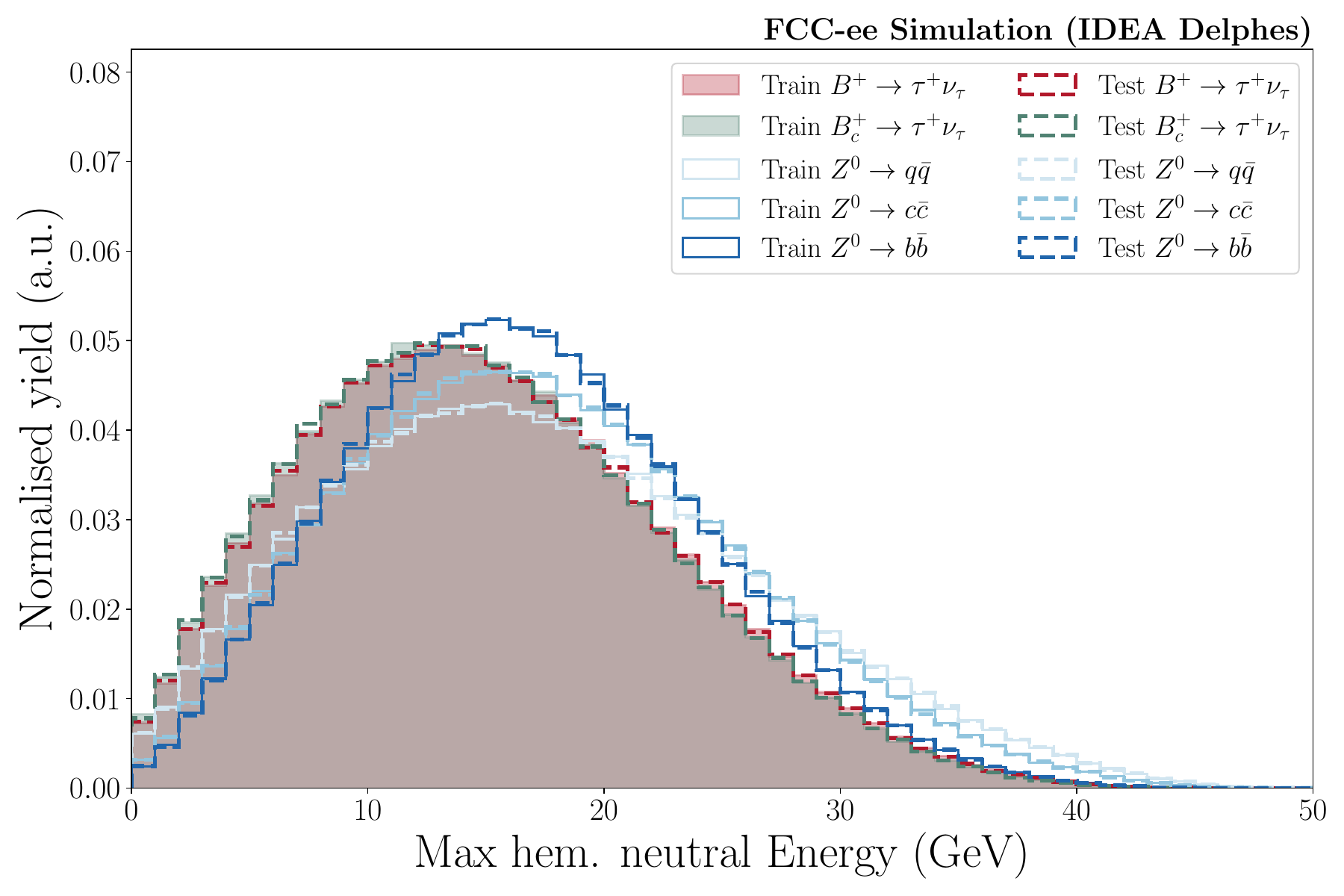} \\
    \includegraphics[width=0.3\textwidth]{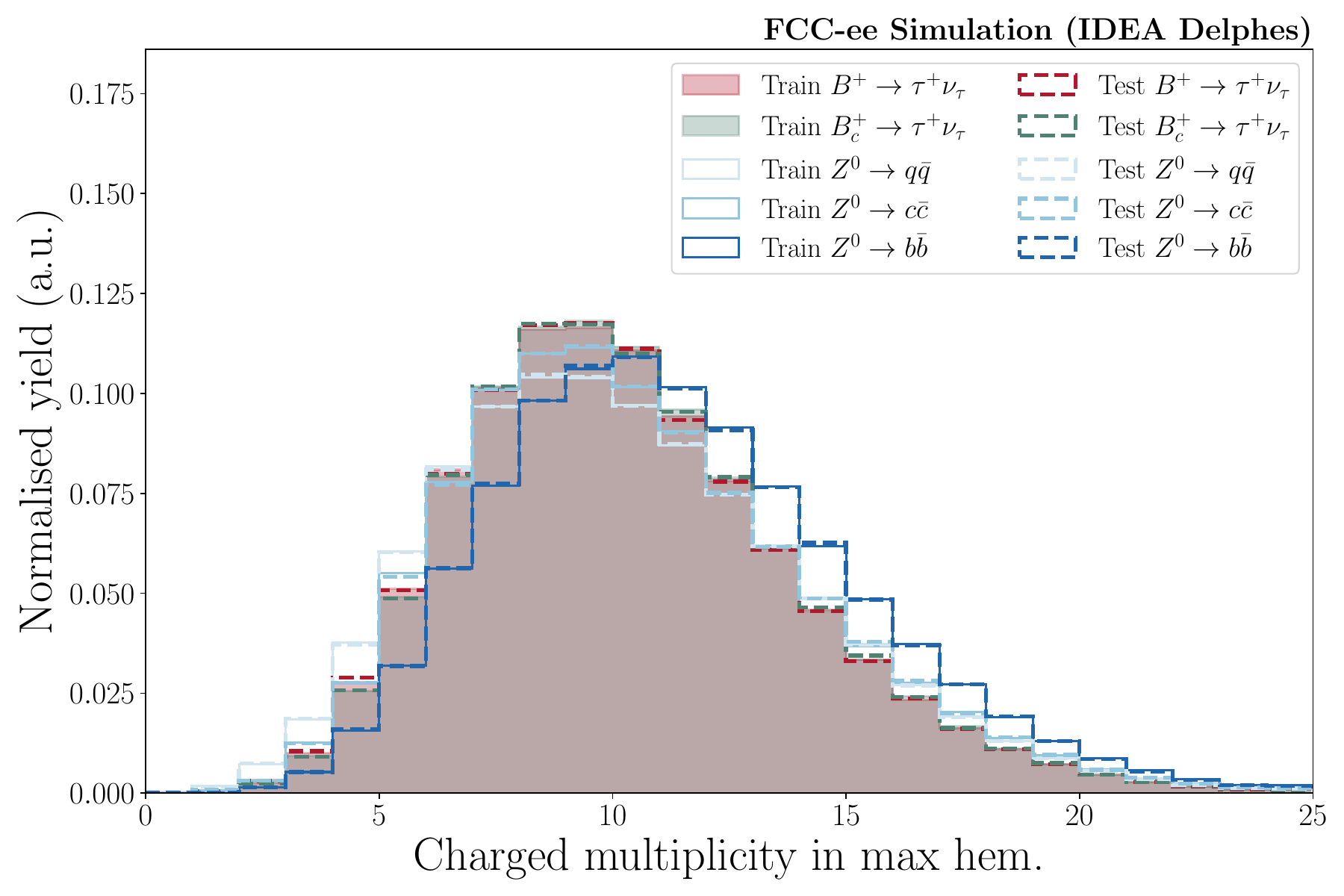}
    \includegraphics[width=0.3\textwidth]{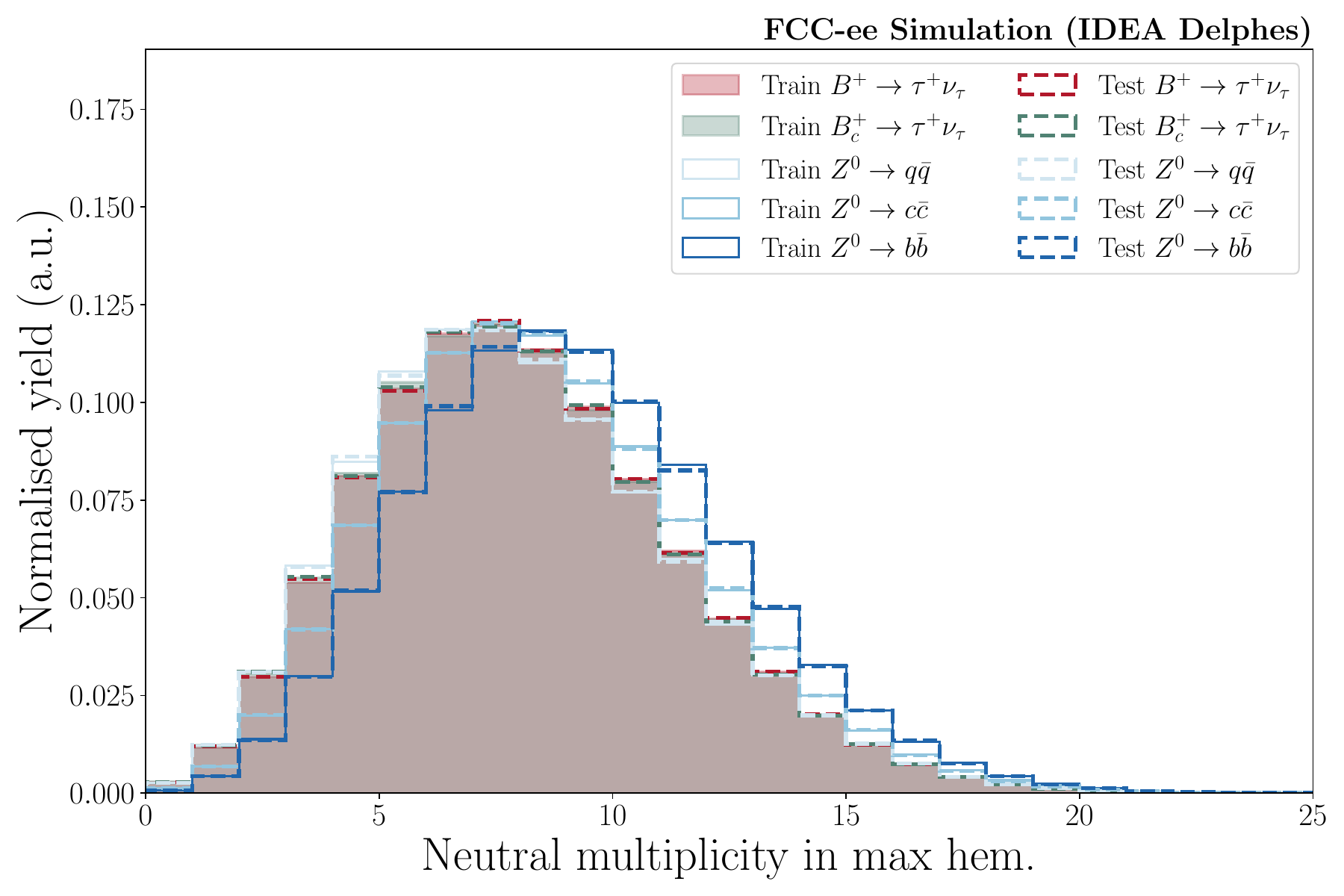}
    \includegraphics[width=0.3\textwidth]{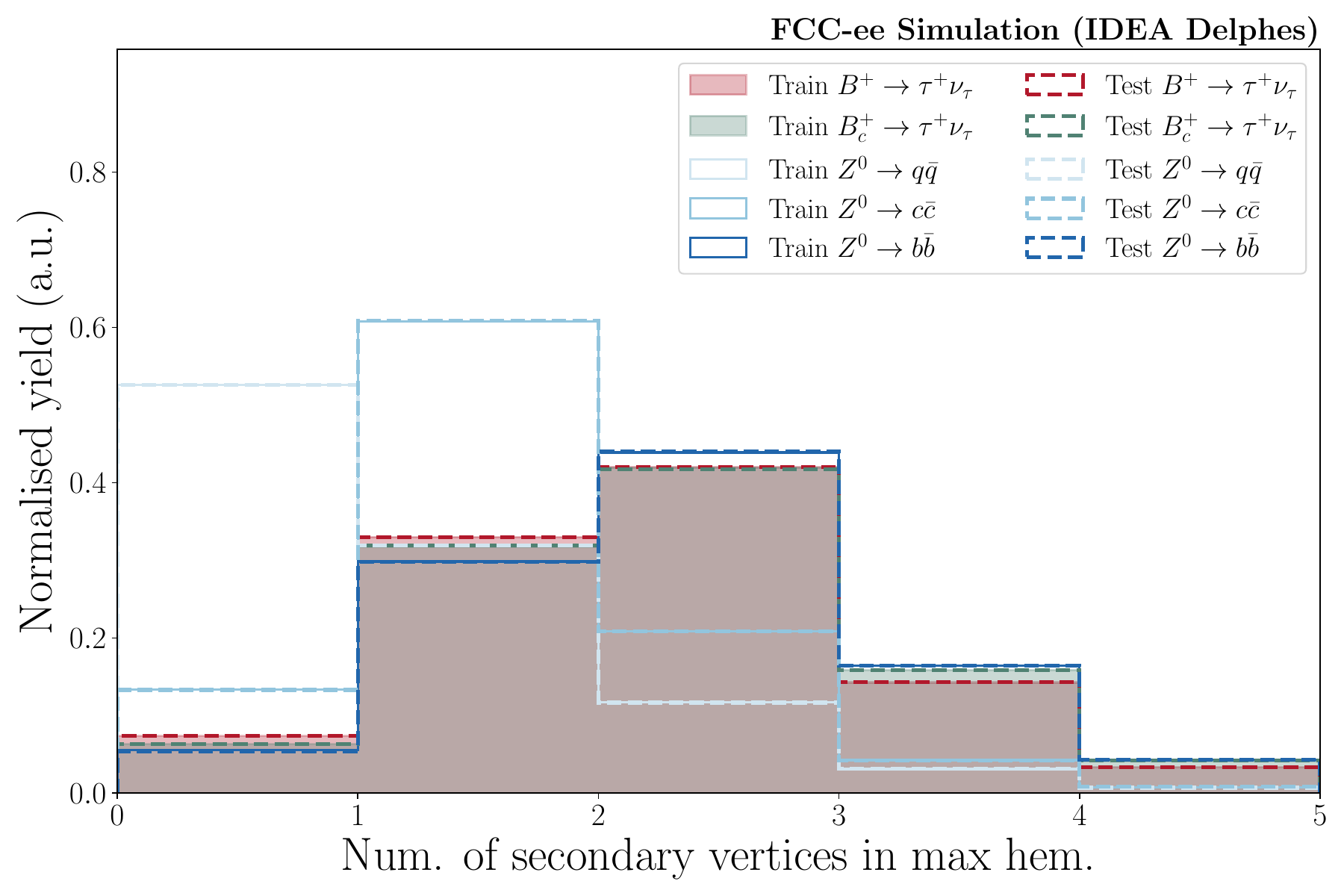} \\
    \includegraphics[width=0.3\textwidth]{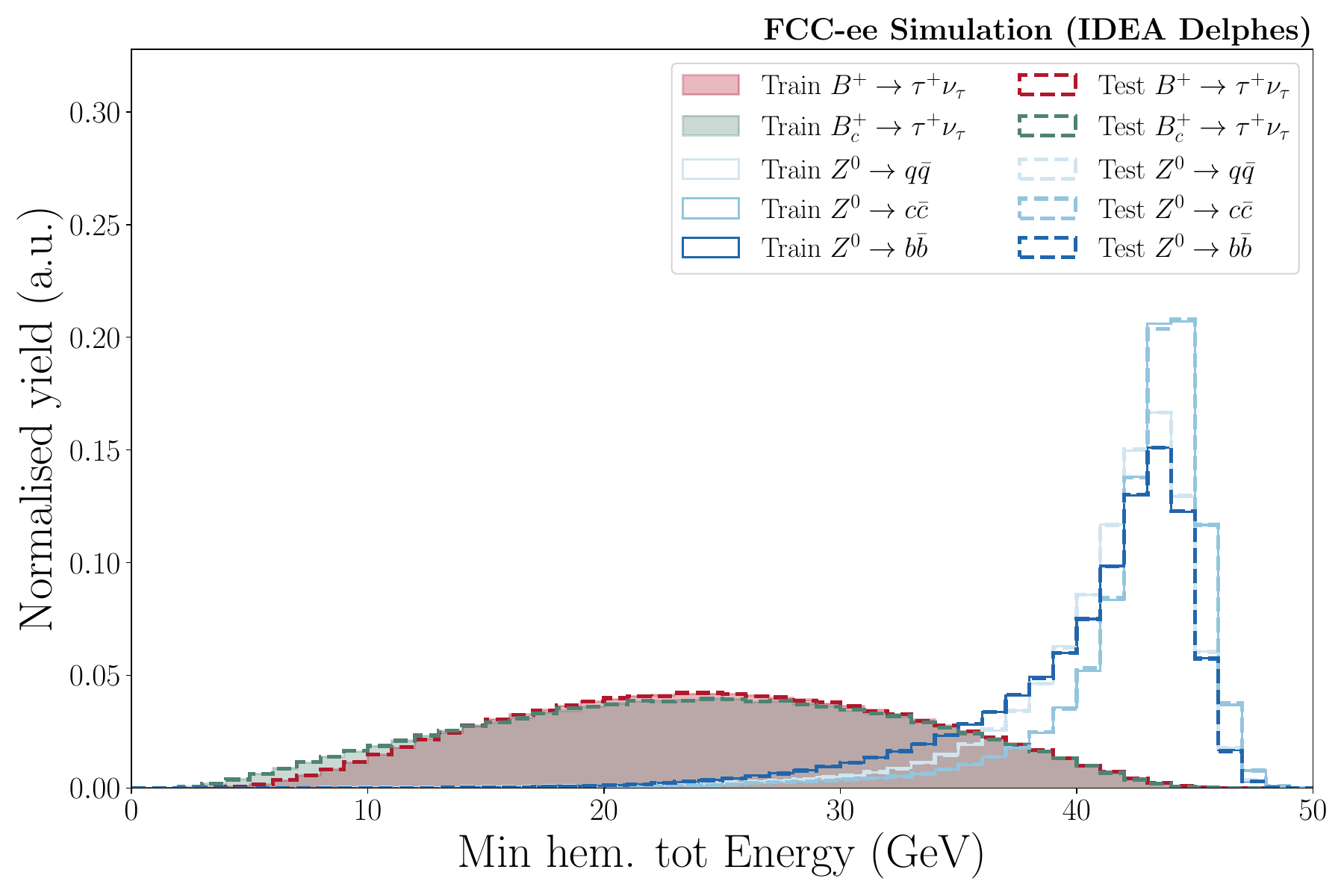}
    \includegraphics[width=0.3\textwidth]{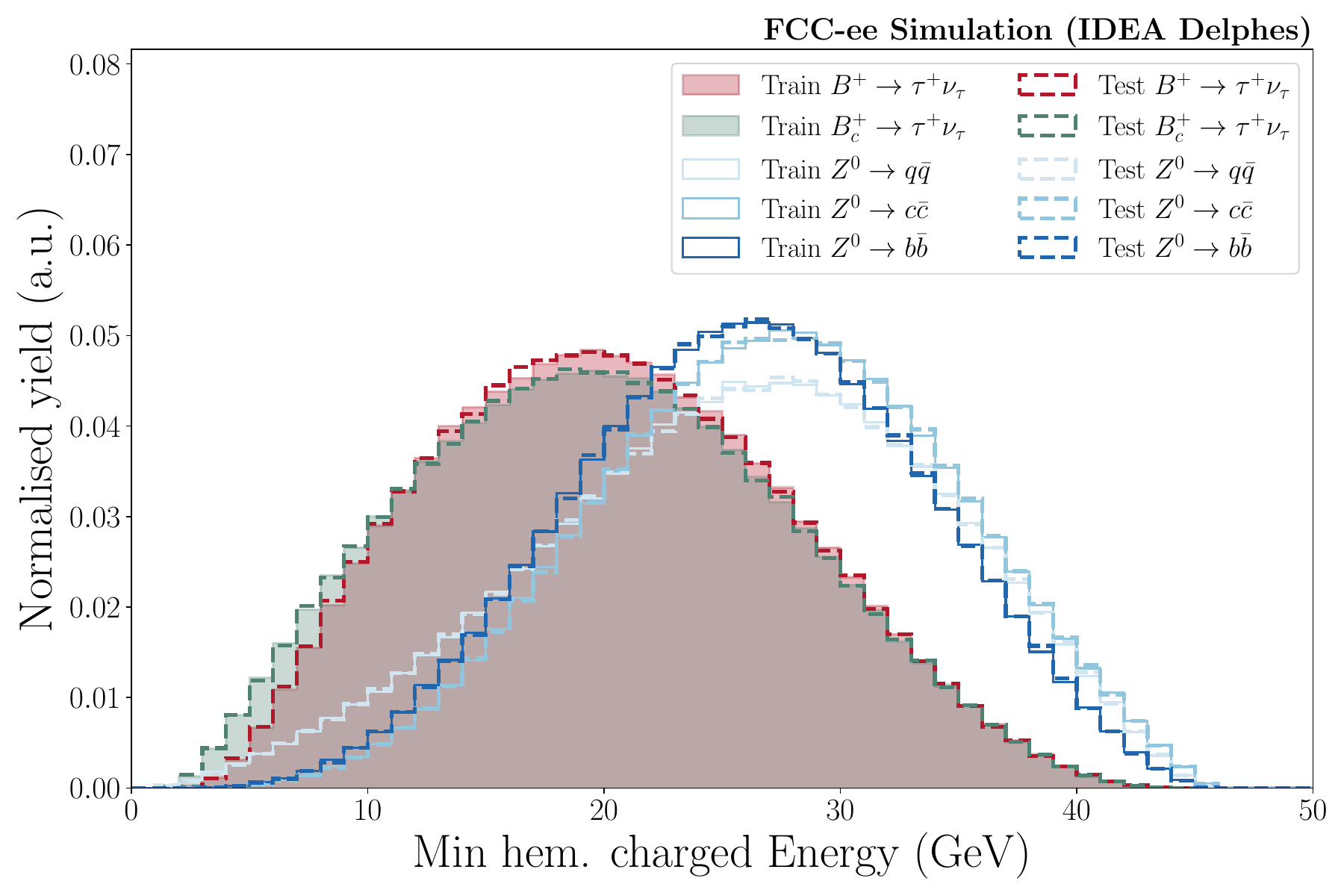}
    \includegraphics[width=0.3\textwidth]{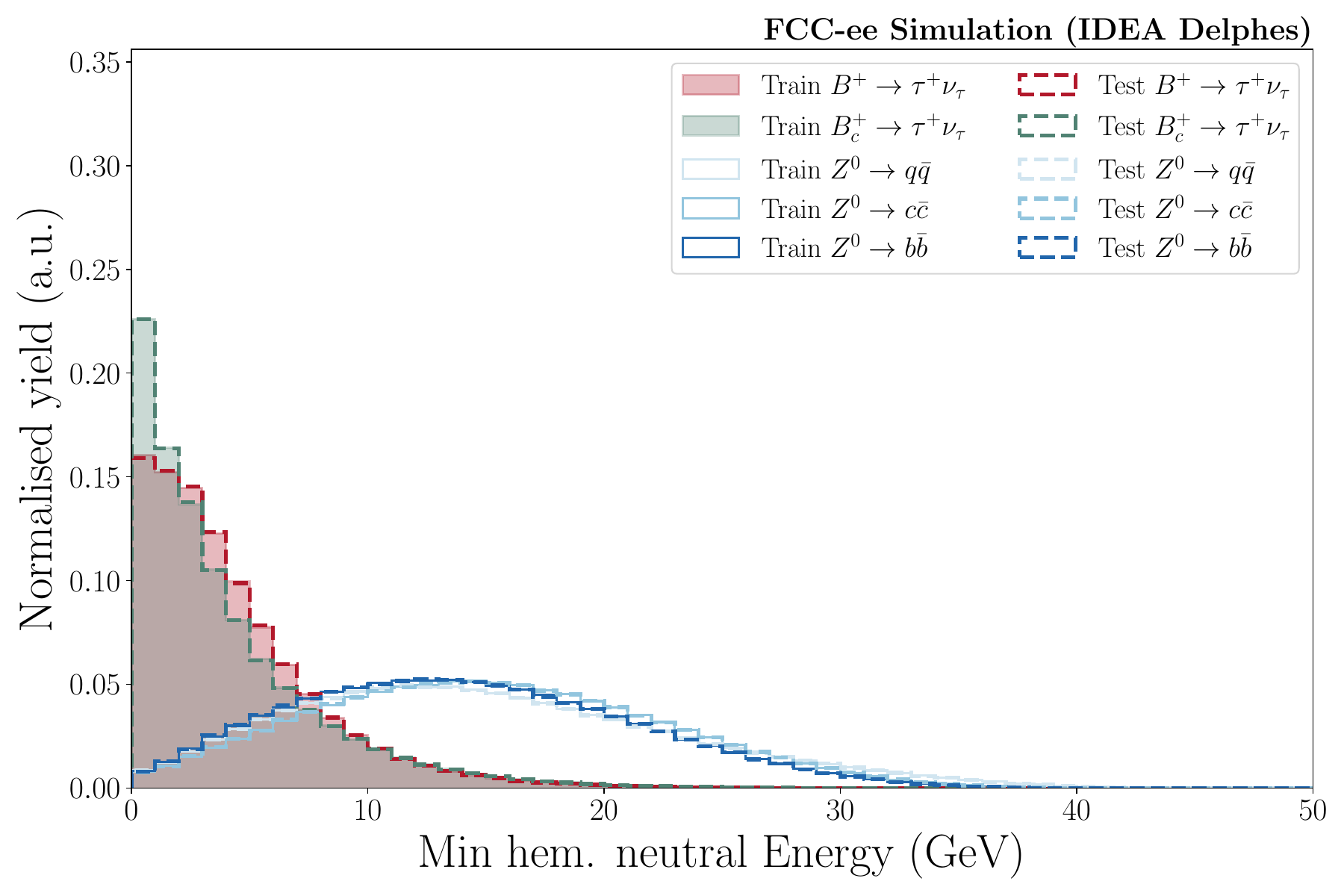} \\
    \includegraphics[width=0.3\textwidth]{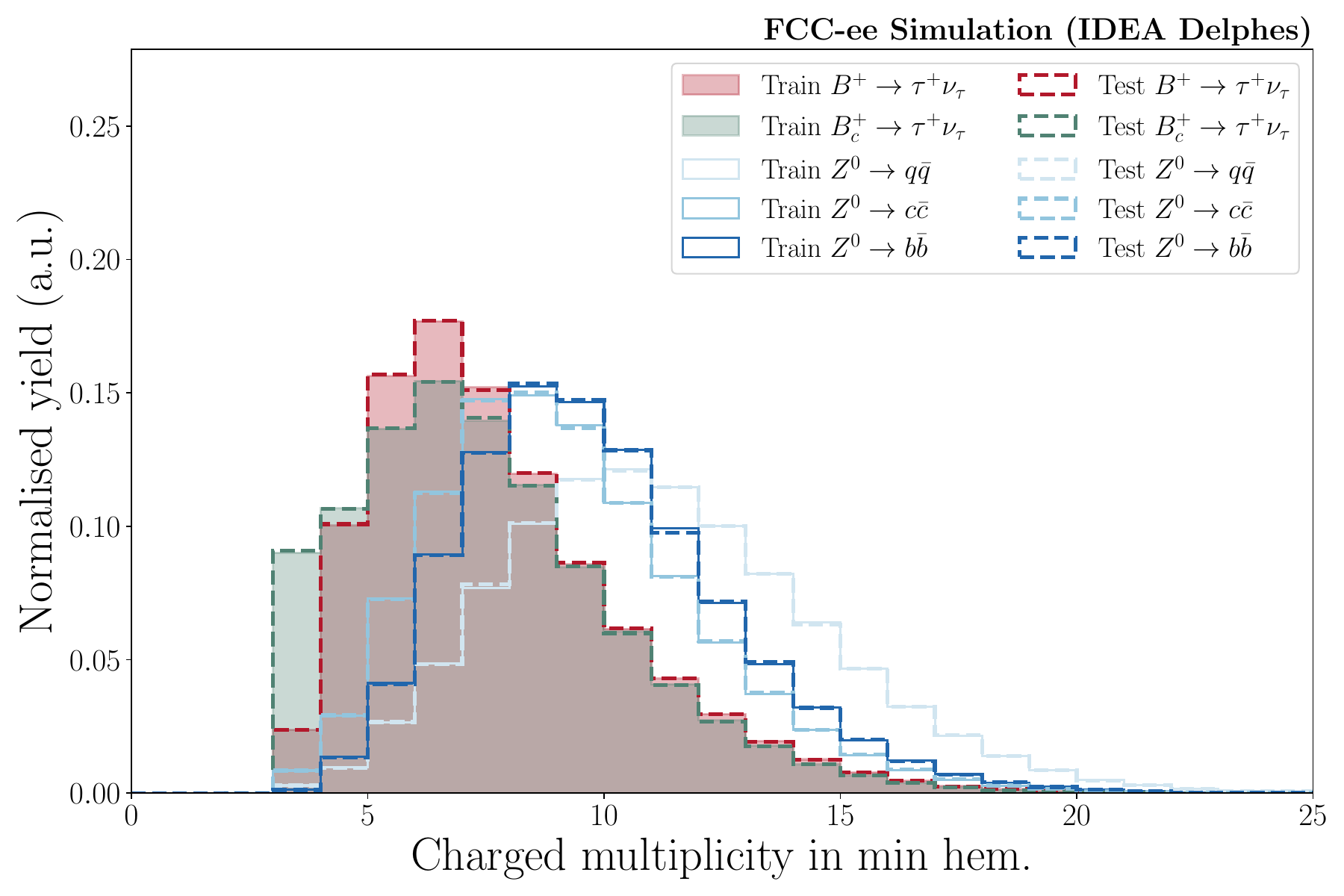}
    \includegraphics[width=0.3\textwidth]{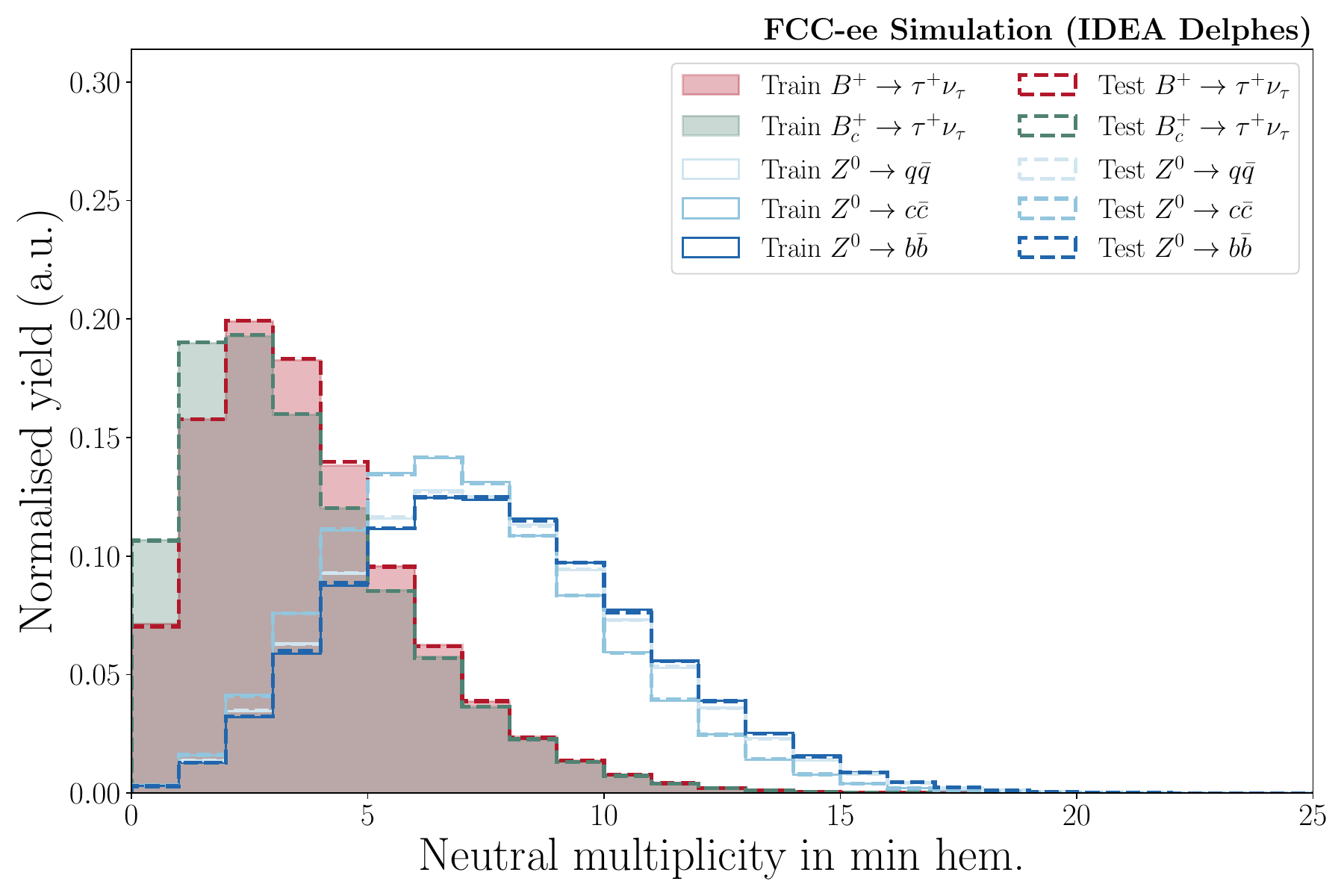}
    \includegraphics[width=0.3\textwidth]{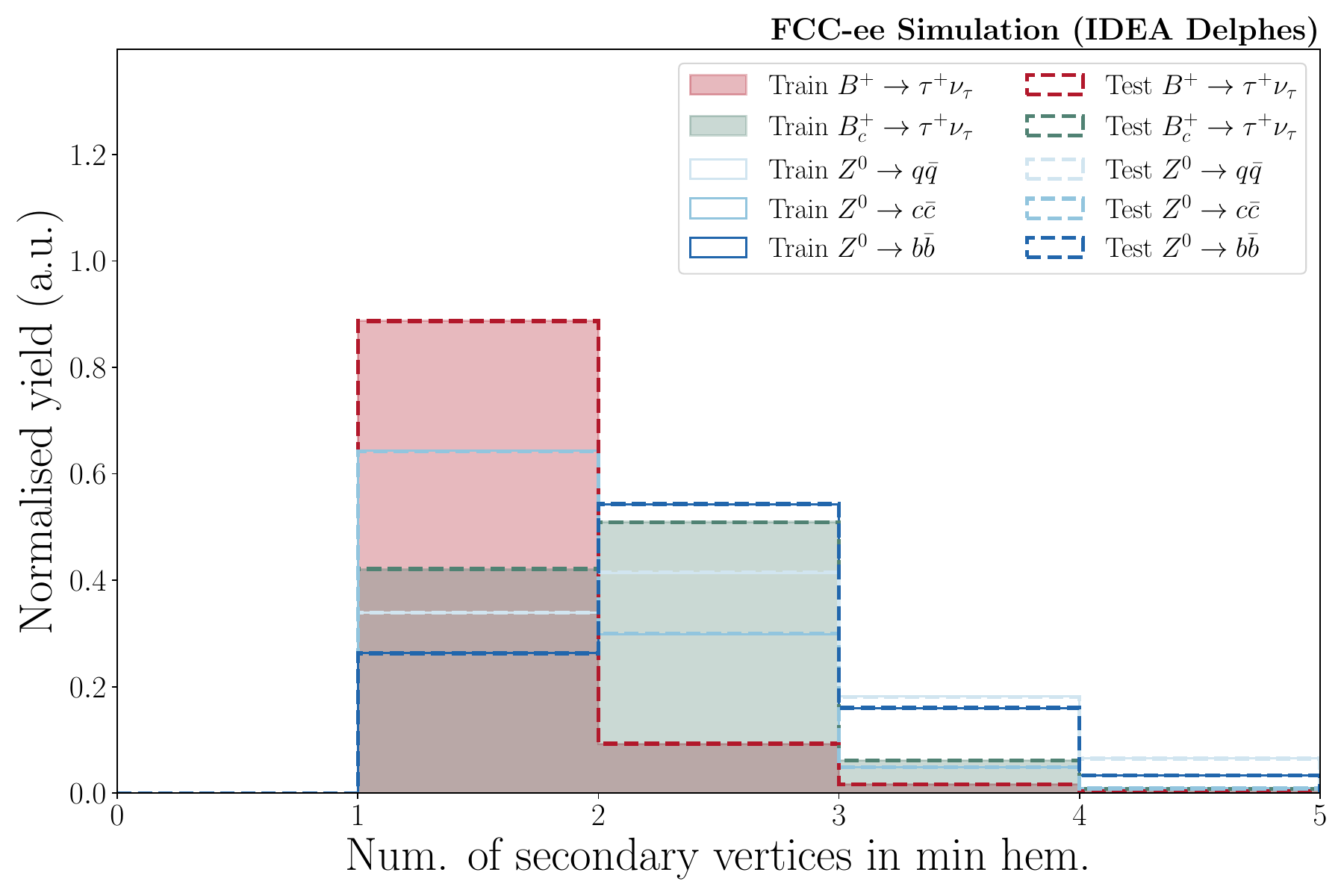} \\
    \includegraphics[width=0.3\textwidth]{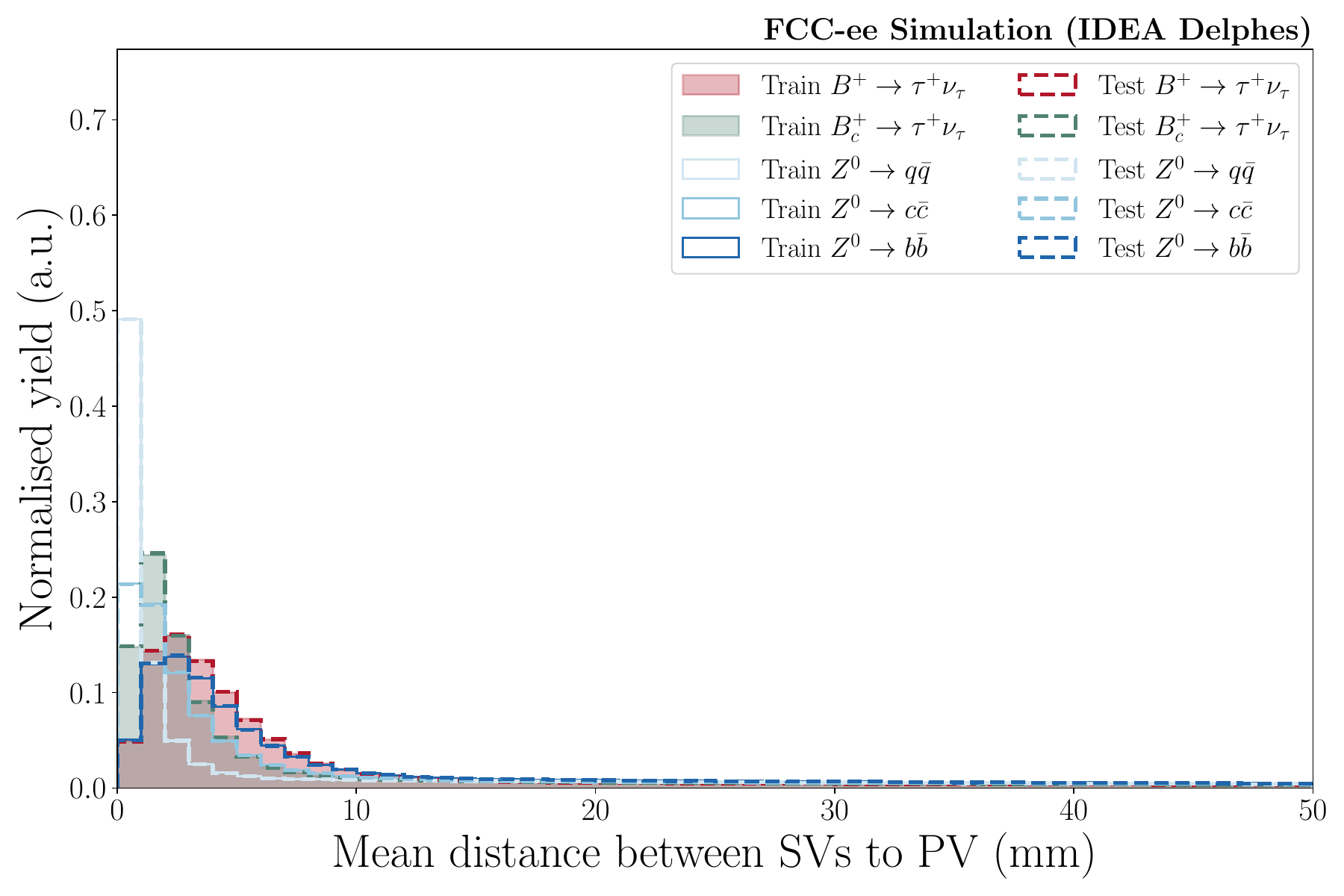}
    \includegraphics[width=0.3\textwidth]{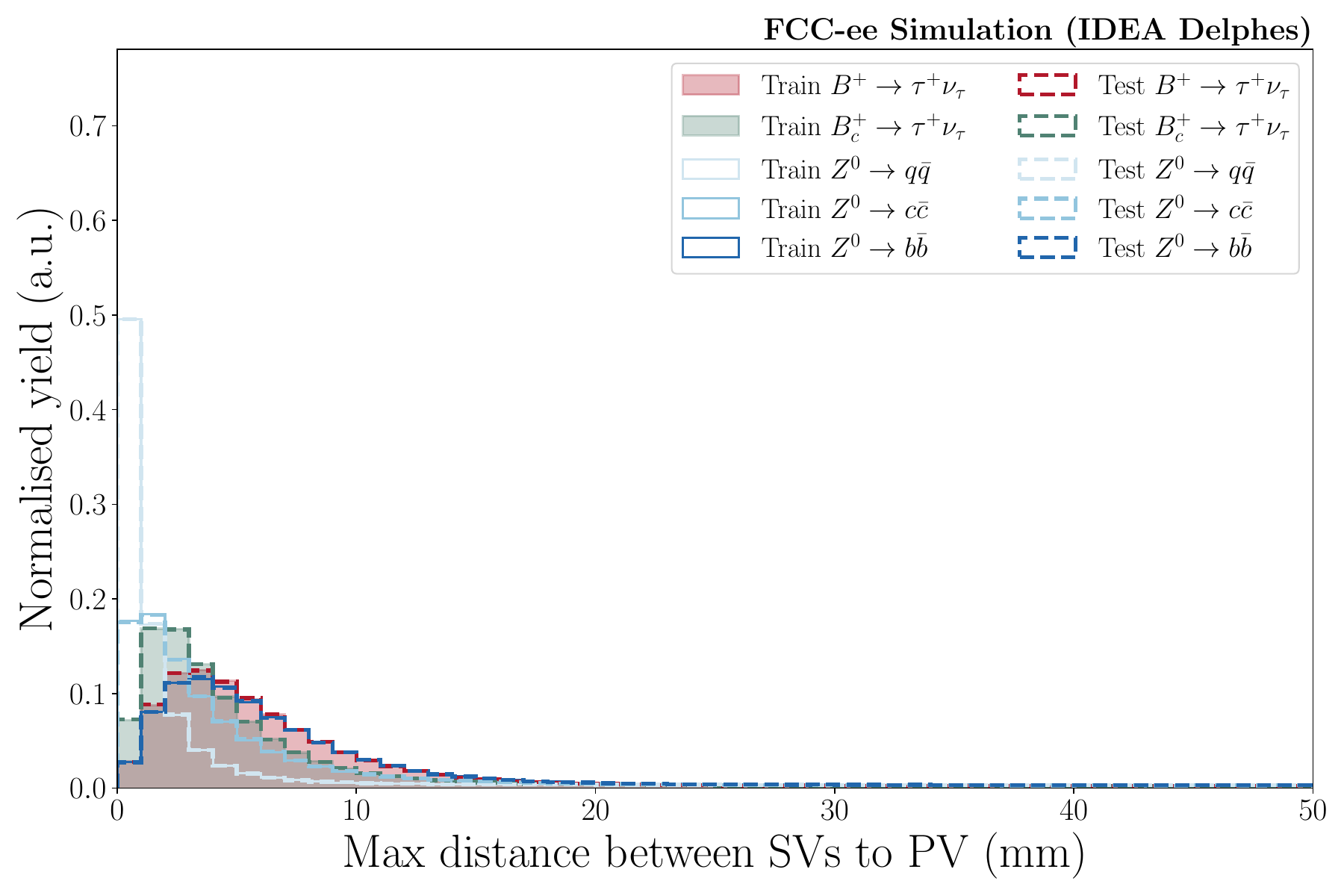}
    \includegraphics[width=0.3\textwidth]{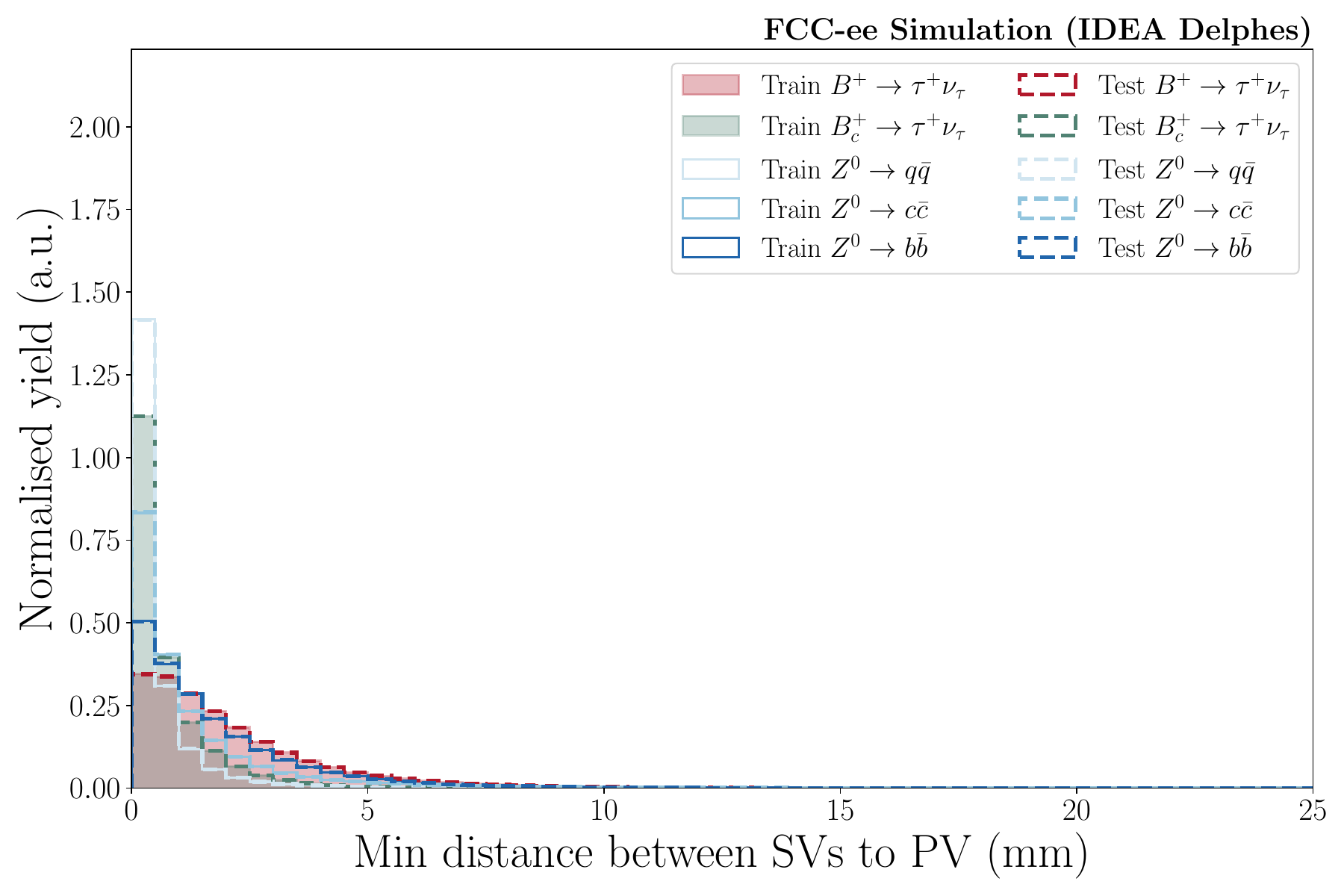} \\
    \caption{First-stage BDT training variable distributions in training samples (solid lines) and testing samples (dashed lines).}
    \label{fig:BDT1_vars}
\end{figure}

\begin{figure}[h!]
    \centering
    \includegraphics[width=0.24\textwidth]{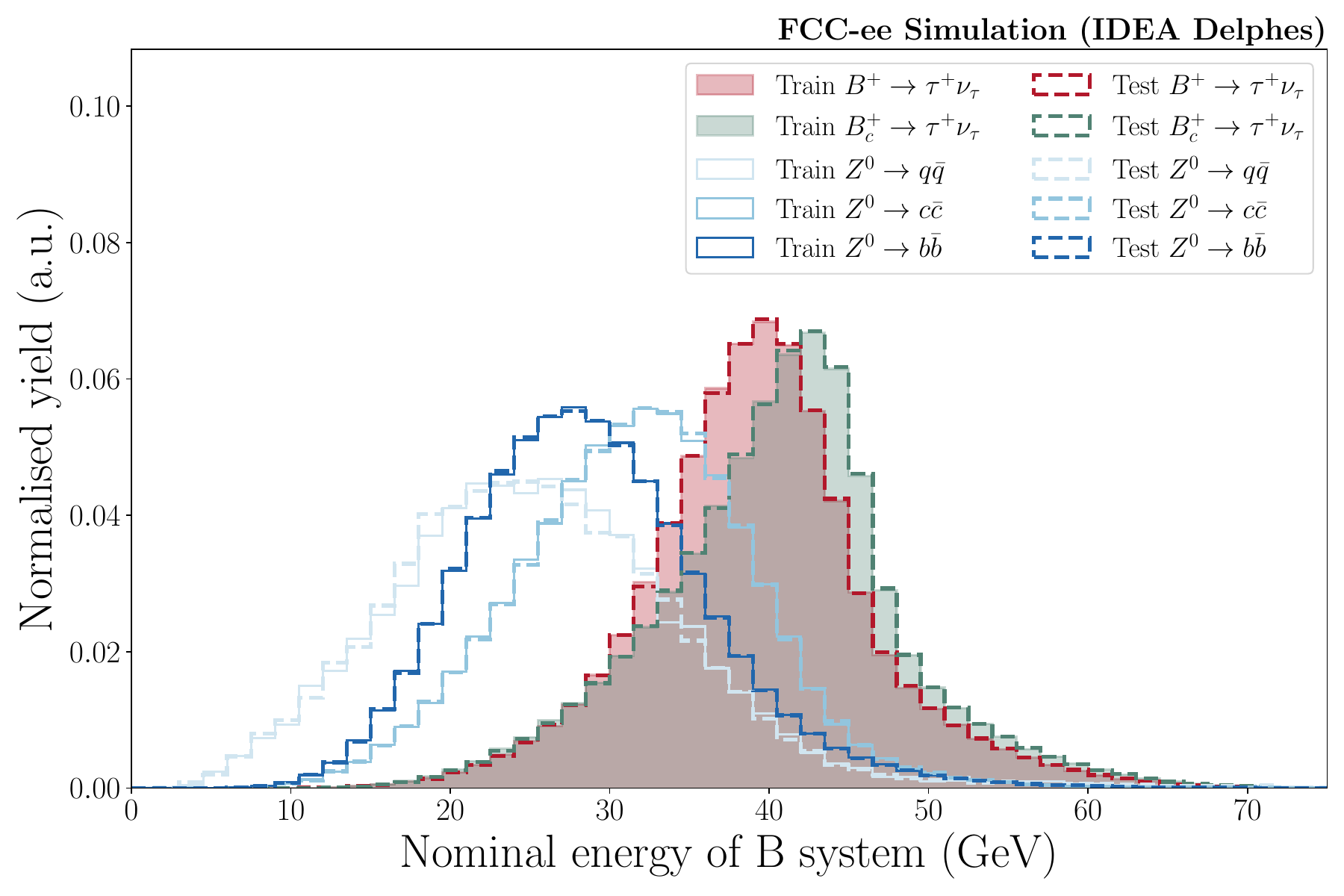}
    \includegraphics[width=0.24\textwidth]{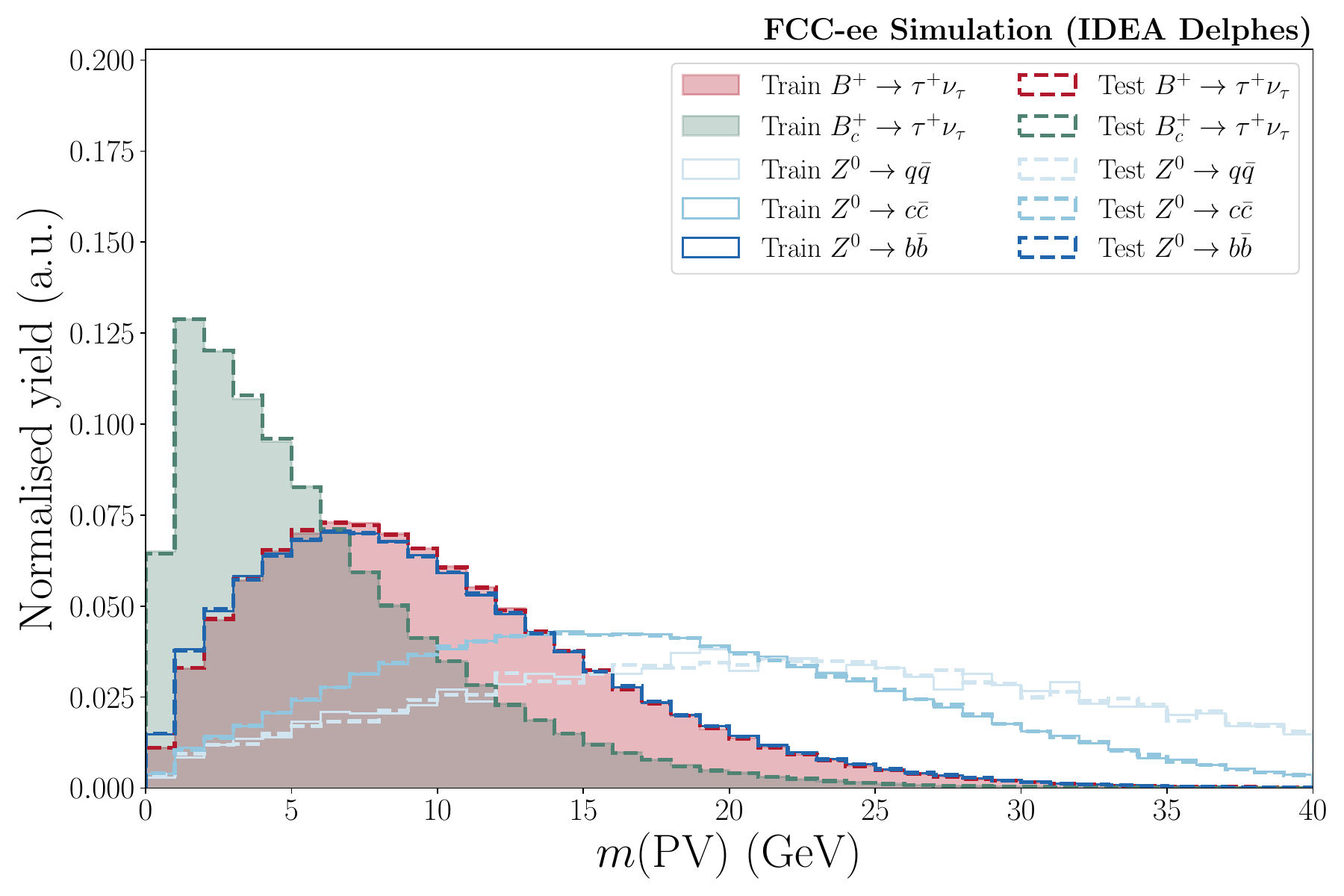}
    \includegraphics[width=0.24\textwidth]{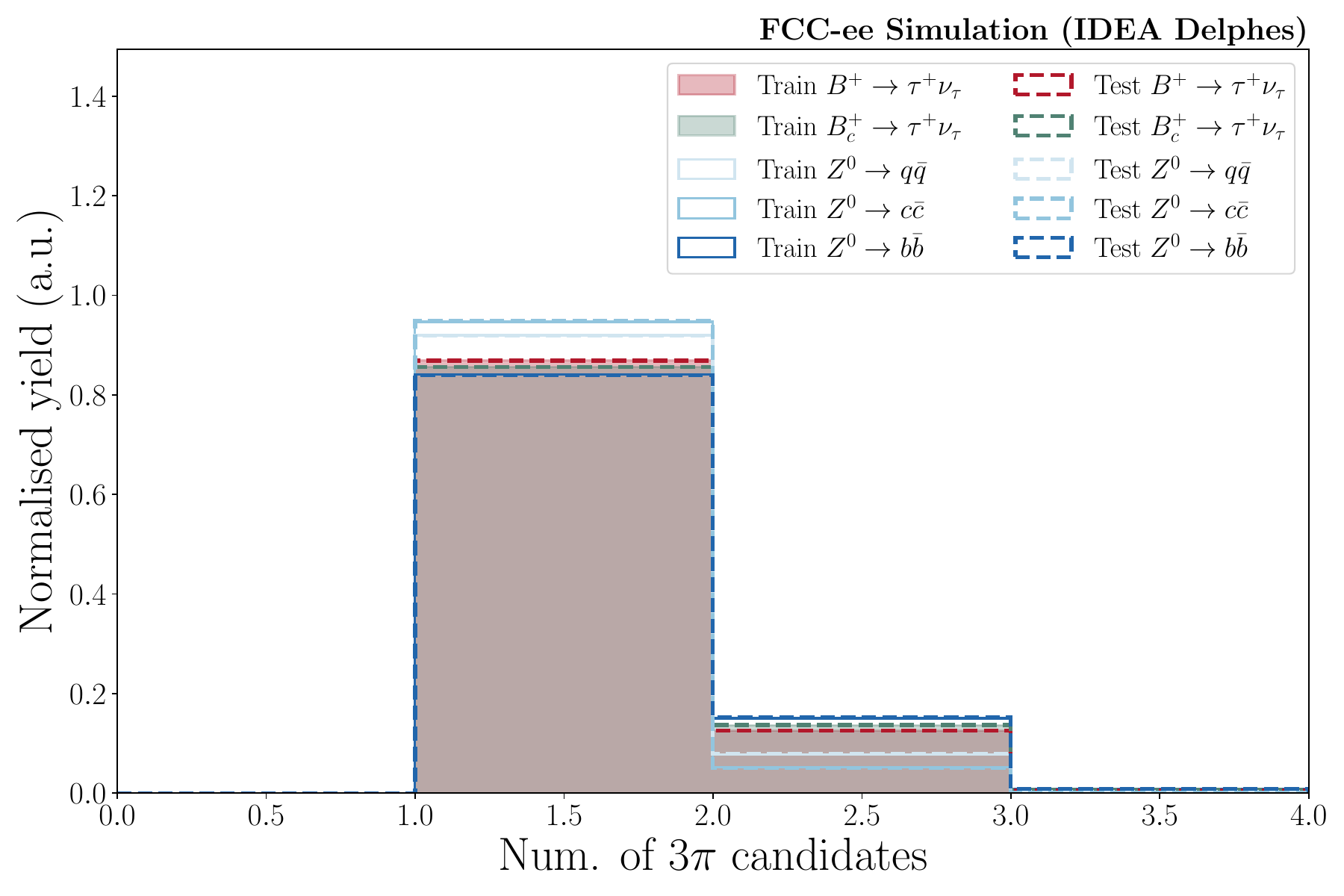} 
    \includegraphics[width=0.24\textwidth]{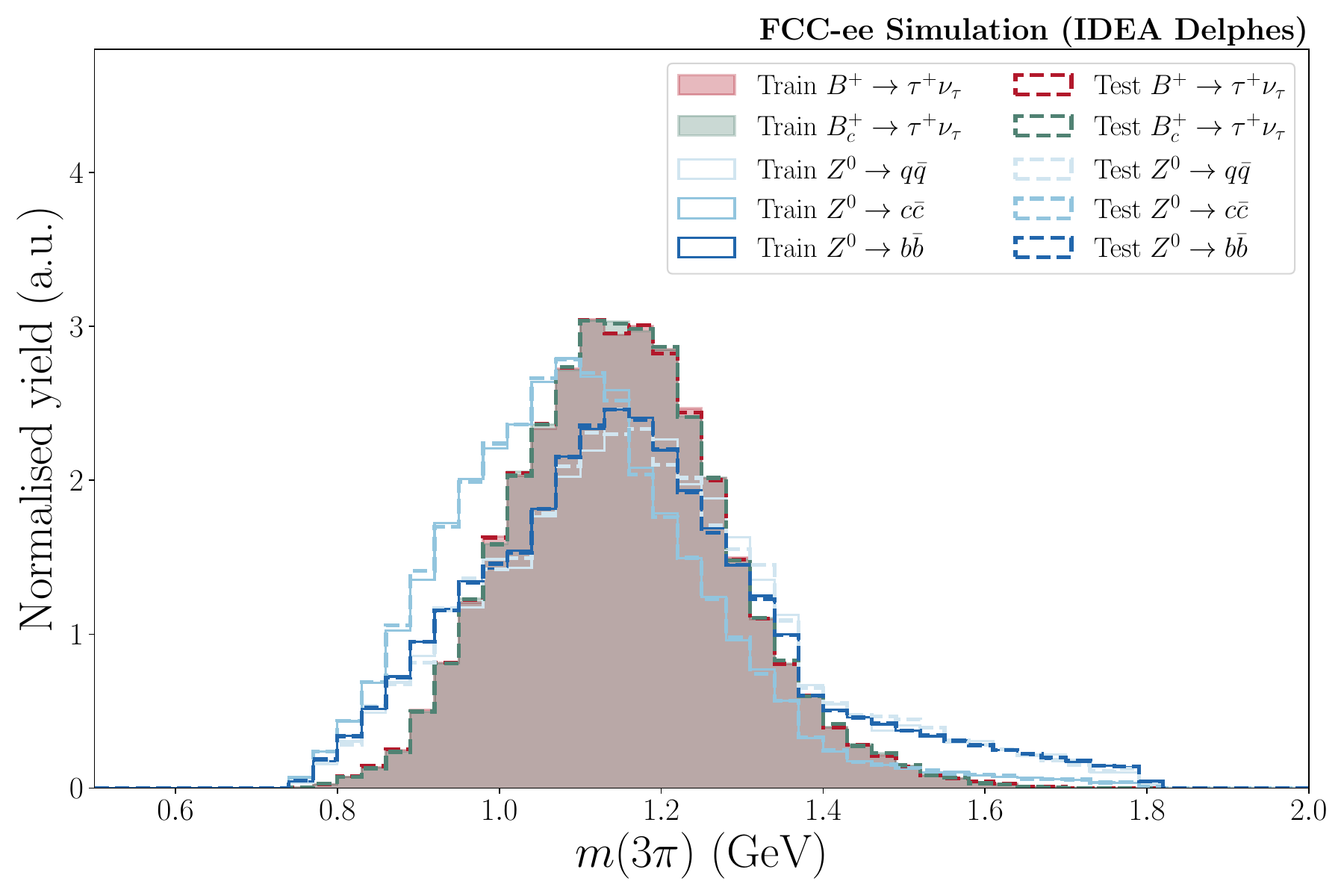} \\
    \includegraphics[width=0.24\textwidth]{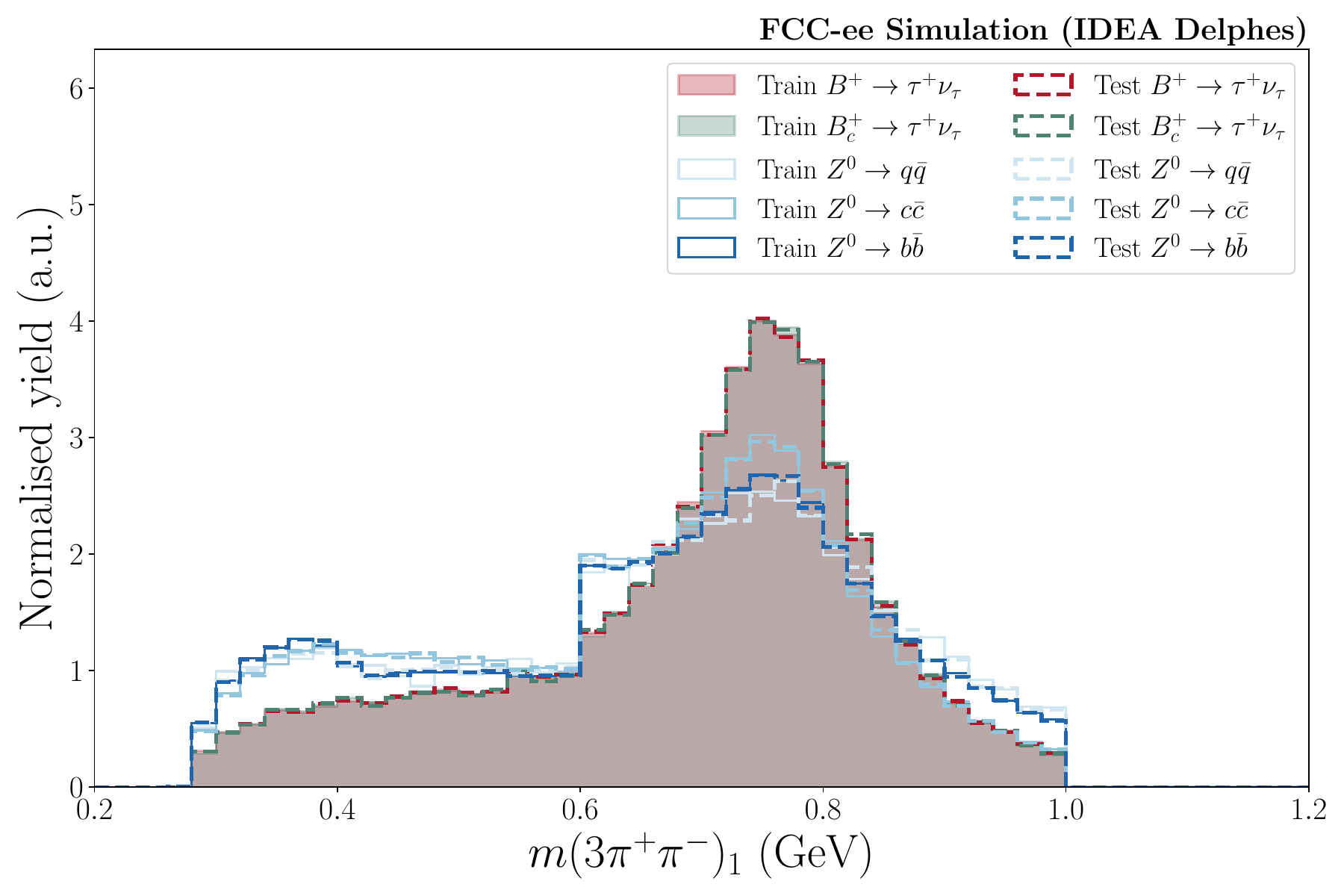}
    \includegraphics[width=0.24\textwidth]{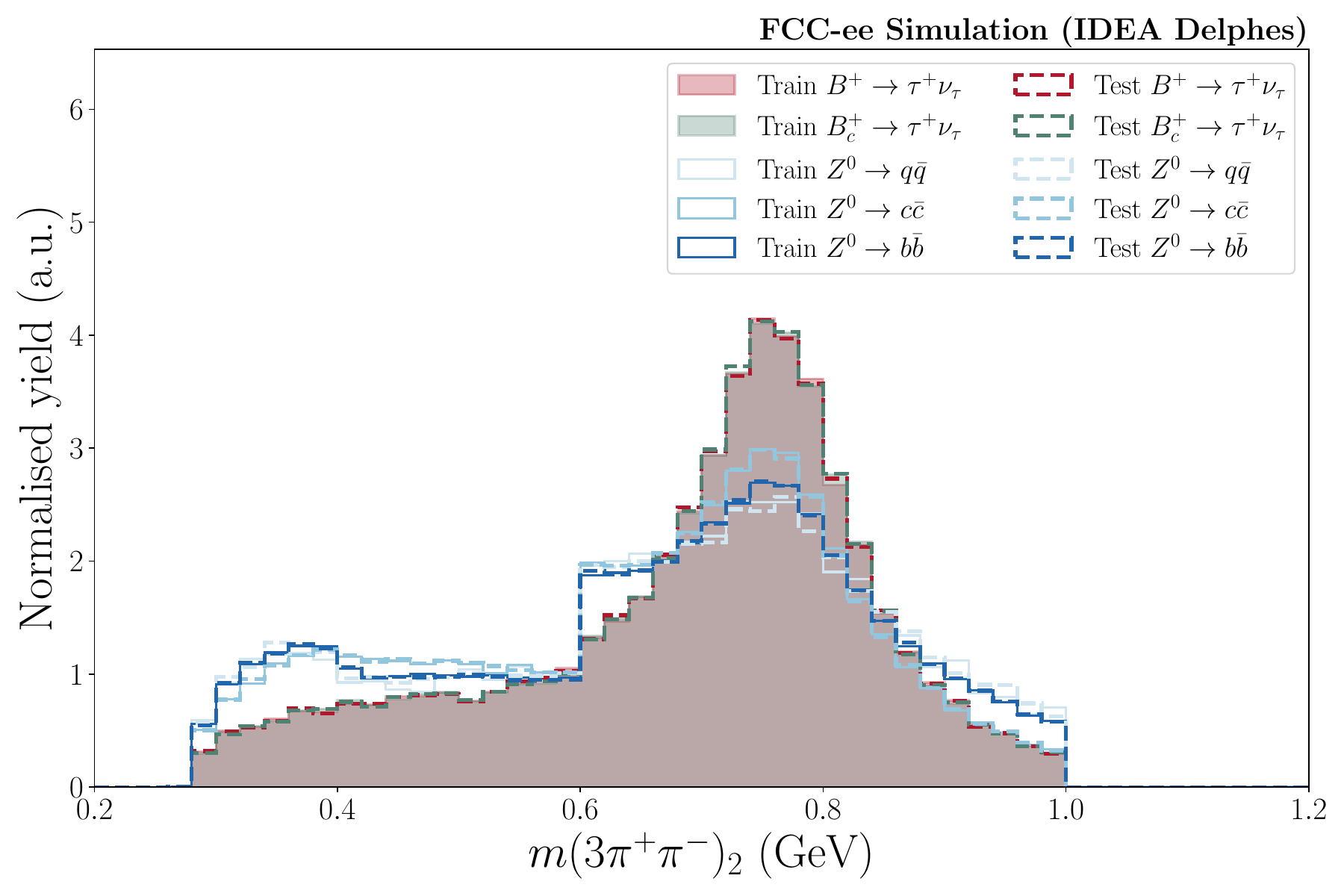} 
    \includegraphics[width=0.24\textwidth]{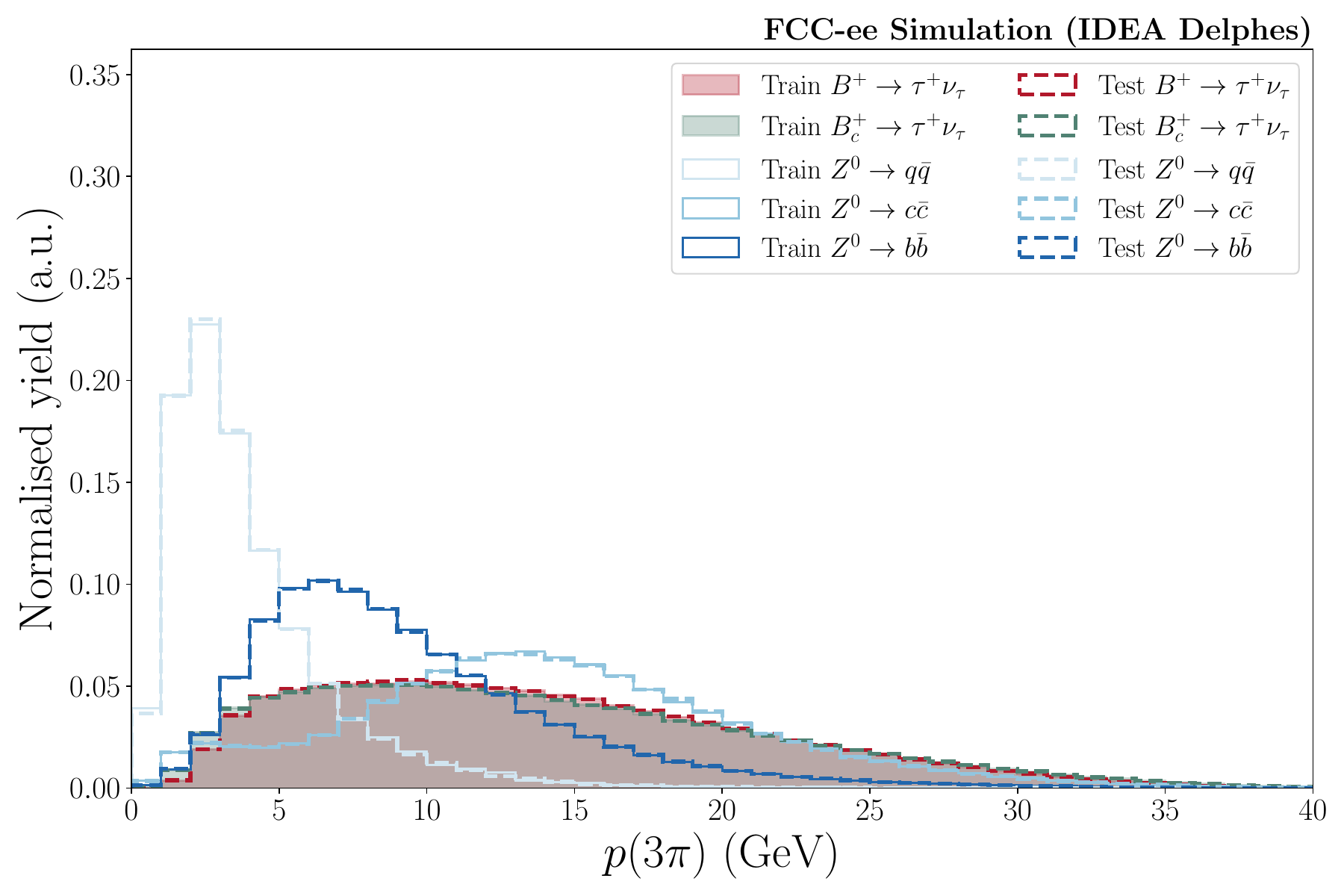}
    \includegraphics[width=0.24\textwidth]{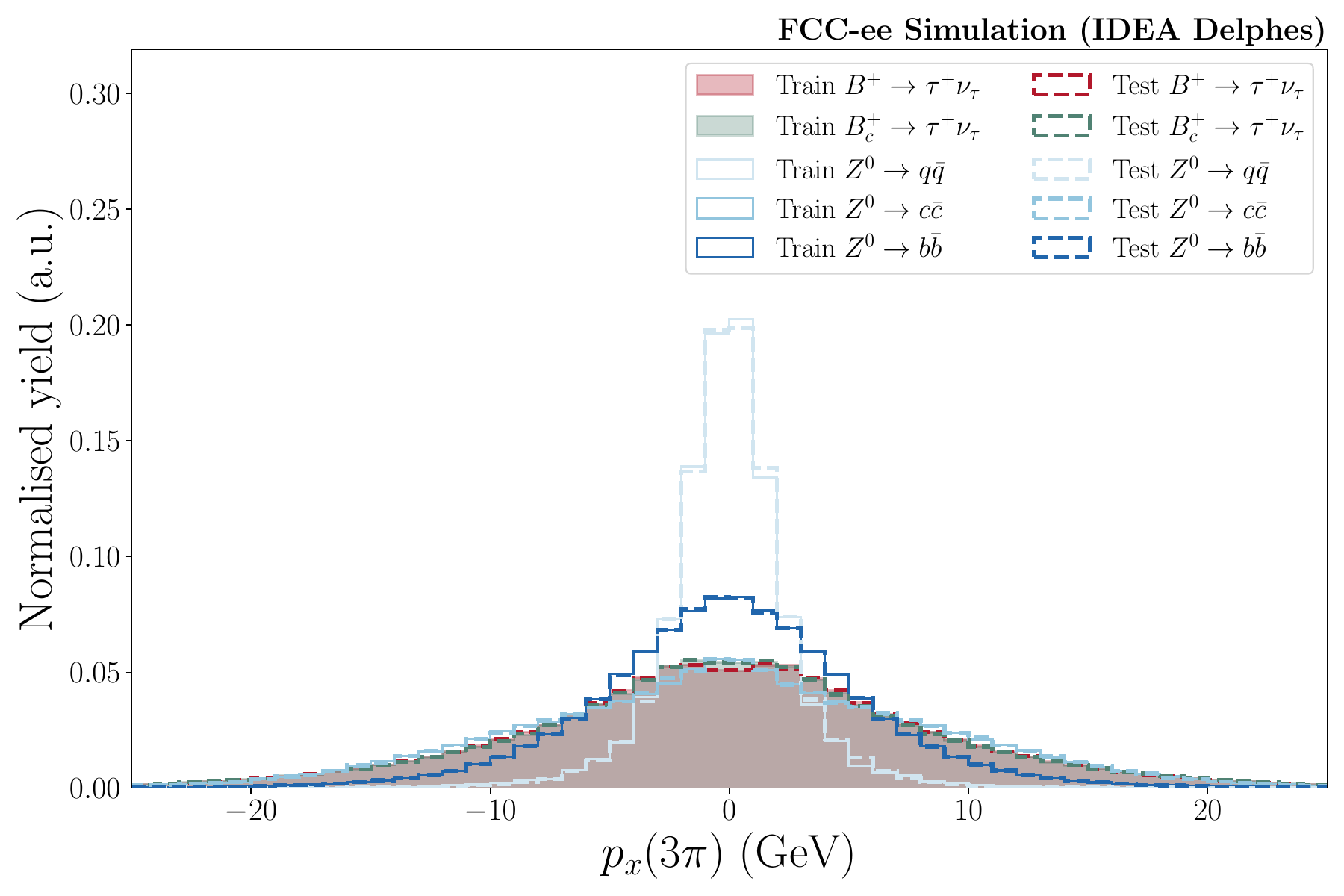} \\
    \includegraphics[width=0.24\textwidth]{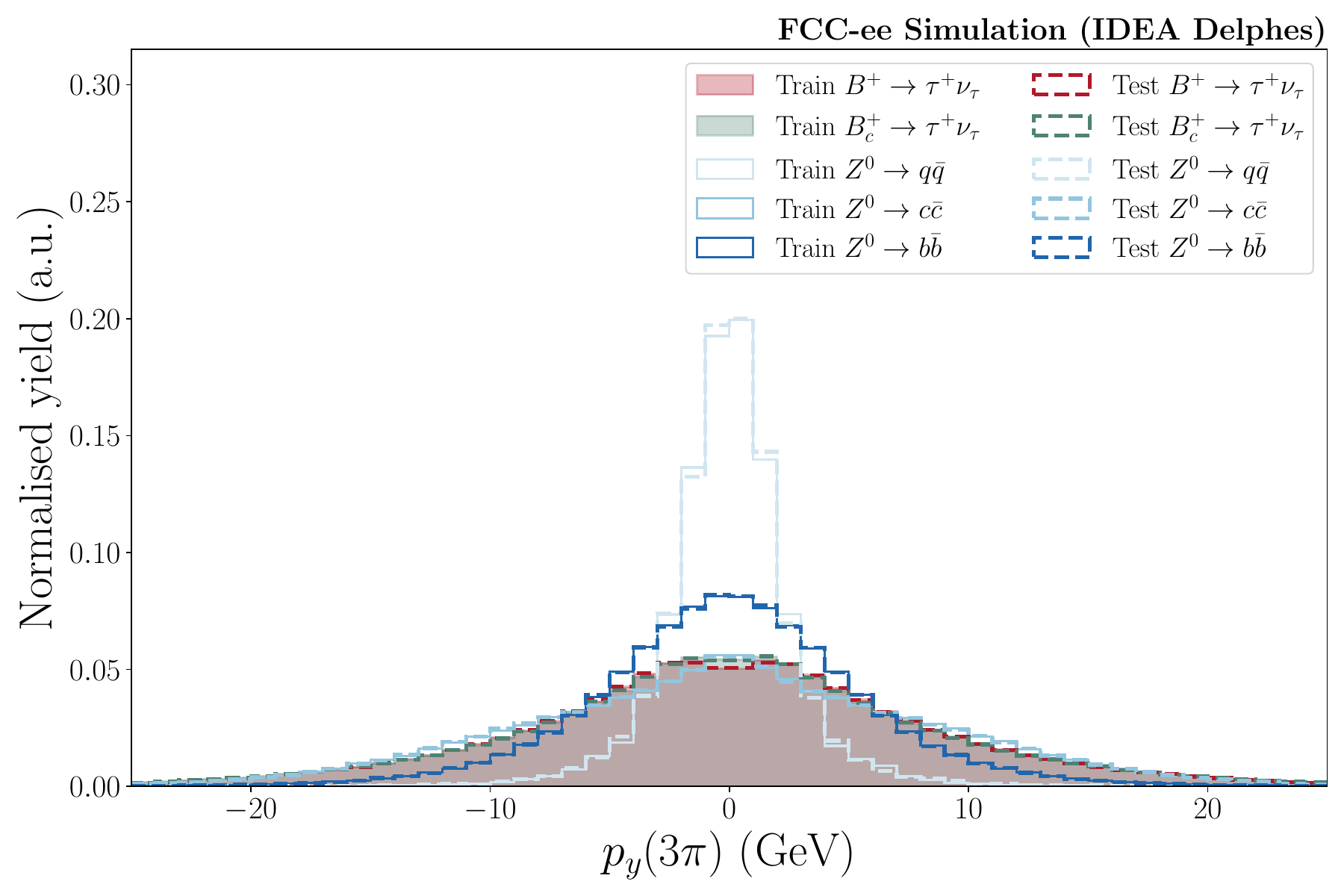} 
    \includegraphics[width=0.24\textwidth]{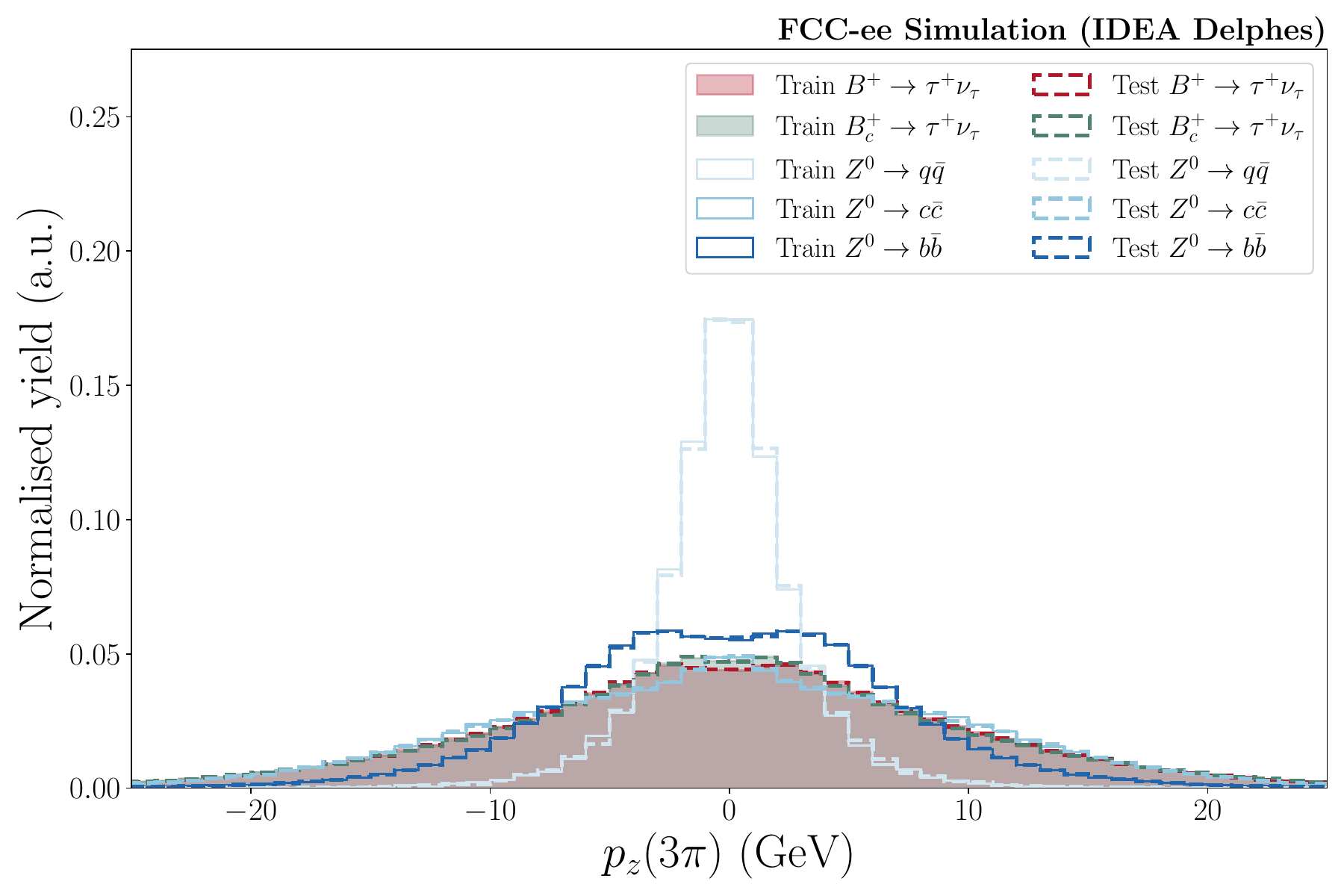}
    \includegraphics[width=0.24\textwidth]{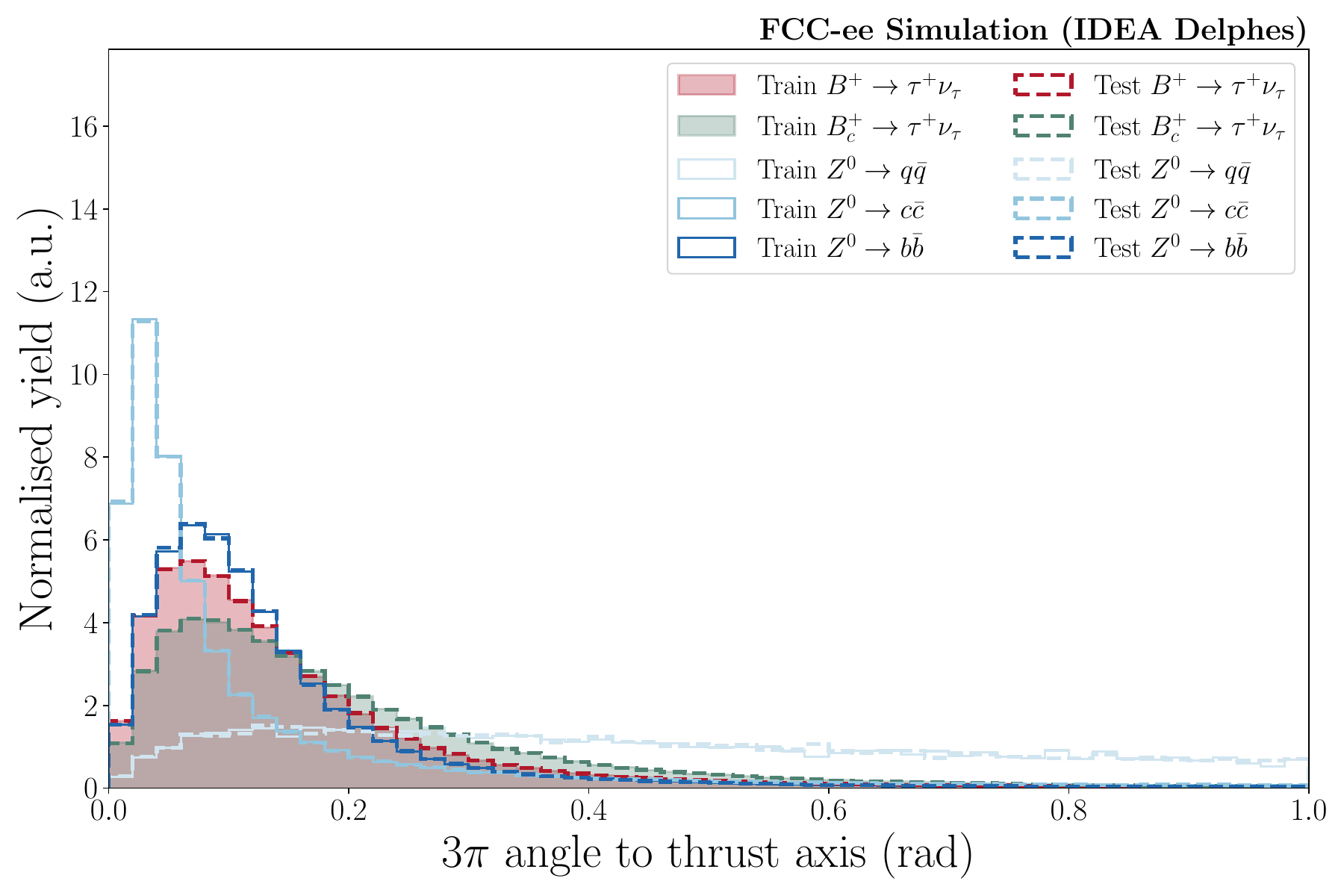}
    \includegraphics[width=0.24\textwidth]{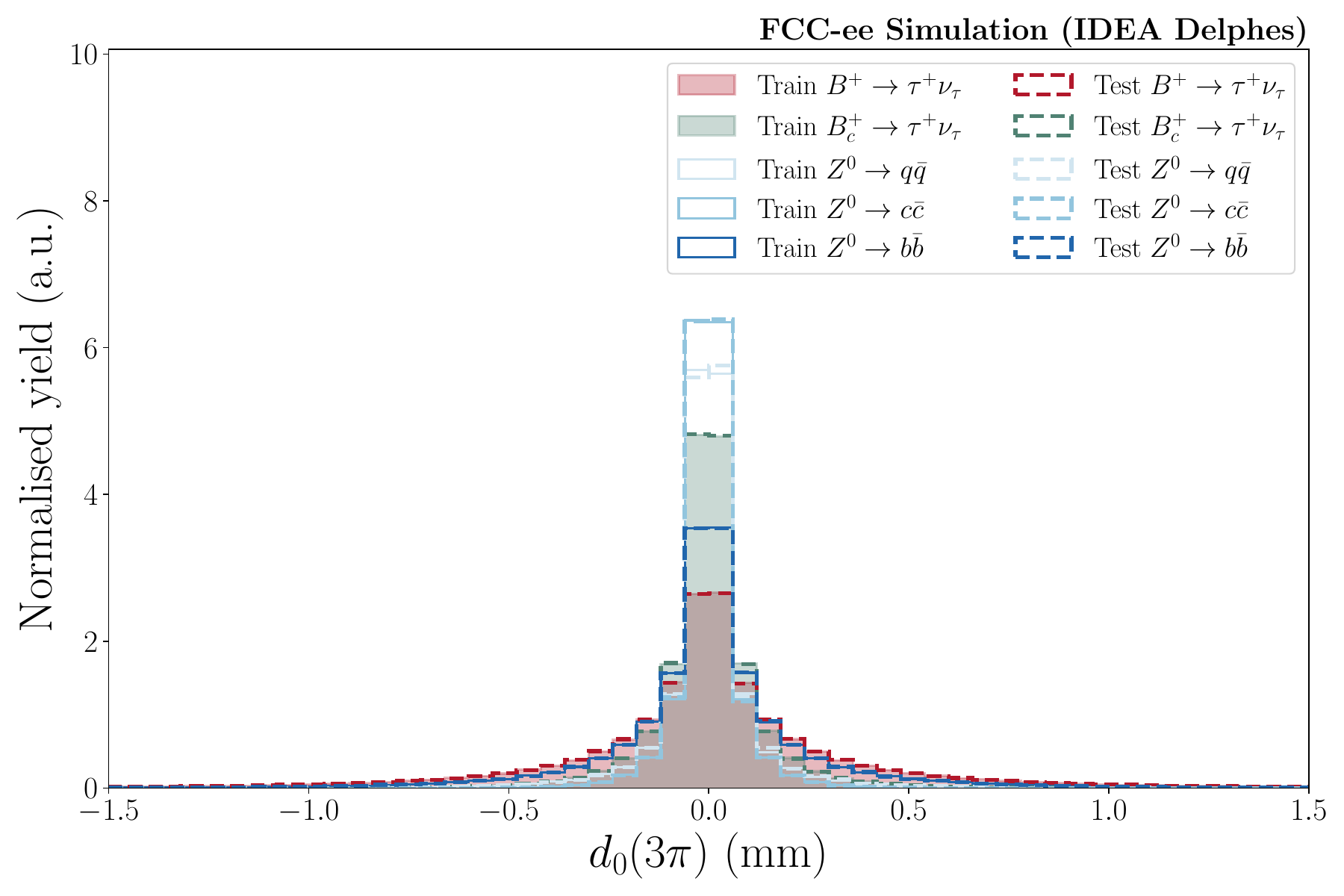} \\
    \includegraphics[width=0.24\textwidth]{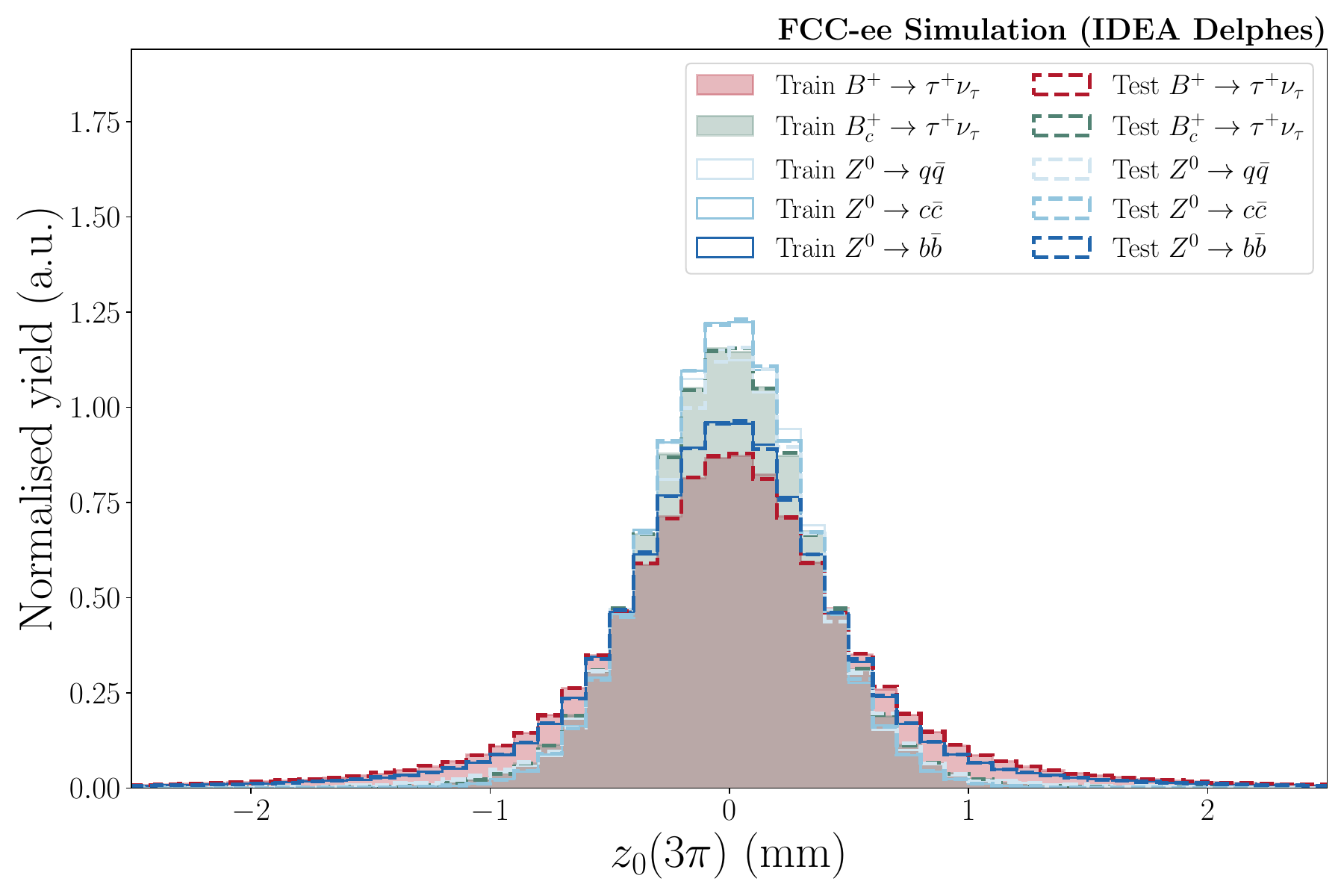}
    \includegraphics[width=0.24\textwidth]{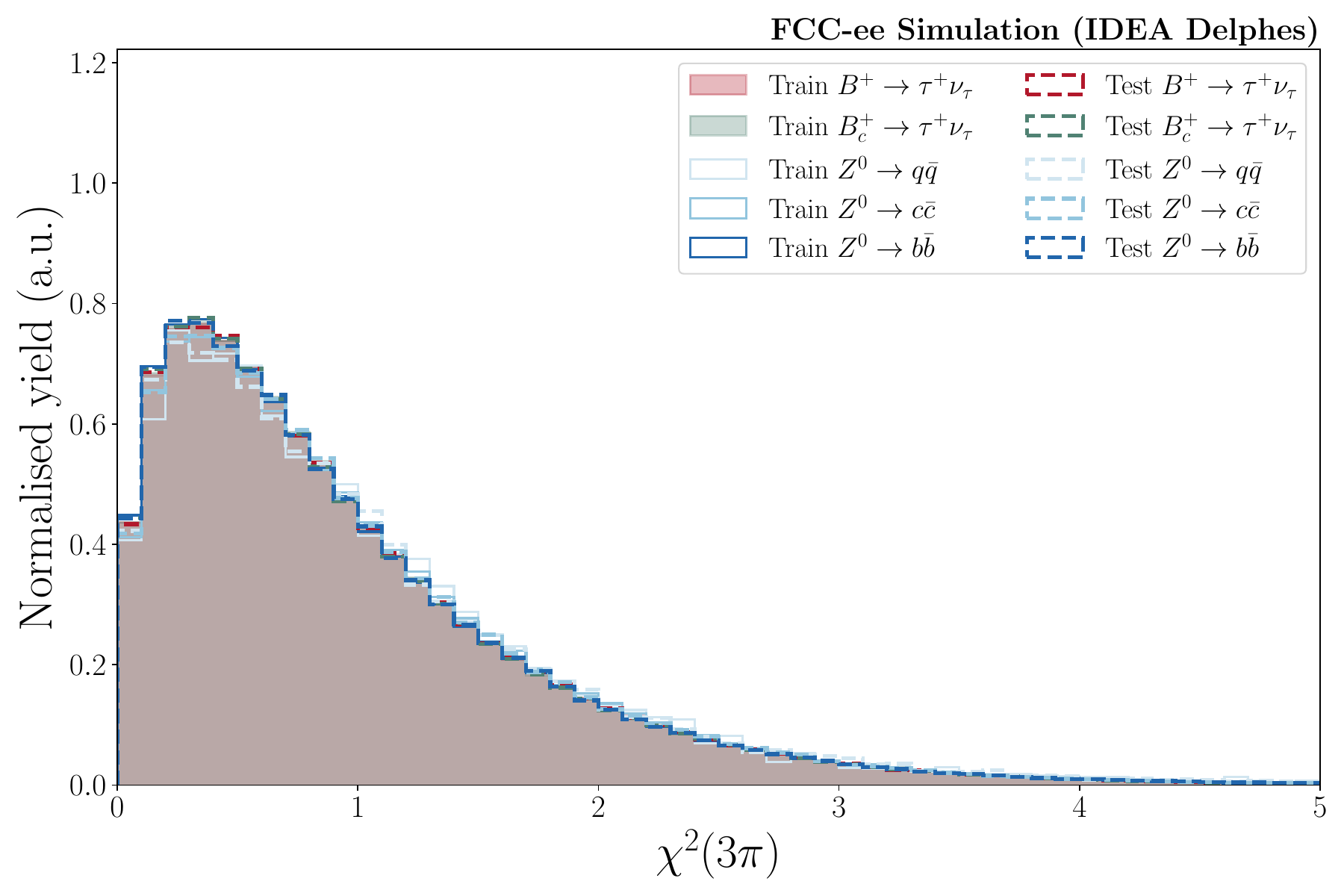}
    \includegraphics[width=0.24\textwidth]{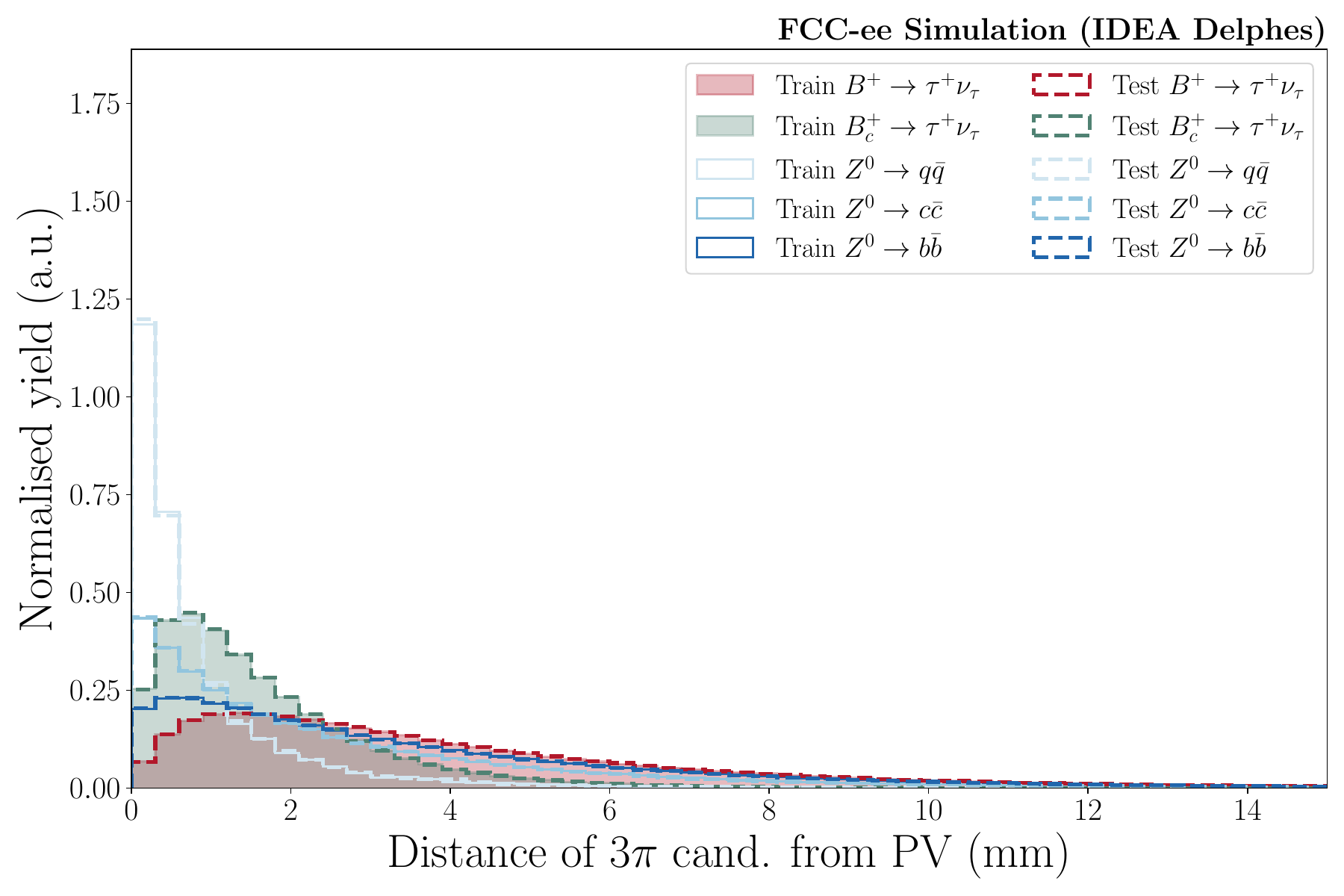} 
    \includegraphics[width=0.24\textwidth]{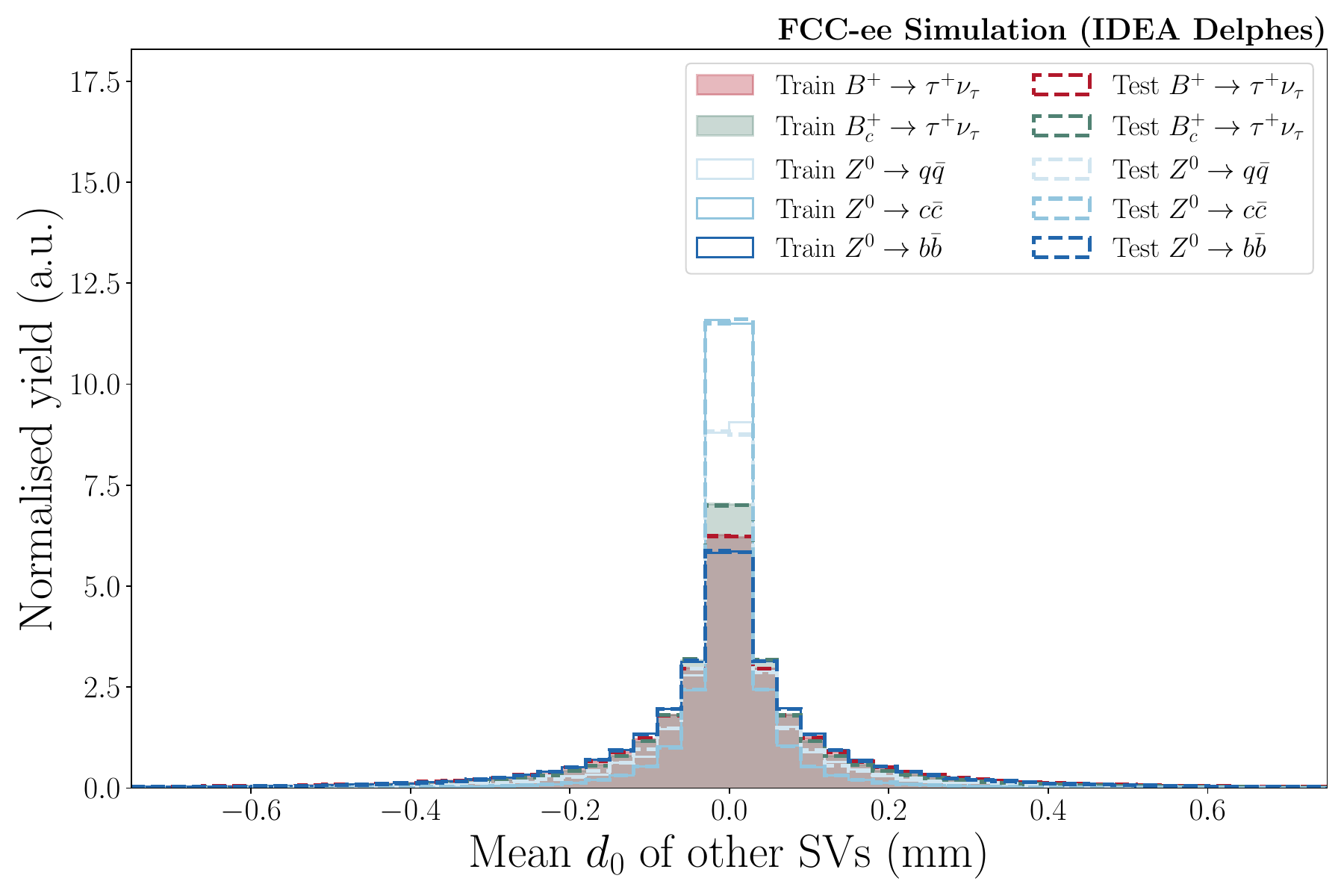} \\
    \includegraphics[width=0.24\textwidth]{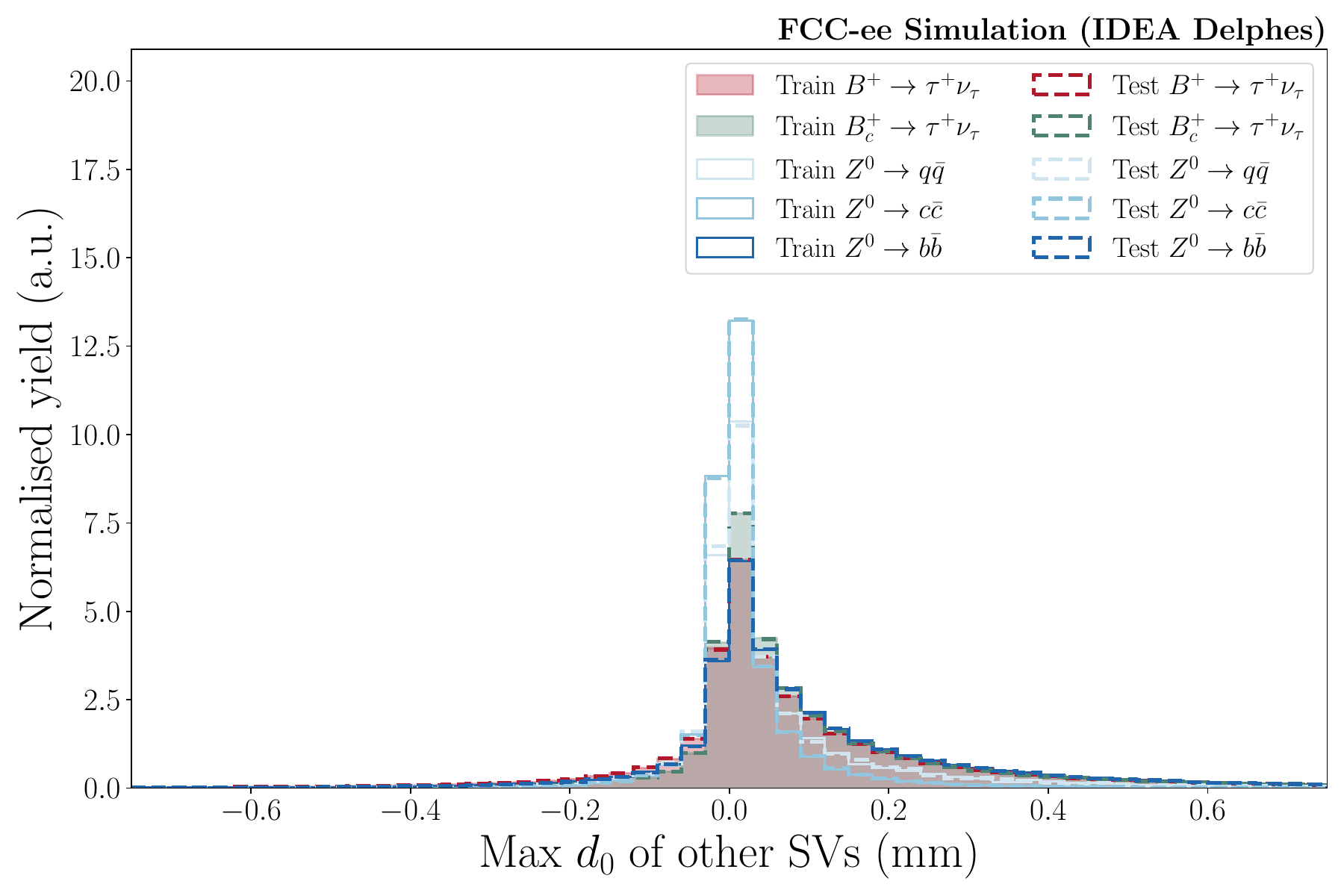}
    \includegraphics[width=0.24\textwidth]{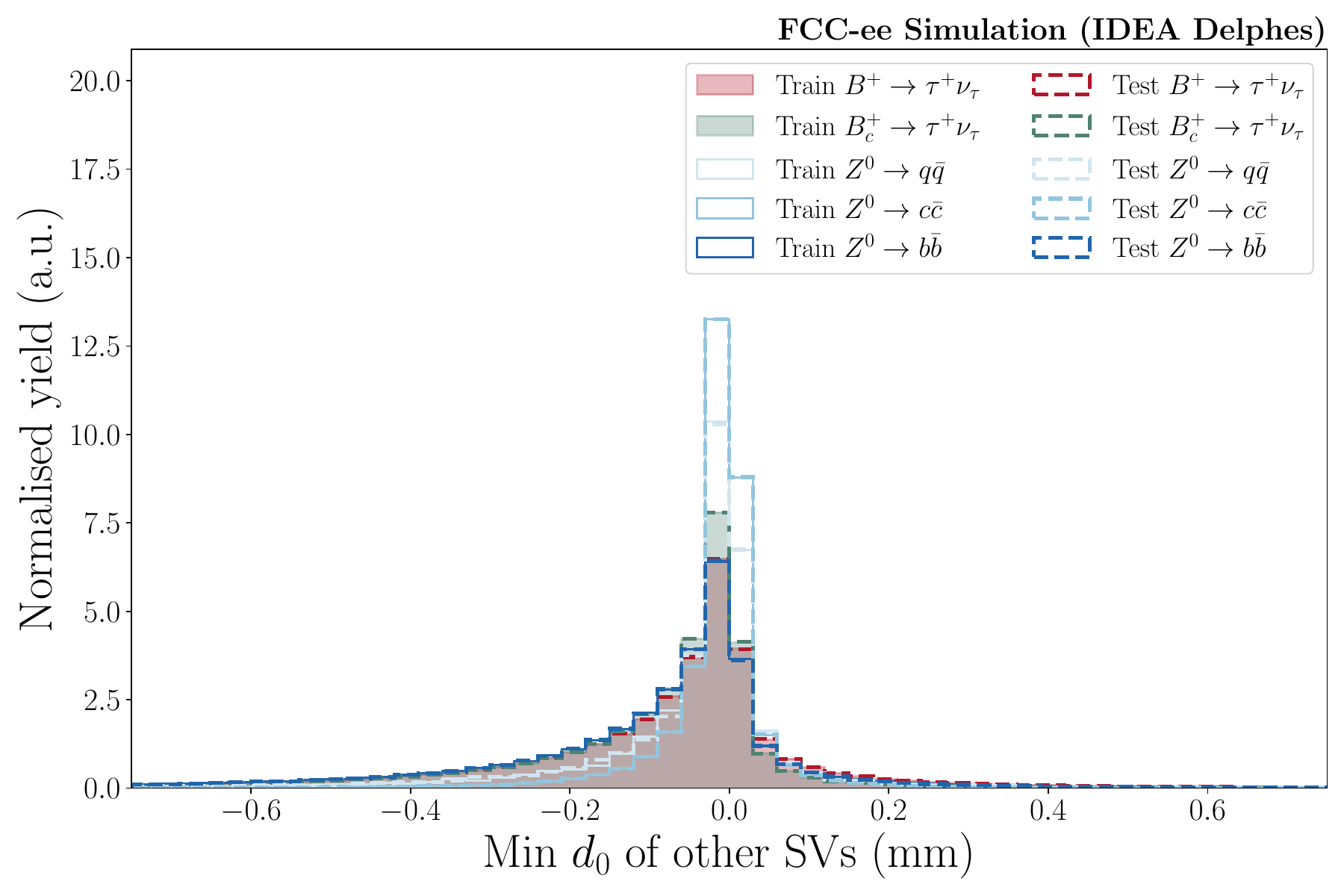} 
    \includegraphics[width=0.24\textwidth]{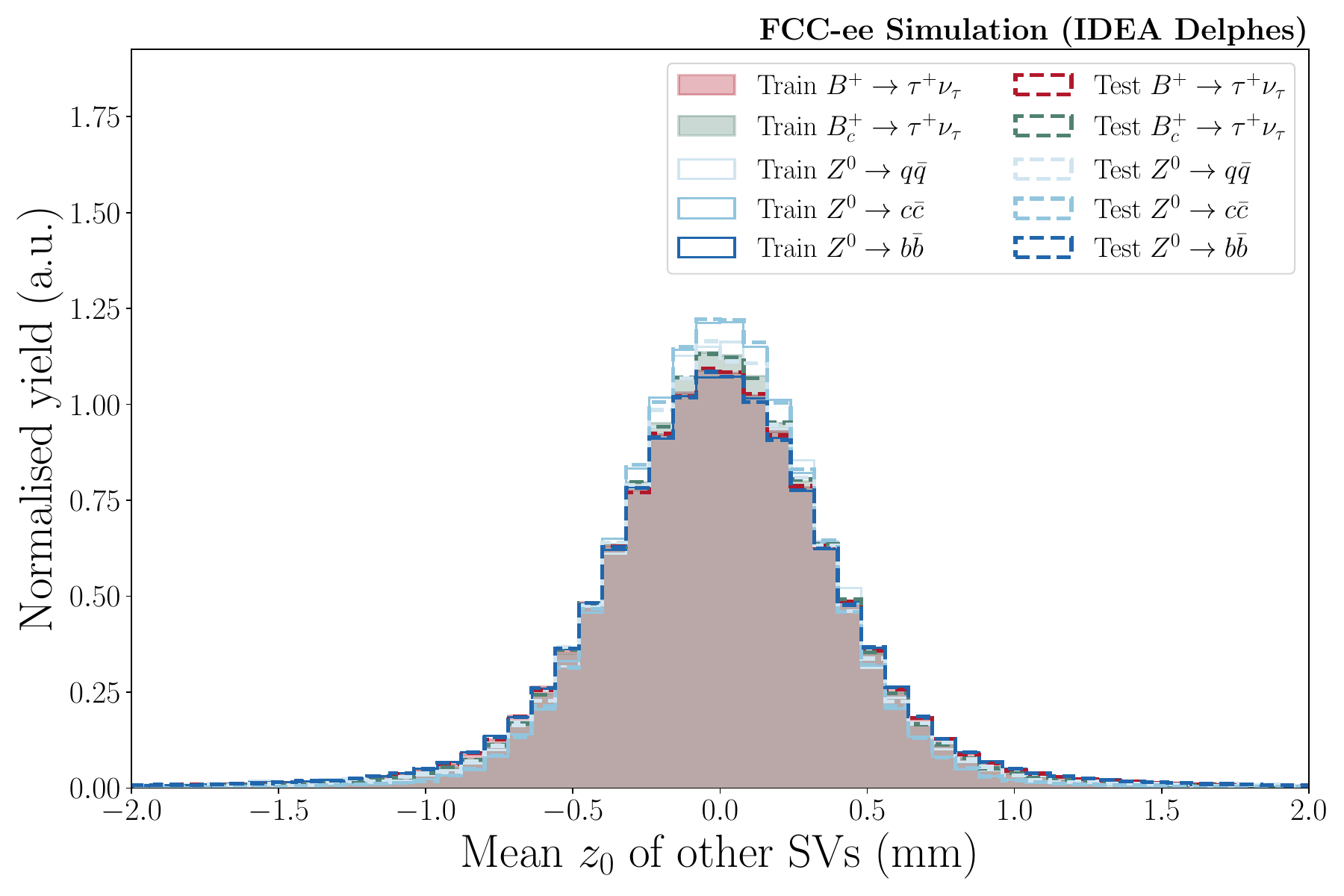}
    \includegraphics[width=0.24\textwidth]{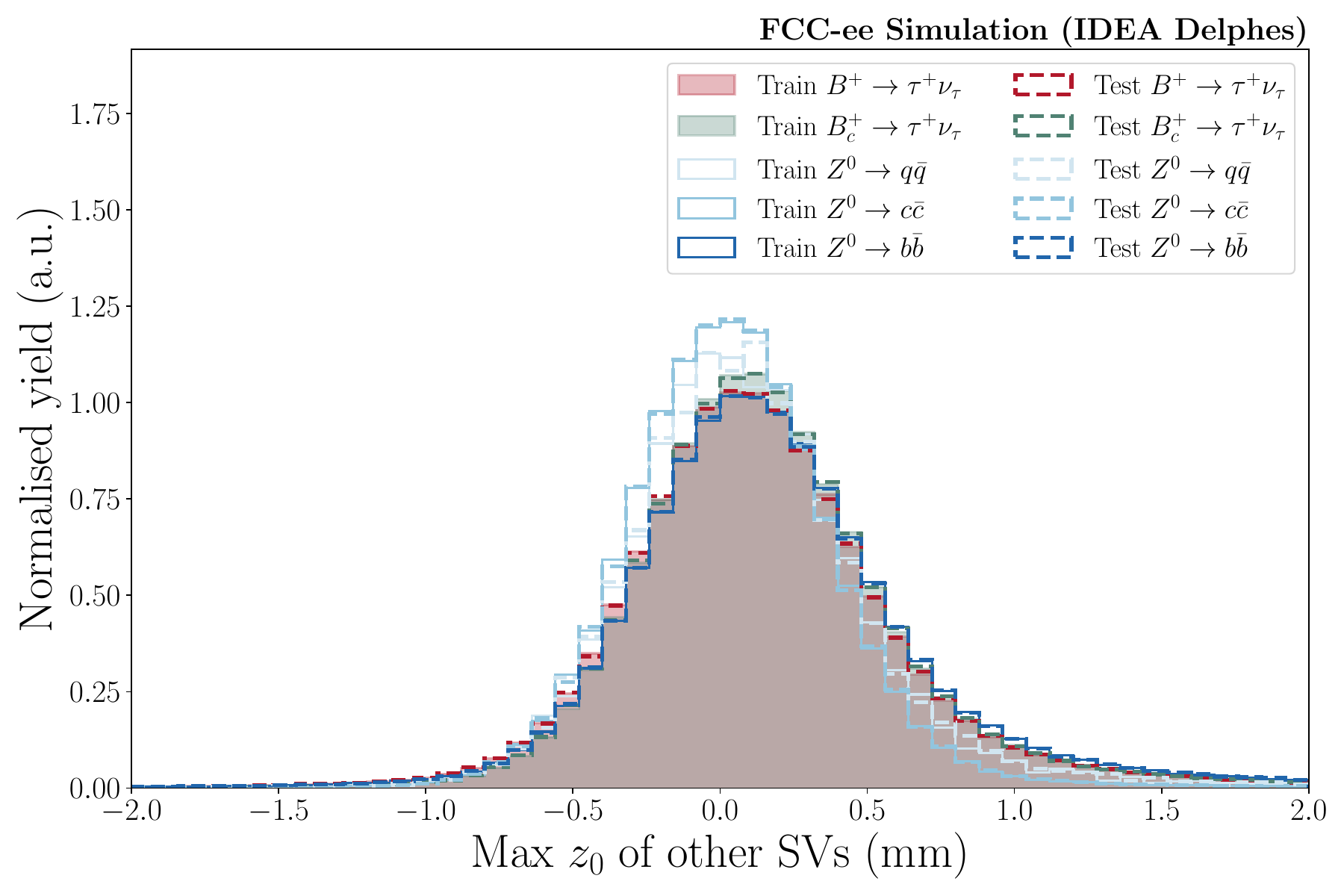} \\
    \includegraphics[width=0.24\textwidth]{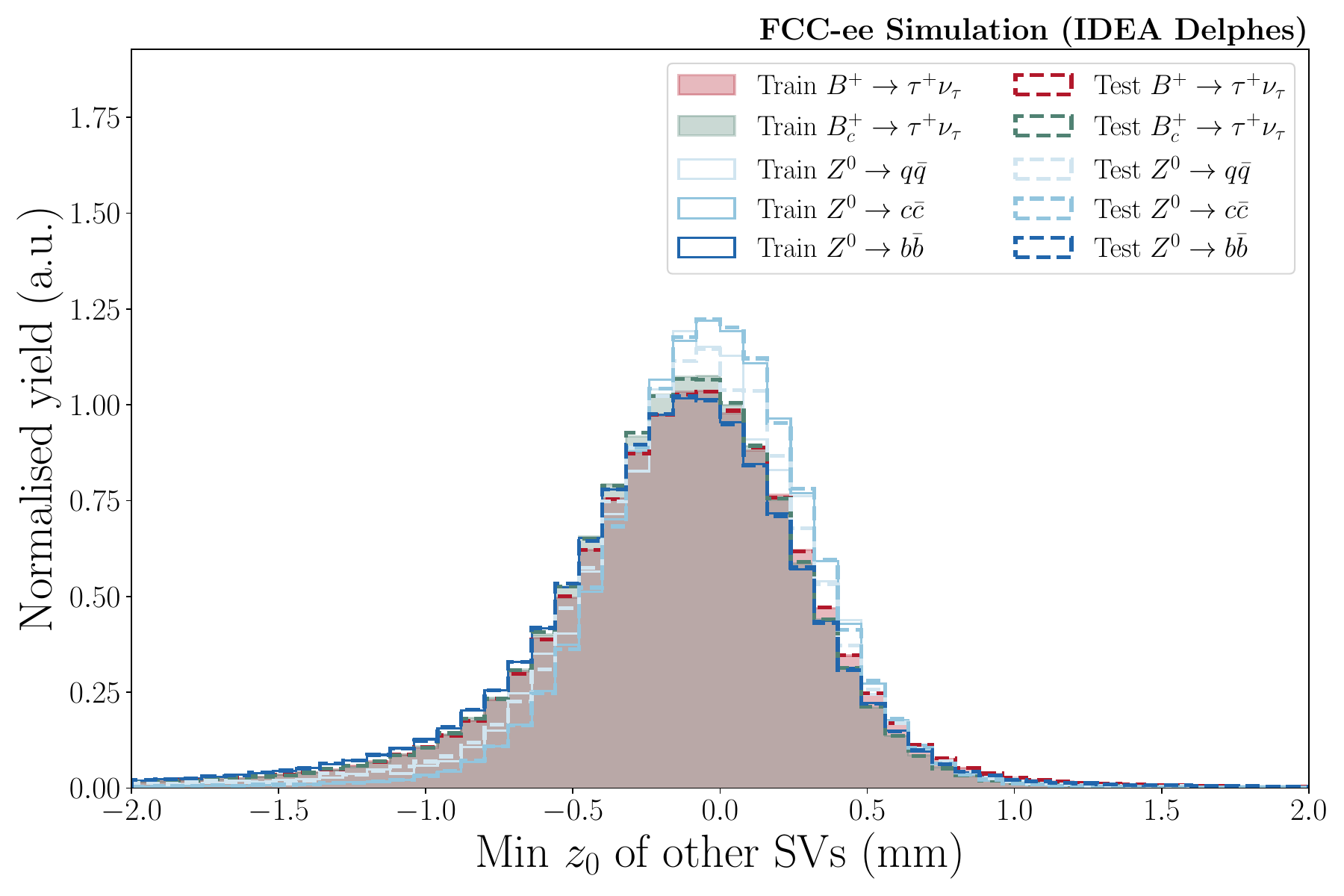} 
    \includegraphics[width=0.24\textwidth]{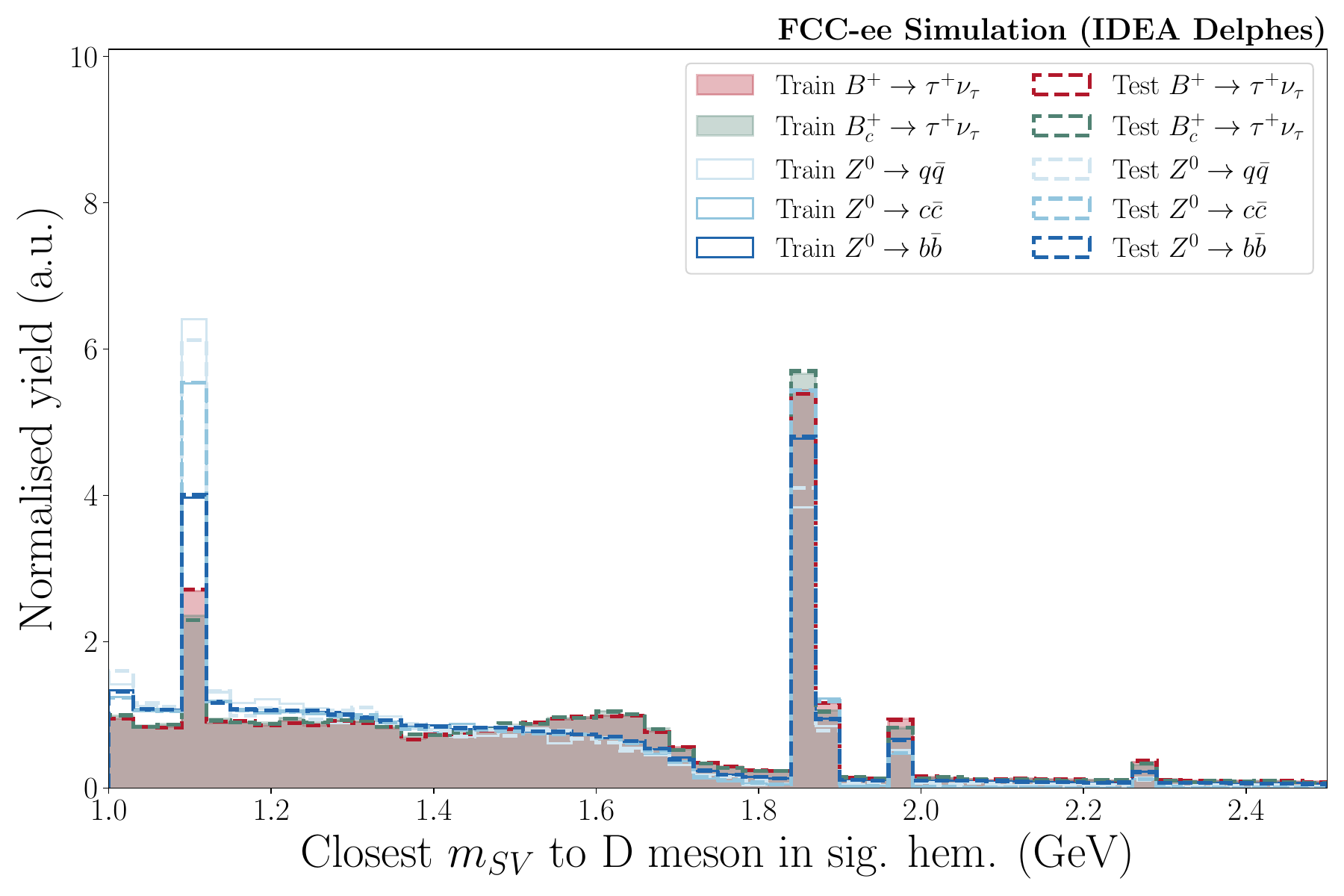}
    \includegraphics[width=0.24\textwidth]{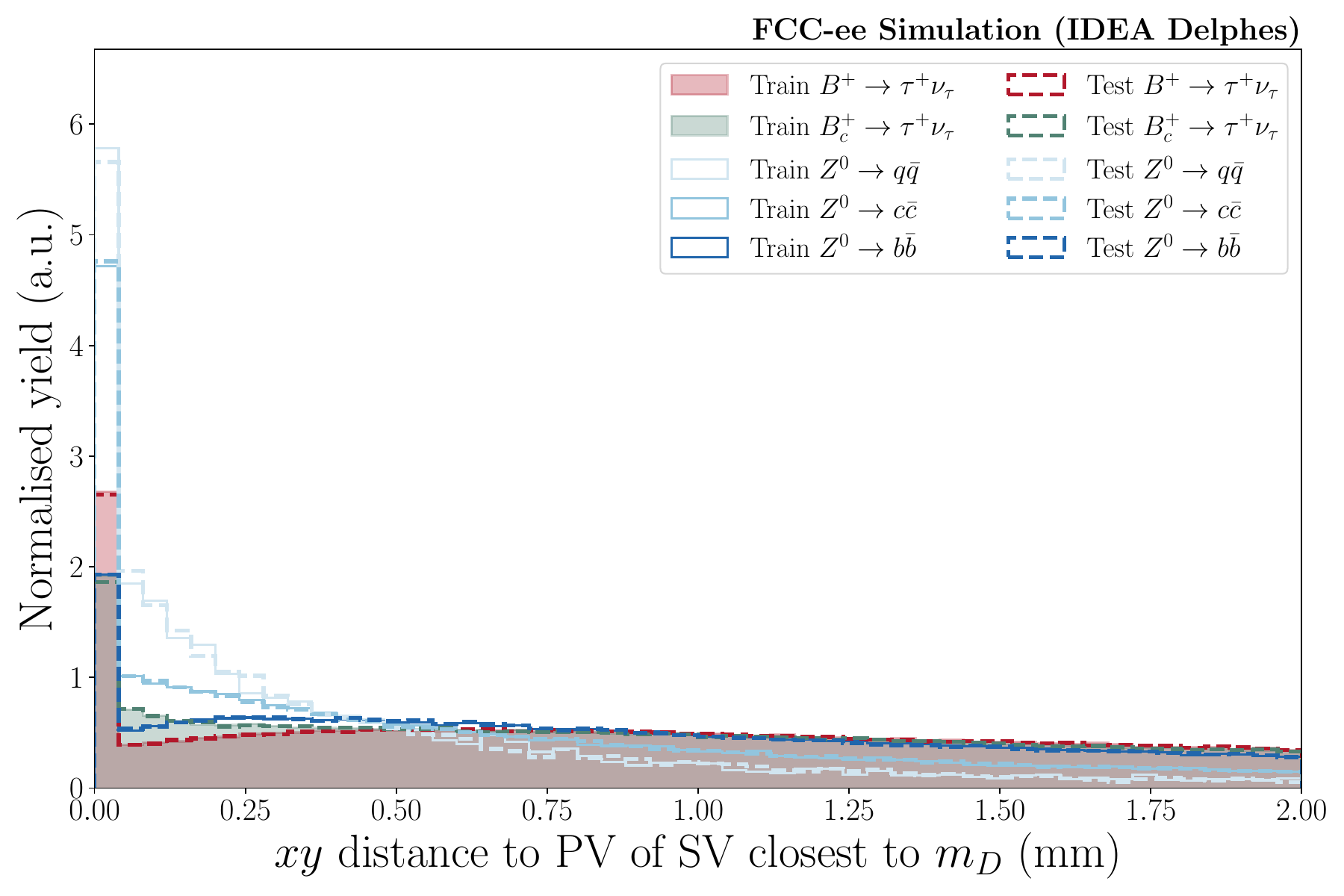}
    \includegraphics[width=0.24\textwidth]{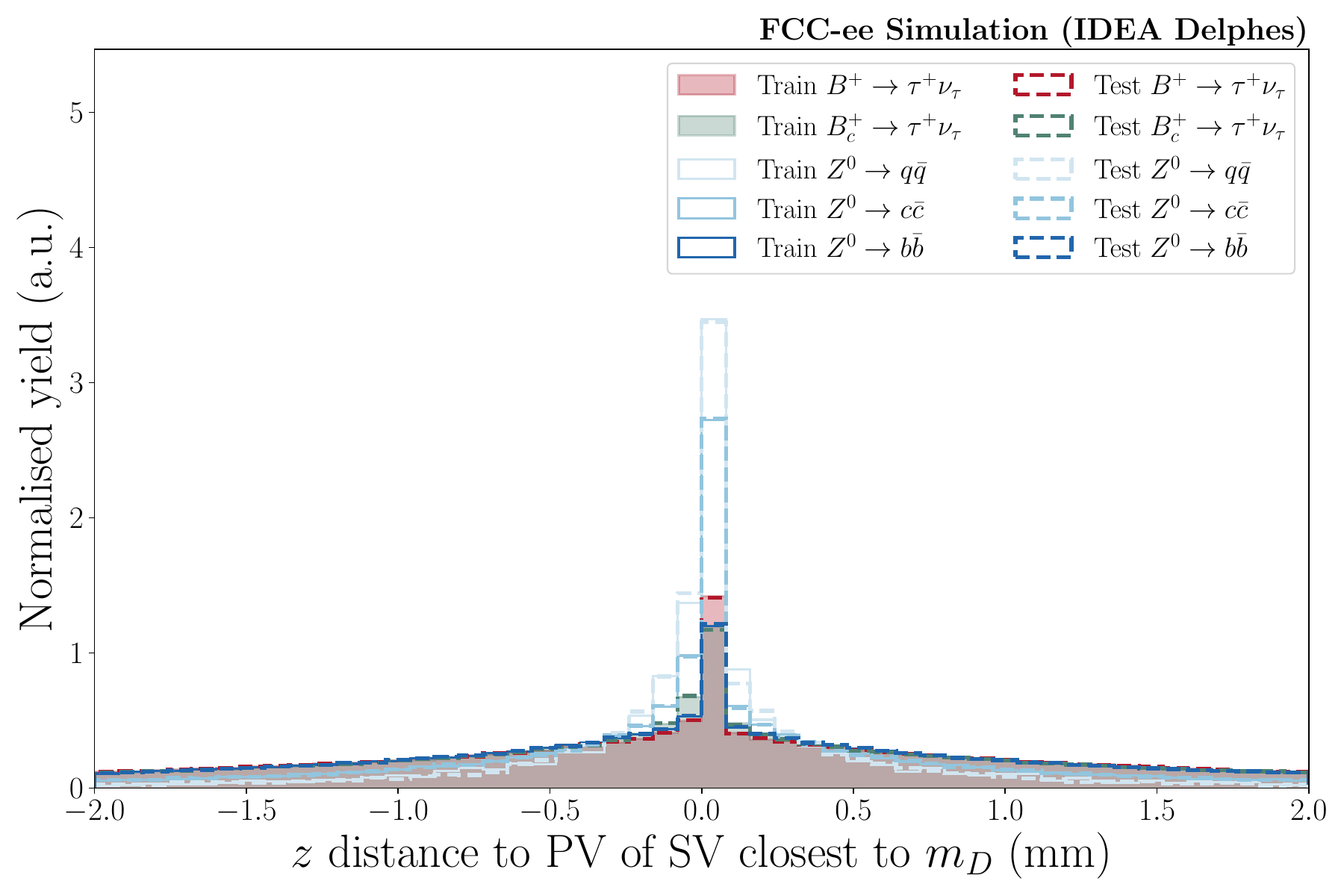} \\
    \caption{Second-stage BDT training variable distributions in training samples (solid lines) and testing samples (dashed lines).}
    \label{fig:BDT2_vars}
\end{figure}

Figures~\ref{fig:BDT1_vars} and~\ref{fig:BDT2_vars} summarise the distributions of input variables to the first-stage and second-stage BDTs.
All features are compared between different signal and background processes, as well as between training samples and testing samples.
These distributions are profiled with the same set of events as those used to study BDT performance in Section~\ref{sec:BDT1} and~\ref{sec:BDT2}.
A high level of consistency is observed in the distributions of all variables between training and testing samples, demonstrating no statistical bias in the training of the BDTs.

\FloatBarrier

\section{Choice of fit variable}\label{app:fit_var}

\begin{figure}[h!]
    \centering
    \includegraphics[width=0.42\textwidth]{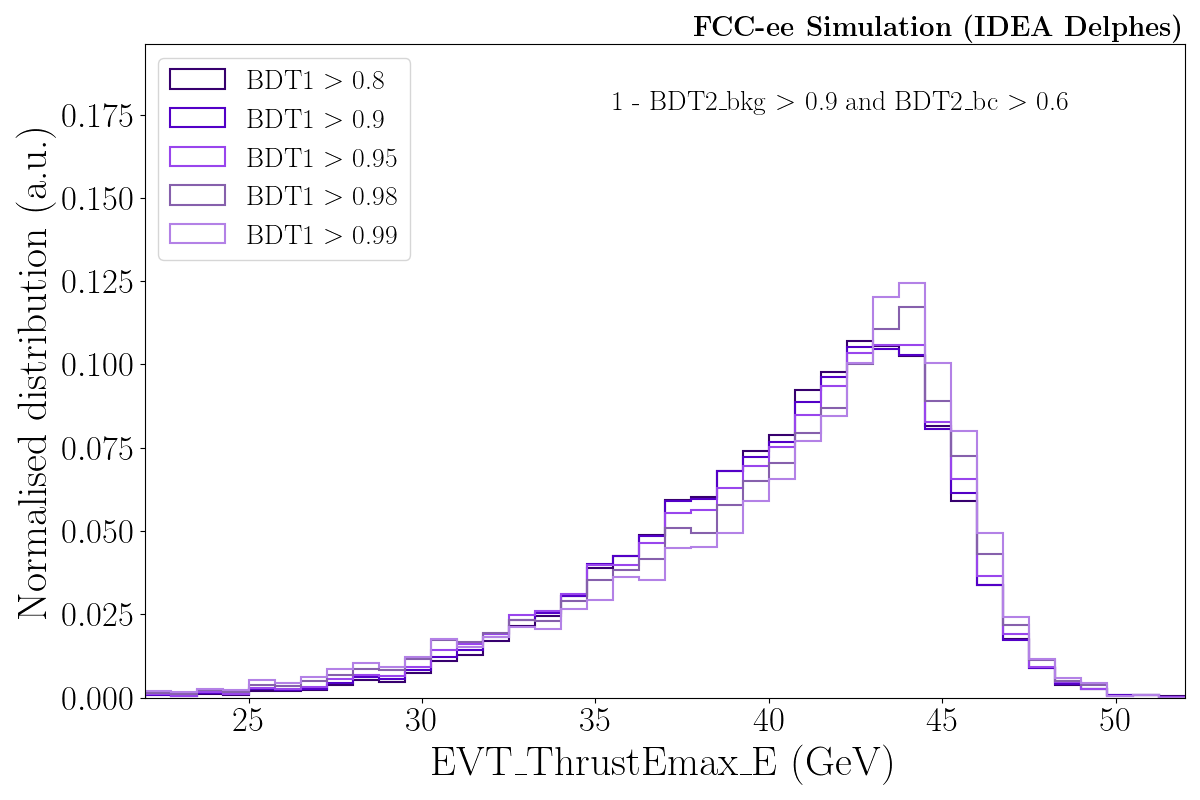}
    \includegraphics[width=0.42\textwidth]{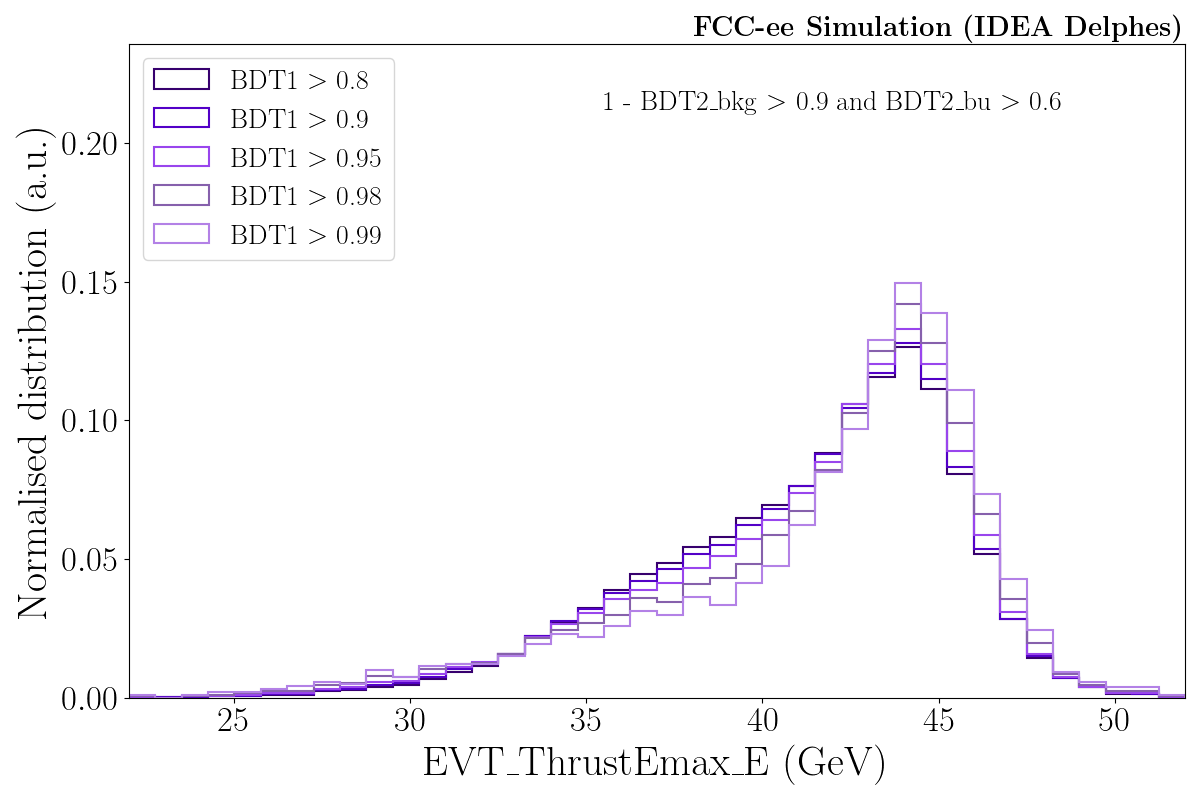}
    \includegraphics[width=0.42\textwidth]{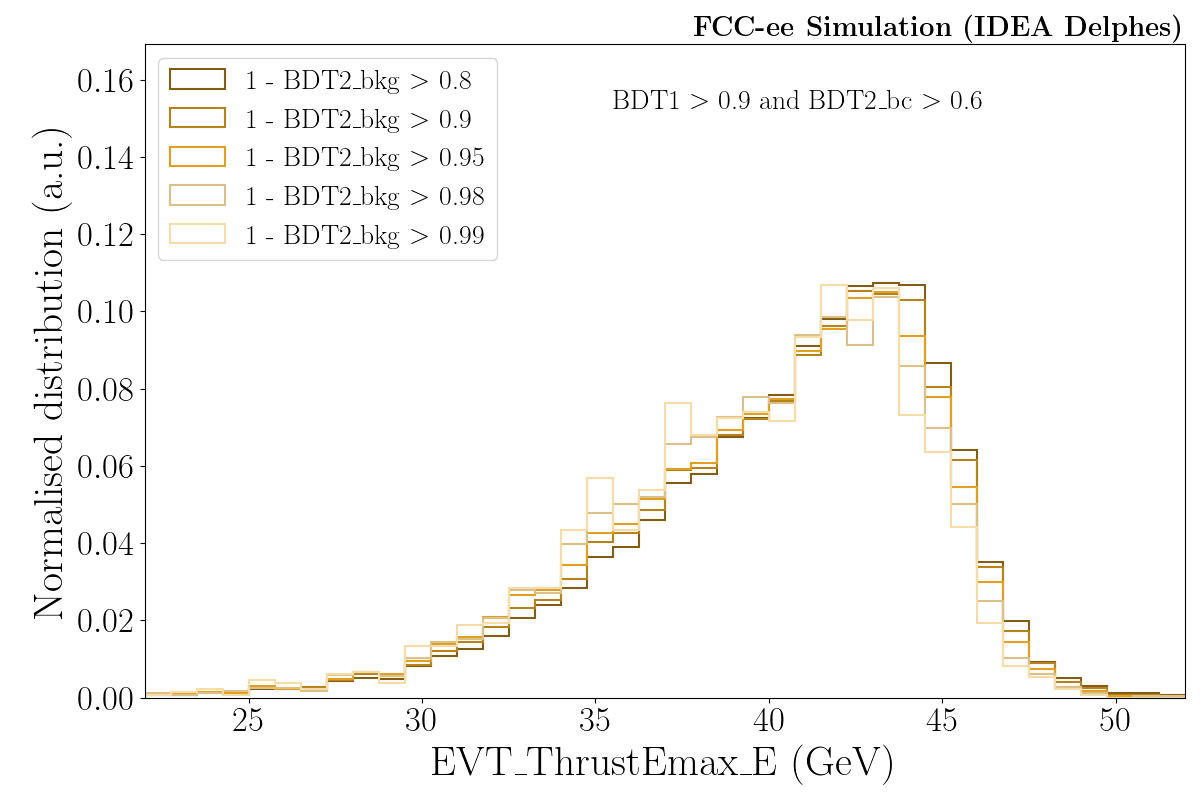}
    \includegraphics[width=0.42\textwidth]{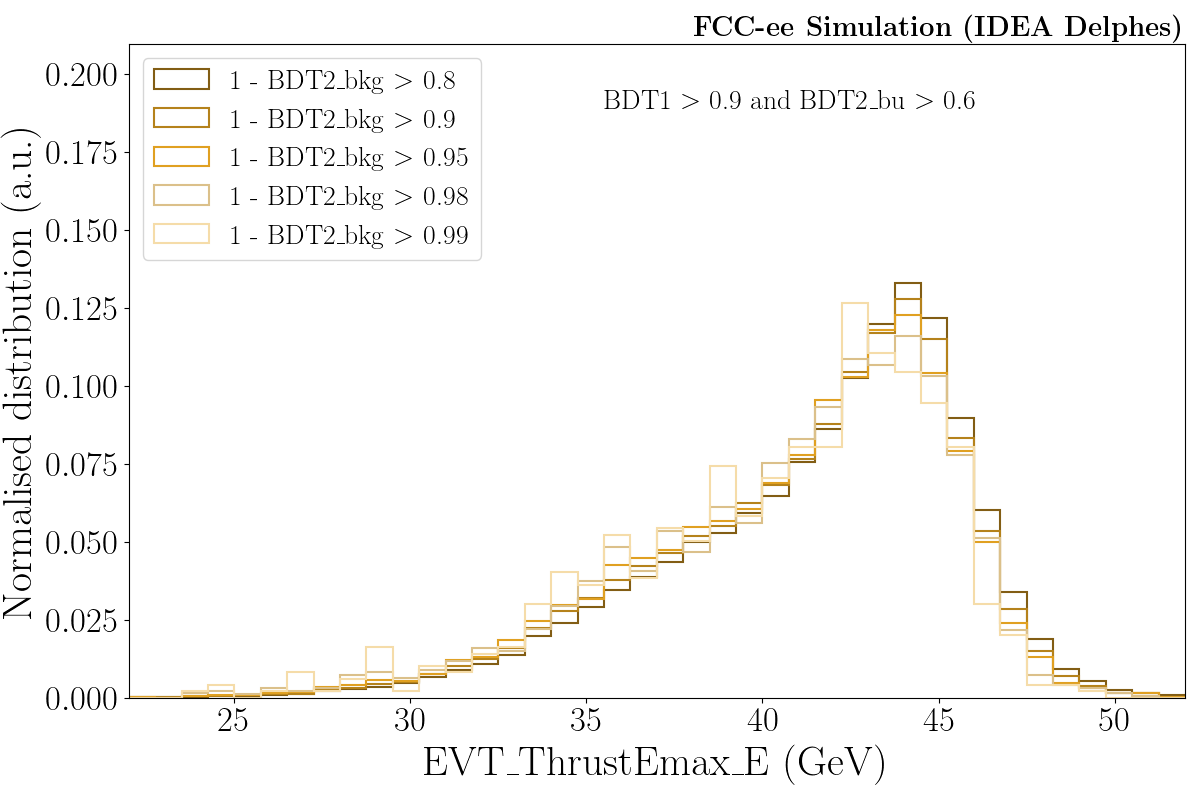}
    \caption{An example of maximum hemisphere energy dependence on MVA selections \revision{for background events}. The left column show the distributions in the Bc category and the right column show the distributions in the Bu category. The top row is the shape variation with regard to the BDT1 cuts and the bottom row is the shape variation with regard to the BDT2 bkg cuts. \revision{The label at the top right of each plot is the starting selection and the legend at the top left corner is the series of tigher selections.} The energy in the maximum hemisphere is independent from MVA cuts from the baseline selection to very tight selections.}
    \label{fig:fit_var_Emax}
\end{figure}

\begin{figure}[htb!]
    \centering
    \includegraphics[width=0.42\textwidth]{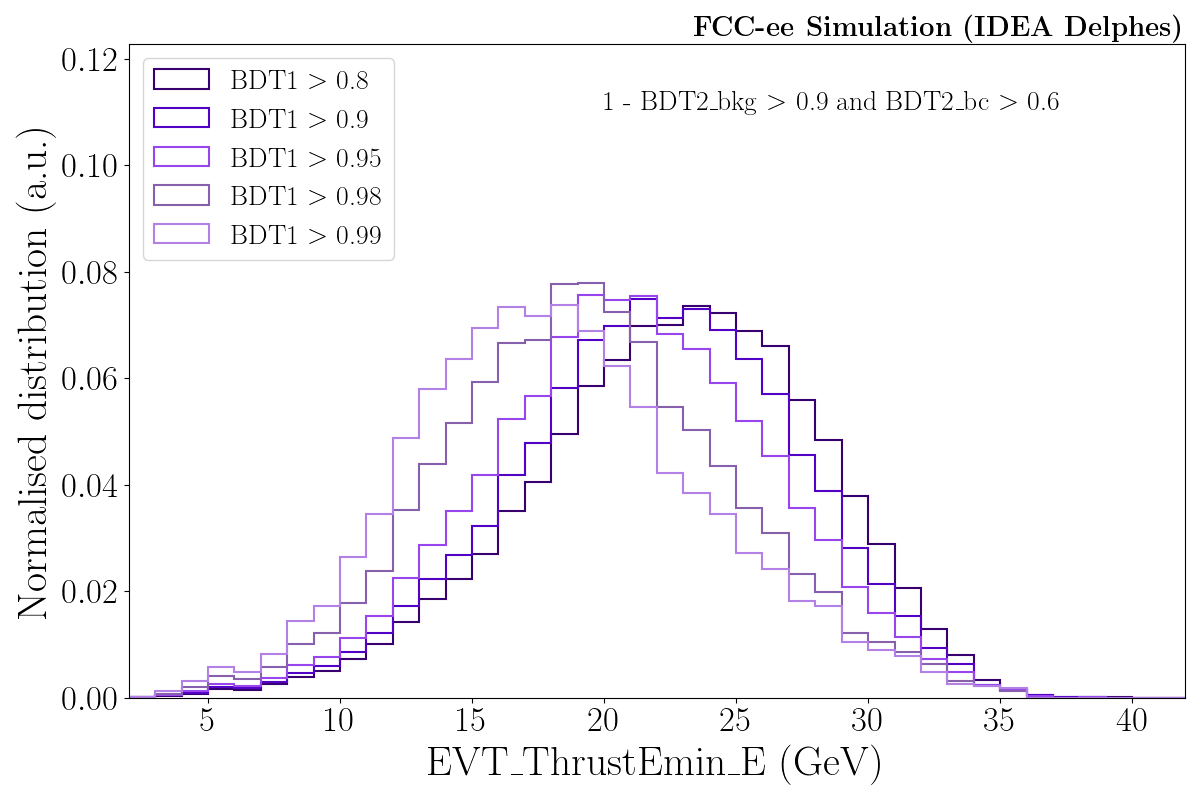}
    \includegraphics[width=0.42\textwidth]{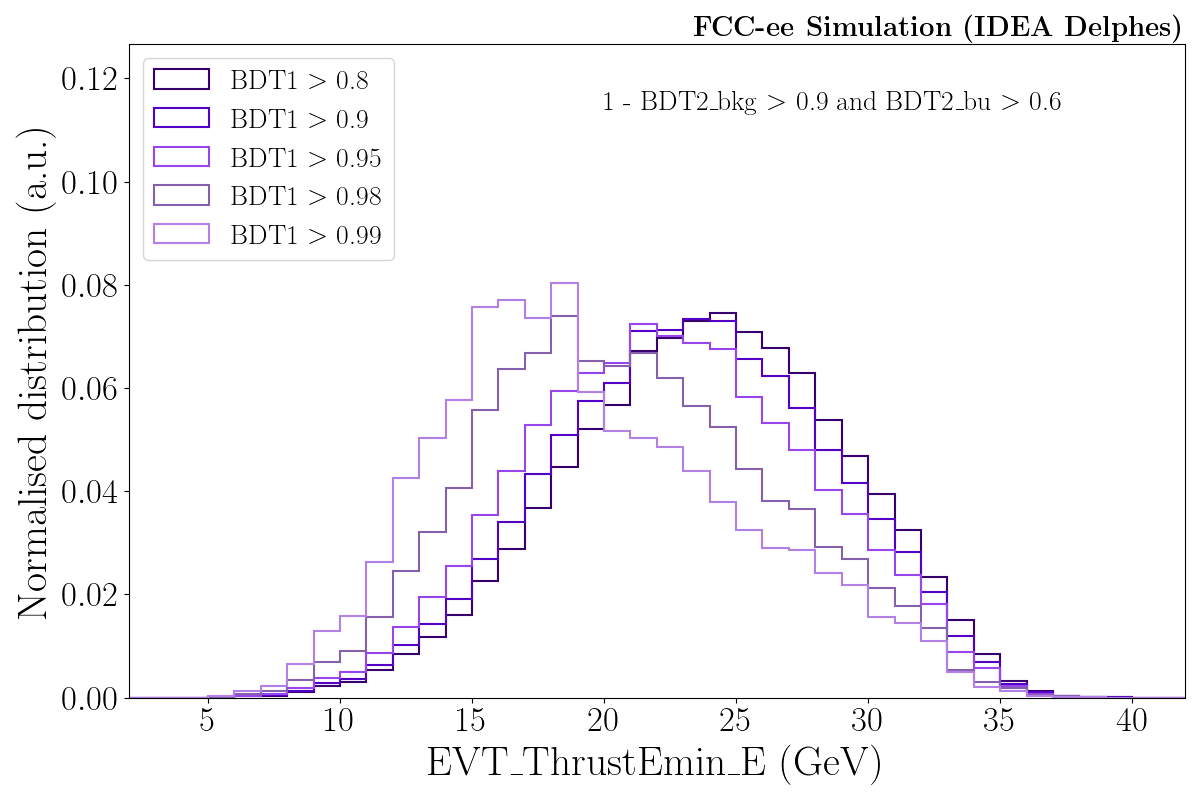}
    \includegraphics[width=0.42\textwidth]{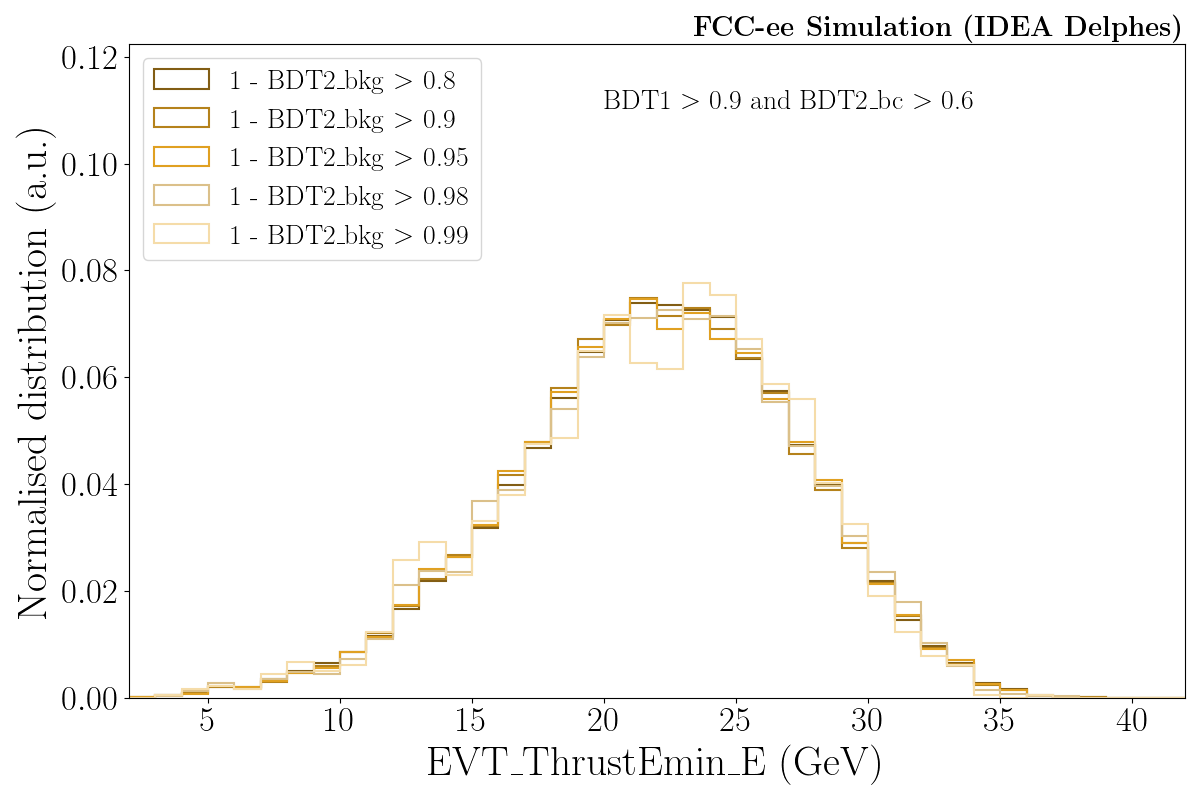}
    \includegraphics[width=0.42\textwidth]{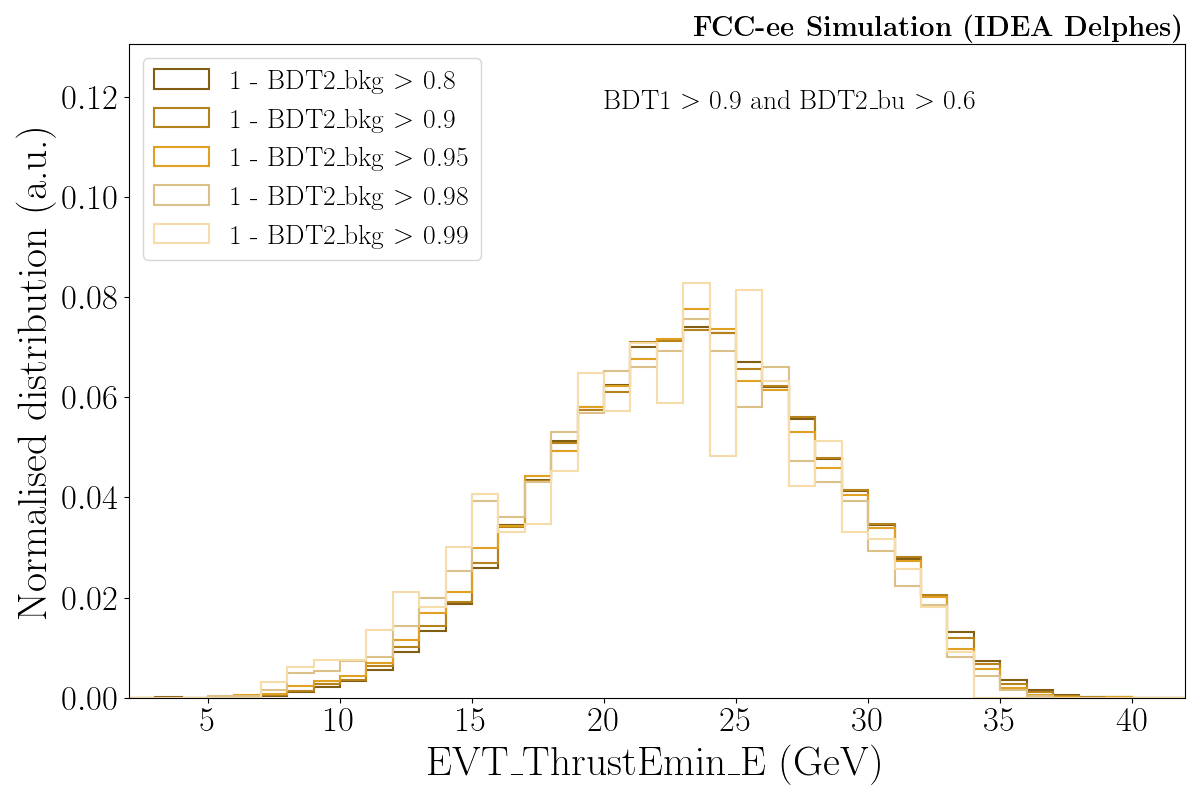}
    \caption{An example of minimum hemisphere energy dependence on MVA selections \revision{for background events}. The left column show the distributions in the Bc category and the right column show the distributions in the Bu category. The top row is the shape variation with regard to the BDT1 cuts and the bottom row is the shape variation with regard to the BDT2 bkg cuts. \revision{The label at the top right of each plot is the starting selection and the legend at the top left corner is the series of tigher selections.} In both categories, the energy in the minimum hemisphere is significantly sculpted by the BDT1 cuts from the baseline selection to very tight selections.}
    \label{fig:fit_var_Emin}
\end{figure}

As explained in Section~\ref{sec:sel_opt}, there are not enough events in background simulation to model the background shape after the final selection, therefore background distributions are extracted with the baseline selection.
To validate the shapes extracted with this approach, the variable for the final fit needs to be uncorrelated with both MVAs.
Figures~\ref{fig:fit_var_Emax} and~\ref{fig:fit_var_Emin} summarises the variations of the energy shape \revision{of background events} in the maximum hemisphere and minimum hemisphere, respectively, with regarding to different MVA selection criteria. 
It is shown that the distribution of the maximum hemisphere energy is stable against MVA selections in both categories, while the minimum hemisphere energy exhibits a dependence on the BDT1 selection. 
Therefore the energy in the maximum hemisphere is chosen as the variable for final fits.

\revision{
\section{Background shape and yield variations from $B\to DD$ decays}\label{app:BDD}
As discussed in Section~\ref{sec:syst}, the background efficiency estimate depends on a set of samples, which includes $B \to DD$ processes. The inclusive background sample is also generated assuming PDG values~\cite{PDG} for the $B \to DD$ branching ratios.
To date, such branching ratios are not measured to very high precision.
Here we apply an uncertainty of 30\% to the $B \to DD$ branching ratios and check for the consequent impacts on the signal sensitivity. }

\begin{figure}[h!]
    \centering
    \includegraphics[width=0.42\textwidth]{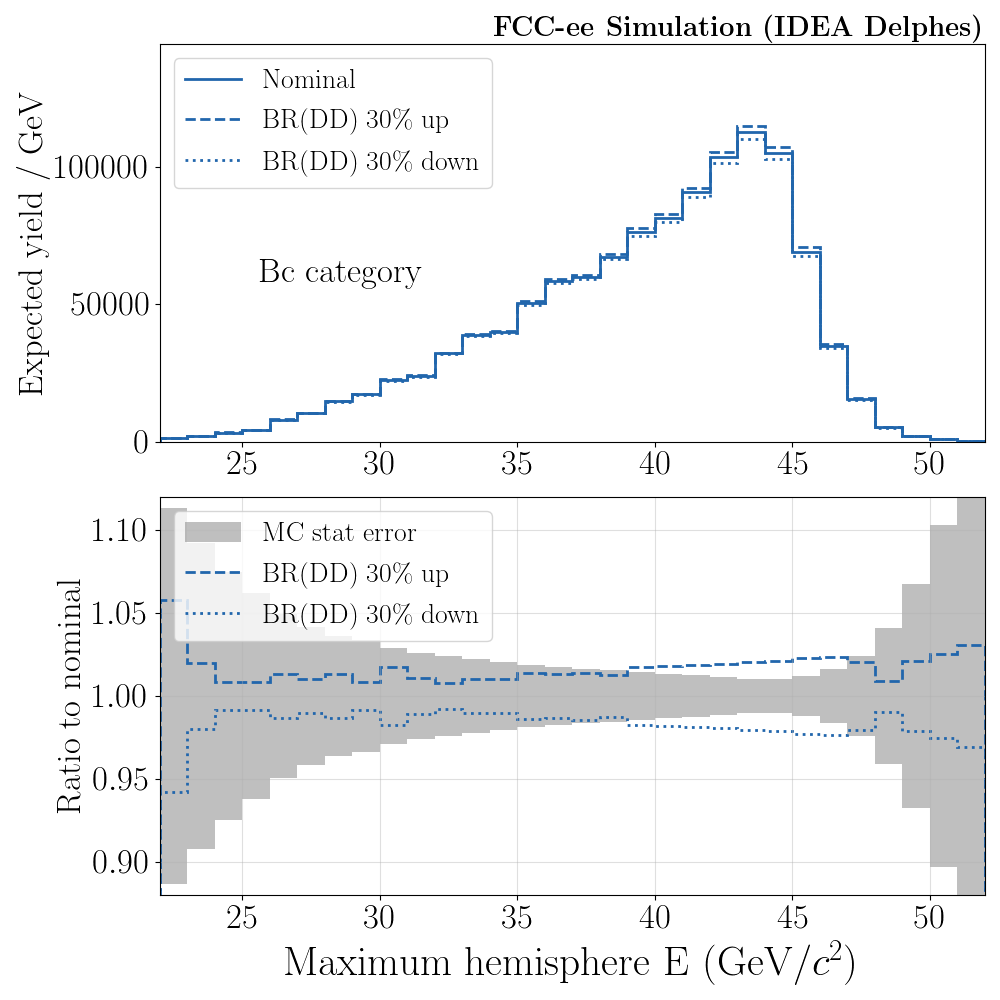}
    \includegraphics[width=0.42\textwidth]{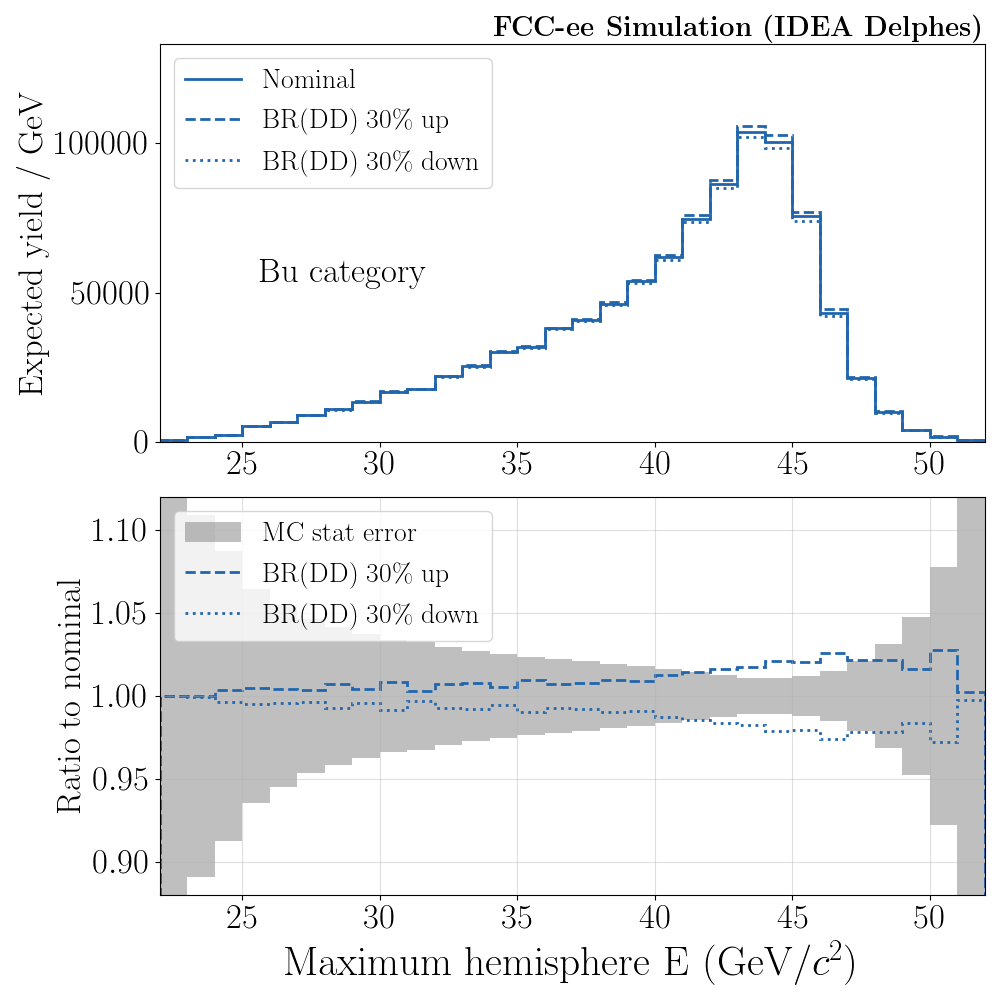}
    \caption{\revision{Systematic variations in background shapes from the $B \to DD$ branching ratio uncertainty. Left plot for the background shape in Bc category and right plot for Bu category. The grey band is the statistical uncertainty based on the size of the available simulation samples. Note that these plots are made with exclusive background samples for illustration purposes.}}
    \label{fig:BDD_shape}
\end{figure}

\revision{
Figure~\ref{fig:BDD_shape} shows the background shape variation from this uncertainty. The variation is in general less than 5\%, as the $B \to DD$ processes only make a small fraction of backgrounds after baseline selections. The up/down shifts in the ratio panel are taken as a background shape variation in the final fit. This leads to a change in signal sensitivities that is only visible in the third or fourth significant digit, and is therefore deemed negligible.
}

\begin{figure}[h!]
    \centering
    \includegraphics[width=0.3\textwidth]{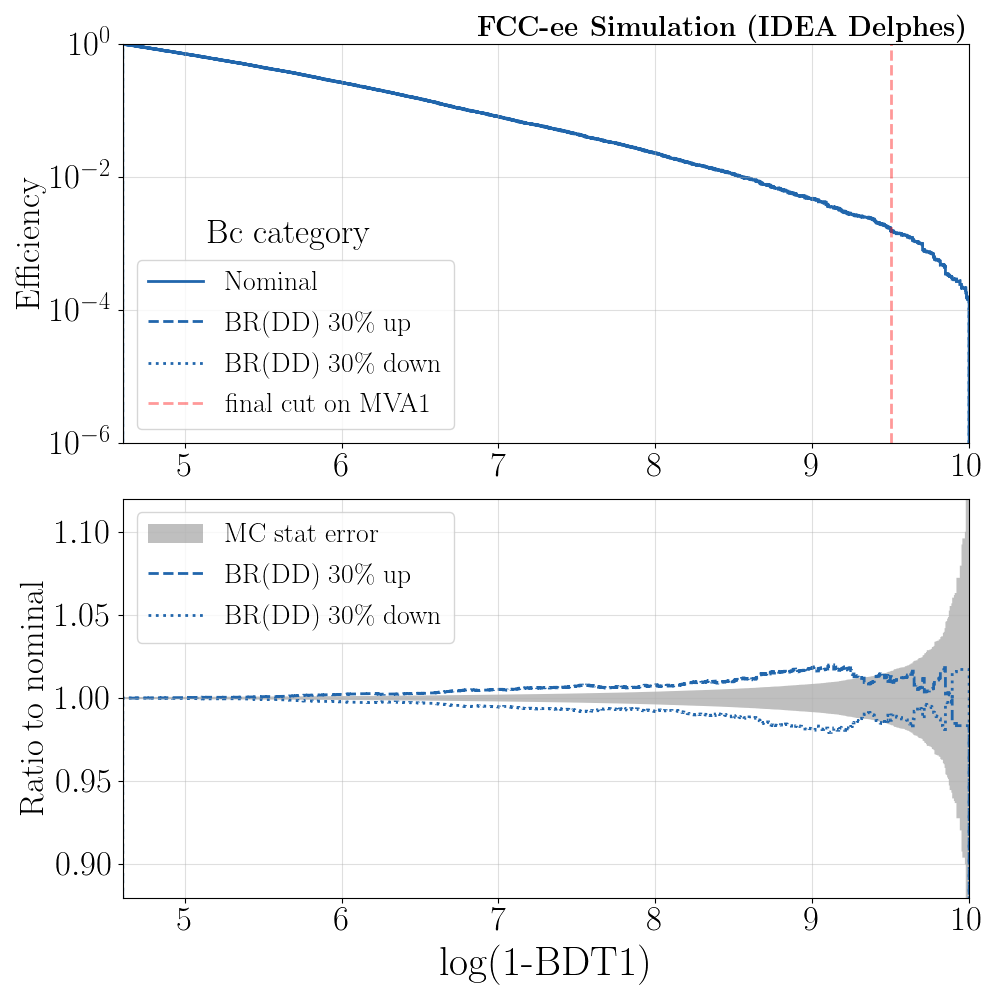}
    \includegraphics[width=0.3\textwidth]{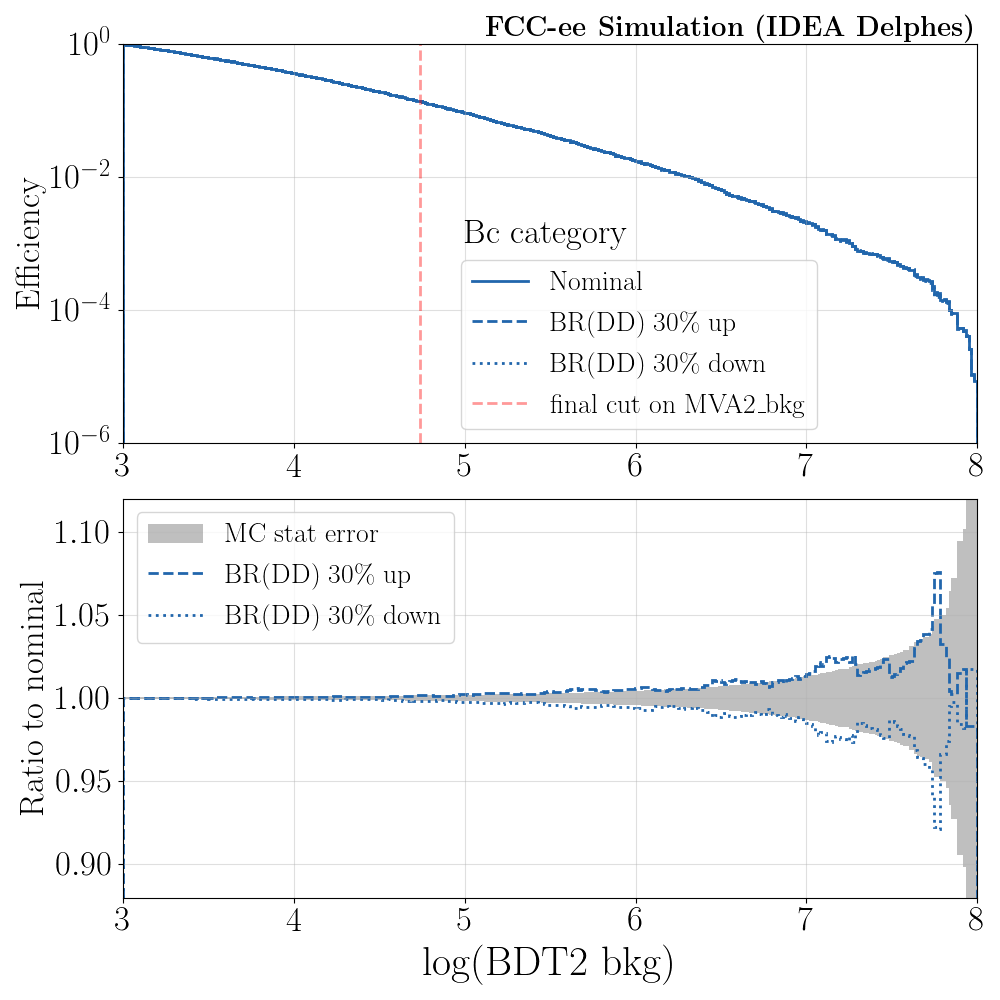}
    \includegraphics[width=0.3\textwidth]{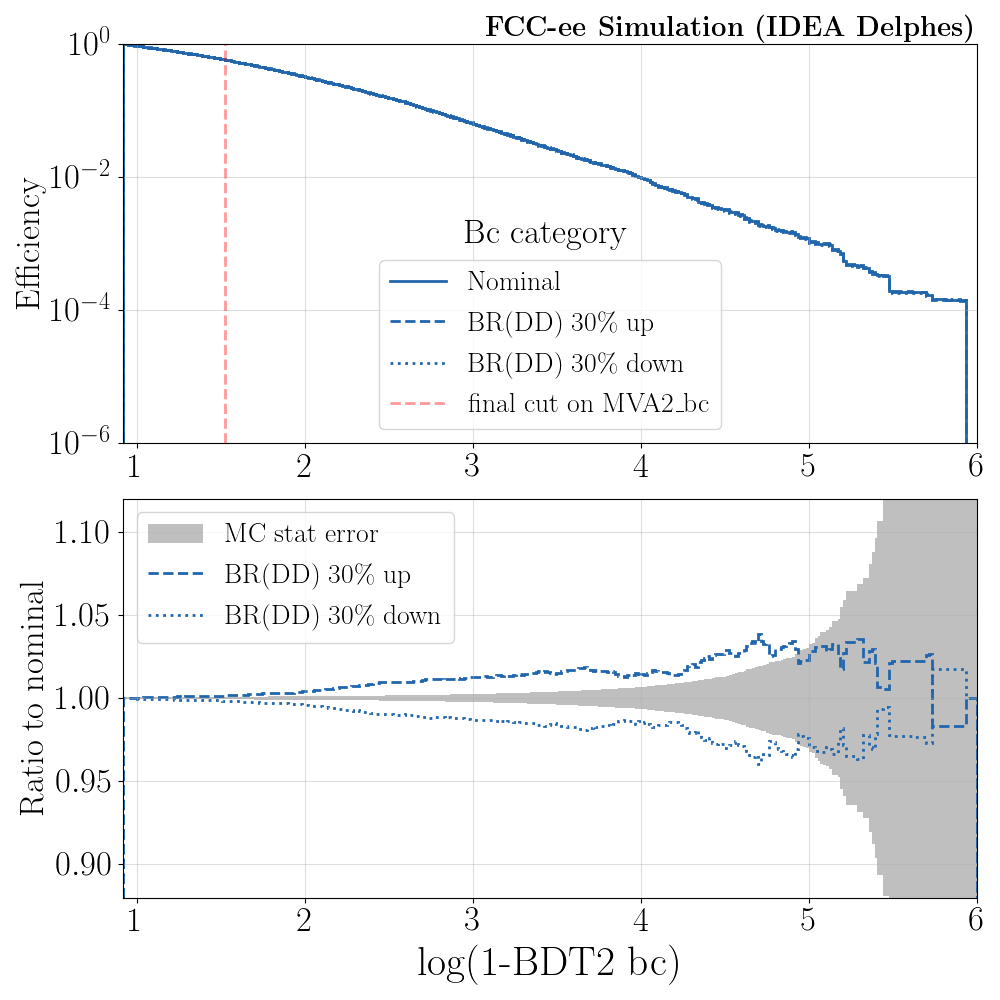} 
    \includegraphics[width=0.3\textwidth]{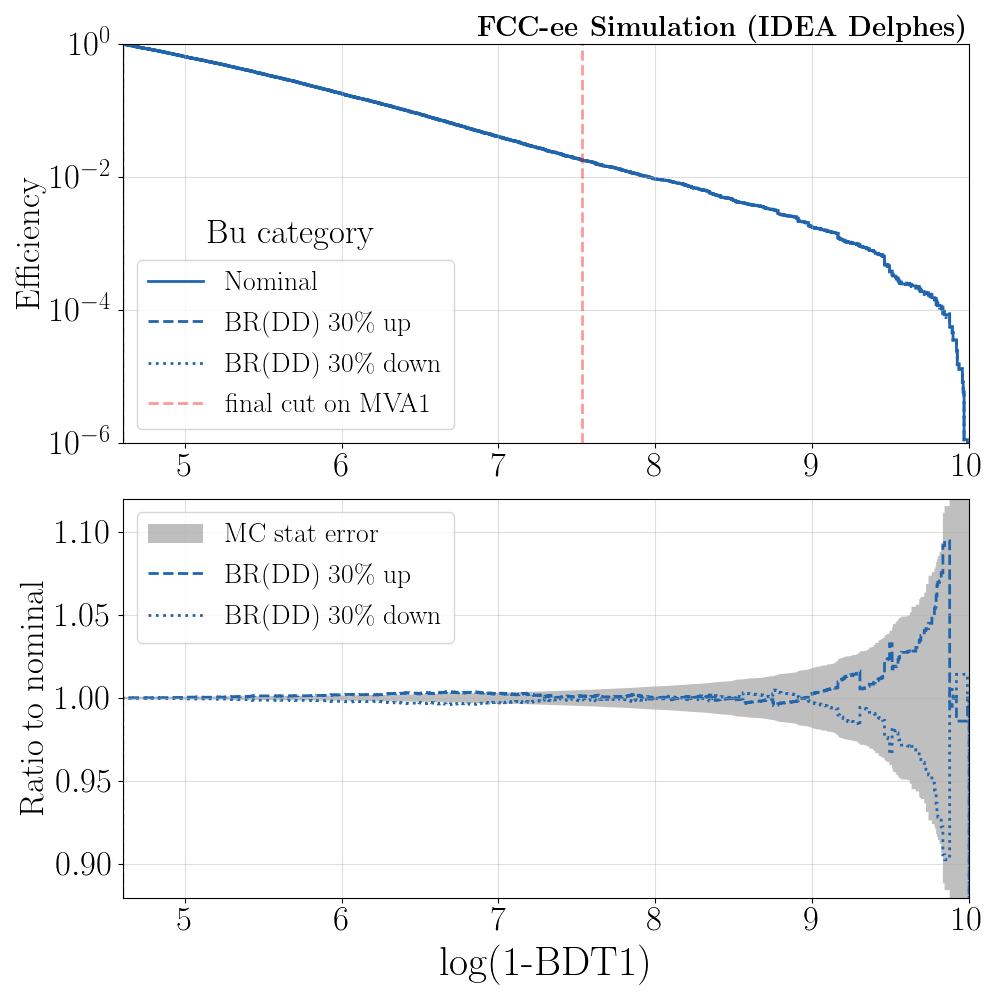}
    \includegraphics[width=0.3\textwidth]{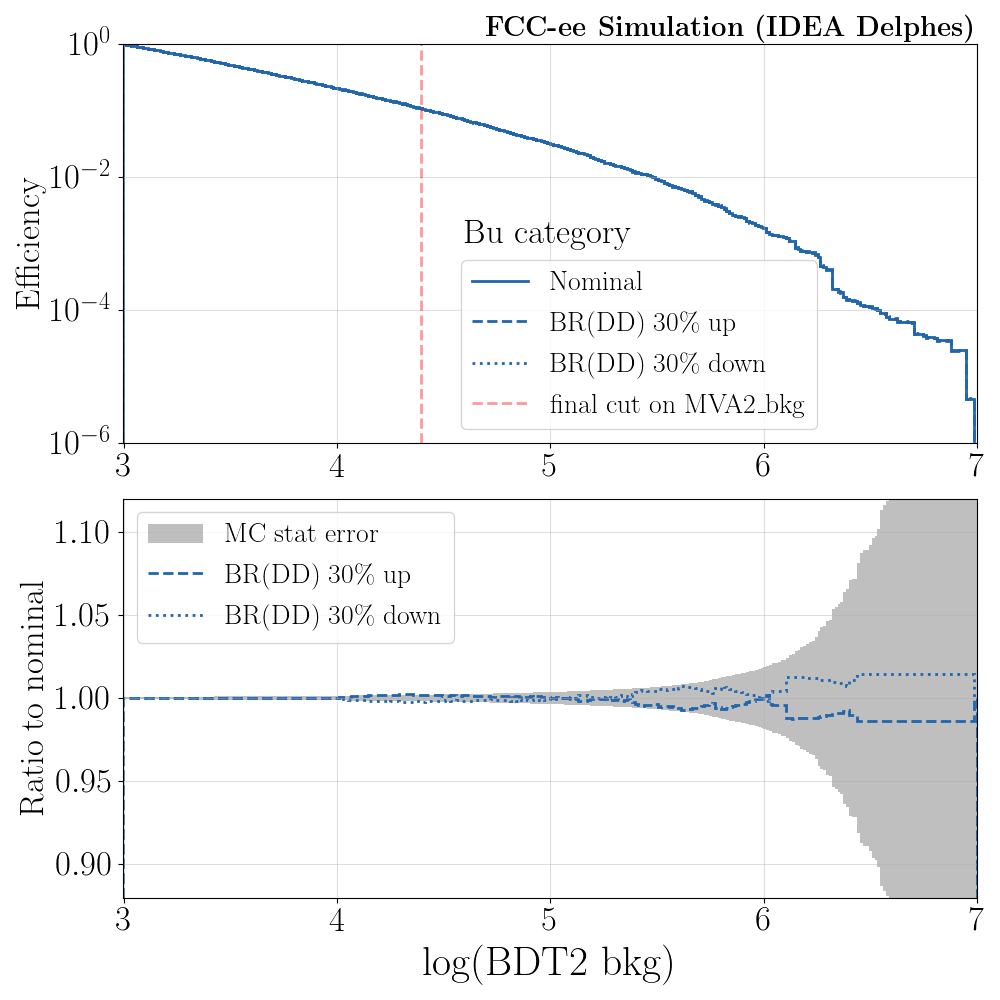}
    \includegraphics[width=0.3\textwidth]{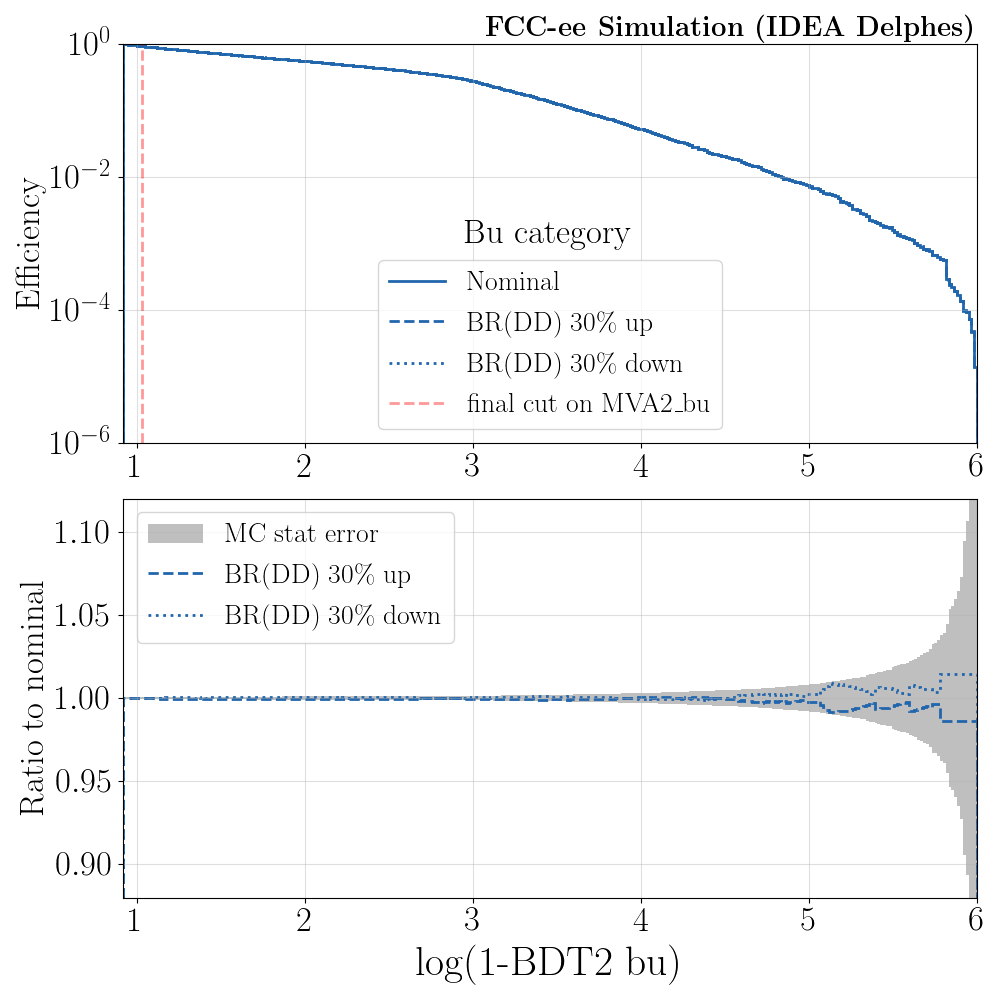} 
    \caption{\revision{Systematic variations in background efficiencies from the $B \to DD$ branching ratio uncertainty. 
    Each plot is a efficiency scan for different BDT selections. Left column for BDT1, middle column for BDT2\_bkg, and right column for BDT2\_sig. The top row is for Bc category and bottom row for Bu category. The grey band is the statistical uncertainty based on the size of the available simulation samples.}}
    \label{fig:BDD_yield}
\end{figure}

\revision{Figure~\ref{fig:BDD_yield} shows the background efficiency variation from the $B\to DD$ variation. It is in general small and within statistical uncertainty. This variation does not have any impact on the signal sensitivity as the background yields are free parameters in the fit.}

\FloatBarrier